\documentclass[pdflatex,sn-mathphys]{sn-jnl}

\usepackage{tikz}
\usetikzlibrary{positioning,decorations.pathreplacing,calc,arrows.meta}
\tikzset{
  myarrow/.style={
    -{To[line width=0.6pt, length=1.1mm,width=1.8mm]}
  }
}

\newcommand{\randest}[5]
{
\tikz[baseline=(k.base)]{
\node (p) {$#1$};
\node (k) [right=10mm of p] {$#2$};
\node (A) [right=5mm of k] {$#3$};
\node (x) [right=10mm of A] {$#4$};
\draw[myarrow] (p) -- node[above] {\text{\footnotesize rnd}} (k);
\draw[myarrow] (k) -- (A);
\draw[myarrow] (A) -- node[above] {\text{\footnotesize meas.}} (x);
\coordinate (a) at ($(k.south)+(0,-2.5ex)$);
\node (kk) at (a -| x) {$k$};
\draw[myarrow] (k.south) -- (a) -- (kk.west);
\coordinate (brtop) at ($(x.east)+(1mm,0)$);
\coordinate (brbot) at (brtop |- kk.east);
\draw[
  decorate,
  line width=0.5pt,
  decoration={brace,amplitude=4pt}
]
      ($(brtop)+(0,1mm)$)
   -- ($(brbot)+(0,-1mm)$);
\coordinate (brmid) at ($(brtop)!0.5!(brbot)$);
\node (xi) [right=9mm of brmid] {
$#5$};
\draw[myarrow] ($(brmid)+(3mm,0)$) -- (xi.west);
}}

\newcommand{\randmeas}[4]
{
\tikz[baseline=(k.base)]{
\node (p) {$#1$};
\node (k) [right=8mm of p] {$#2$};
\node (A) [right=5mm of k] {$#3$};
\node (x) [right=10mm of A] {$#4$};
\draw[myarrow] (p) -- node[above] {\text{\footnotesize rnd}} (k);
\draw[myarrow] (k) -- (A);
\draw[myarrow] (A) -- node[above] {\text{\footnotesize meas.}} (x);
\coordinate (a) at ($(k.south)+(0,-2.5ex)$);
\node (kk) at (a -| x) {$k$};
\draw[myarrow] (k.south) -- (a) -- (kk.west);
}}

\usepackage{amsmath,amssymb,amsfonts}%
\usepackage{mathtools}
\mathtoolsset{showonlyrefs=true/fault}
\usepackage{lineno} 
\usepackage{mathrsfs}
\usepackage{latexsym}
\usepackage[all]{xy}
\usepackage{bm}
\usepackage{comment}
\usepackage{amscd}
\usepackage{amsthm}

\usepackage{xcolor}
\def\red#1{#1} 
\def\rred#1{#1} 
 
\usepackage{blindtext}

\numberwithin{equation}{section} 

\theoremstyle{plain}
\newtheorem{theorem}    {Theorem}[section]
\newtheorem{corollary}  [theorem]{Corollary}
\newtheorem{lemma}      [theorem]{Lemma}
\newtheorem{proposition}        [theorem]{Proposition}

\theoremstyle{definition}

\theoremstyle{remark}
\newtheorem{remark}	[theorem]{Remark}
\newtheorem{example}    [theorem]{Example}

\newenvironment{proofs}%
    {\par\noindent{\bf Proof}\quad}{\hfill$\Box$\bigskip}
\newenvironment{proofsvar}[1]%
    {\par\noindent{\bf Proof #1}}{\hfill$\Box$\bigskip}

\newcommand{\be}{\begin{equation}}
\newcommand{\ee}{\end{equation}}
\newcommand{\noret}{\nonumber \\}

\newcommand{\any}{\forall}
\newcommand{\some}{\exists}

\newcommand{\cL}{{\mathcal L}}
\newcommand{\cLh}{{\mathcal L}_{{\rm h}}}
\newcommand{\cLhone}{{\cLh}_{,1}}
\newcommand{\cLhzero}{{\cLh}_{,0}}

\newcommand{\cU}        {{\mathcal E}}
\renewcommand{\cU} {{\mathcal U}}
\newcommand{\cH}        {{\mathcal H}}
\newcommand{\cP}        {{\mathcal P}}
\newcommand{\cV} {{\mathcal V}}
\newcommand{\cA} {{\mathcal A}}
\newcommand{\cB} {{\mathcal B}}
\newcommand{\sSB} {\sS^{\cB}}
\newcommand{\cE}{{\mathcal E}}
\newcommand{\cD}{{\mathcal D}}
\newcommand{\cR}{{\mathcal R}}
\newcommand{\cQ}{{\mathcal Q}}

\newcommand{\mS}{S} 
\newcommand{\mM}{M}  
\newcommand{\mN}{N}  

\newcommand{\sS}{{\mathcal S}} 
\newcommand{\sM}{{\mathcal M}}  
\newcommand{\sN}{{\mathcal N}}  

\newcommand{\func}{{\mathcal F}} 
\newcommand{\vf}{{\mathscr X}} 
\newcommand{\para}{\Phi} 
\newcommand{\vfparae}{{\vf}_\text{e-par}}
\newcommand{\vfparanabla}{{\vf}_\text{$\nabla$-par}}

\newcommand{\torsion} {{\mathcal T}}

\newcommand{\ap}{autoparallel}
\newcommand{\apty}{autoparallelity}

\newcommand{\e}{{\rm e}}
\newcommand{\m}{{\rm m}}
\newcommand{\nablae}{\nabla^{(\e)}}
\newcommand{\nablam}{\nabla^{(\m)}}

\newcommand{\iidn}{k} 
\newcommand{\iidtensor}{\otimes \iidn}

\newcommand{\disp} {\displaystyle}

\newcommand{\Ra}	{\Rightarrow}

\newcommand{\inprod}[2]{\left\langle #1 , #2 \right\rangle}

\newcommand{\expect}[1]{\left\langle #1 \right\rangle}

\newcommand{\bu}        {{\mbox{\boldmath$u$}}}
\newcommand{\bw}        {{\mbox{\boldmath$w$}}}
\newcommand{\transpose}[1]{{#1}{}^T}
\newcommand{\tbu}       {\transpose{\bu}}
\newcommand{\bv}        {\mbox{\boldmath$v$}}

\newcommand{\bbe}        {{\mbox{\boldmath$e$}}}

\newcommand{\uvec}{\vec{u}}
\newcommand{\vvec}{\vec{v}}
\newcommand{\xvec}{\vec{x}}

\newcommand{\rvec}{\vec{r}}
\newcommand{\svec}{\vec{s}}
\newcommand{\avec}{\vec{a}}
\newcommand{\bvec}{\vec{b}}
\newcommand{\cvec}{\vec{c}}
\newcommand{\ellvec}{\ell}
\newcommand{\sigmavec}{\vec{\sigma}}
\newcommand{\bcdot}{{\boldsymbol{\cdot}}}

\newcommand{\bR}        {\mathbb{R}}
\newcommand{\bC}        {\mathbb{C}}

\newcommand{\Tr}        {{\rm Tr}}
\newcommand{\tr}        {{\rm tr}\,}

\newcommand{\eps}	{{\varepsilon}}

\newcommand{\wgt}  {W}

\newcommand{\xihat}{\hat{\xi}}

\newcommand{\filtPi}{\vec{\Pi}}

\newcommand{\ket}[1]{\left| #1 \right\rangle}
\newcommand{\bra}[1]{\left\langle #1 \right|}

\DeclareMathOperator*{\Motimes}{\text{\raisebox{0.25ex}{\scalebox{0.8}{$\bigotimes$}}}}

\begin{document}

\title[Article Title]{Autoparallelity of Quantum Statistical Manifolds 
in Light of Quantum Estimation Theory}

\author*[1]{\fnm{Hiroshi} \sur{Nagaoka}}\email{nagaoka@is.uec.ac.jp}

\author[2, 3]{\fnm{Akio} \sur{Fujiwara}}\email{fujiwara@math.sci.osaka-u.ac.jp}

\affil*[1]{
\orgname{The University of Electro-Communications}, 
\orgaddress{\street{Chofu}, \city{Tokyo} \postcode{182-8585}, 
\country{Japan}}}
\affil[2]{\orgdiv{Department of Mathematics}, \orgname{The University of Osaka}, \orgaddress{\street{Toyonaka}, \city{Osaka}  \postcode{560-0043}, \country{Japan}}}
\affil[3]{\orgdiv{Center for Quantum Information and Quantum Biology}, \orgname{The University of Osaka}, \orgaddress{\street{Toyonaka}, \city{Osaka}  \postcode{560-0043}, \country{Japan}}}

\abstract{
In this paper we study the {\apty} w.r.t.\ the e-connection for an information-geometric structure 
called the SLD structure, 
which consists of a Riemannian metric and mutually dual e- and m-connections, 
induced on the manifold of strictly positive density operators. 
Unlike the classical information geometry, the e-connection has non-vanishing torsion, 
which brings various mathematical difficulties. 
The notion of e-{\ap} submanifolds is regarded as a quantum version of 
exponential families in classical statistics, which is known to be characterized as 
statistical models having efficient estimators (unbiased estimators 
uniformly achieving the equality in the Cram\'{e}r-Rao inequality).  
As quantum extensions of this classical result, 
we present two different forms of estimation-theoretical characterizations of the e-{\ap} 
submanifolds. 
We also give 
several results on the e-{\apty}, some of which are valid 
for the {\apty} w.r.t.\ an affine connection in a more general geometric situation.
}




\maketitle

\newpage

\section{Introduction}

\red{
Information geometry has revealed that many fundamental structures in classical statistics, most notably exponential families and efficient estimation, are naturally encoded in the geometry of the statistical manifold, with certain ``straight'' submanifolds singled out by their optimal estimation properties. In quantum statistics, the space  of density operators carries a rich family of information-geometric structures, among which the SLD (symmetric logarithmic derivative) structure is distinguished by its close connection to quantum Cram\'er--Rao bounds, but the corresponding geometric picture is far less understood than in the classical, dually flat setting. The present paper investigates how this geometric notion of ``straightness'' extends to quantum statistical models via autoparallel submanifolds with respect to the SLD structure, and how such submanifolds can be characterized by estimation-theoretic conditions weaker than global efficiency as well as by purely geometric existence conditions for a non-flat, torsionful connection.
} 

To clarify the motivation for our study, we begin with an overview of a result in the classical estimation theory. 
Let  $\cP=\cP(\Omega)$ be the totality of strictly positive probability distributions  (probability mass functions) on a finite set $\Omega$, 
and let $\sM = \{ p_\xi\,|\, \xi
= (\xi^1, \dots , \xi^n) \in\Xi\} \subset \cP$ be a statistical model whose 
elements 
$p_\xi $ are smoothly and injectively parametrized by an $n$-dimensional parameter $\xi$ ranging over 
an open subset  $\Xi$ of $\bR^n$.  As is well known, the variance matrix $V_\xi  \in \bR^{n\times n}$ of an 
arbitrary unbiased estimator for the parameter $\xi$ satisfies the Cram\'{e}r-Rao inequality $V_\xi \geq 
G_\xi^{-1}$, where $G_\xi \in \bR^{n\times n}$ denotes the Fisher information matrix. 
An unbiased estimator achieving the equality $V_\xi = G_\xi^{-1}$ for every $\xi\in \Xi$ 
is called an \emph{efficient estimator} for the parameter $\xi$. 
A statistical model does not have an efficient estimator in general. Indeed, 
the existence of efficient estimator 
 is known to impose strong restrictions on both the set $\sM$ and the parametrization 
$\xi\mapsto p_\xi$. Namely, we have the following theorem (e.g.\ 
\S~\!5a.2 (p.324) of \cite{rao},  Eq.~(7.14) in \S\!~7.2 of \cite{kiefer}, 
Theorem~3.12 of \cite{amanag}).
\begin{theorem}
\label{thm_classical_exponential_family}
For a statistical model $\sM = \{p_\xi\}$, the following conditions are equivalent.
\begin{itemize}
\item[(i)] There exists an efficient estimator for the parameter $\xi$.
\item[(ii)]
$\sM$ is an exponential family, and $\xi$ is an expectation parameter; i.e., there exist functions 
$C(\omega),  F^i(\omega), \theta_i (\xi), \psi (\xi)$ ($i=1, \ldots , n$) such that 
the elements of $\sM$ are represented as 
\be p_\xi (\omega)  = \exp [ C(\omega) + \sum_{i=1}^n \theta_i (\xi) F^i(\omega) - \psi (\xi)]
\label{c:exponential_family}
\ee
and that the parameter $\xi$ satisfies 
\be \xi^i = E_\xi (F^i), 
\label{c:expectation_coordinates}
\ee
where $E_\xi $ denotes the expectation w.r.t.\ the distribution $p_\xi$. 
 \end{itemize}
\end{theorem}

\begin{remark}
\label{remark_index_positions} 
In the notation of  \cite{amanag} as well as of many references in information geometry, 
$\theta_i$, $F^i$ and $\xi^i$  in \eqref{c:exponential_family} and \eqref{c:expectation_coordinates} 
are expressed as $\theta^i$, $F_i$ and $\eta_i$, respectively.  The reason why the upper and lower indices  
are reversed here (and throughout the paper) 
 is that we first treat $\xi$ as an arbitrary coordinate system, 
which has an upper index (superscript) as in the standard notation of differential geometry, 
and then consider the condition for $\xi$ to become an m-affine coordinate system such as 
the expectation coordinate system. 
\end{remark}

What is important about Theorem~\!\ref{thm_classical_exponential_family} for the main concern of this paper is that condition (ii) is expressed in the language of geometry as follows: 
\begin{itemize}
\item[(ii)'] $\sM$ is an e-{\ap} submanifold of $\cP$, and $\xi$ forms an m-affine coordinate system of $\sM$. 
\end{itemize}
We refer the reader to Theorem~3.12 of \cite{amanag} and the later sections of this paper for details. For the present, it suffices to understand only the following points.
\begin{itemize}
\item The {\apty} is a multi-dimensional version (including the $1$-dimensional case in particular) of the notion of geodesic for a manifold equipped with an affine connection, which expresses the condition that a submanifold is ``straight'' according to the criterion defined by the connection.
\item The e-{\apty} means the  {\apty} w.r.t.\ the e-connection of $\cP$ denoted by $\nablae$,  which is one of the most important 
differential-geometric structures in information geometry.
\item  The manifold $\cP$ has two other important structures: an affine connection called the m-connection $\nablam$ and 
a Riemannian metric called the Fisher metric $g$. The triple $(g, \nablae, \nablam)$ makes $\cP$ into what is called a dually flat space. 
\item The parameter $\xi$ is said to be m-affine when it is regarded as an affine coordinate system of $\sM$ w.r.t.\ the affine structure induced from the m-connection of $\sM$ that is induced from $\nablam$ and $g$. 
\end{itemize}
The equivalence (i) $\Leftrightarrow$ (ii)' is regarded as one of the most fundamental results in information geometry, revealing a beautiful harmony between estimation theory and geometry.

Let us now turn to the quantum case and observe how the harmony that prevailed in the classical case breaks down. 
We first note that 
a quantum version of Theorem~\!\ref{thm_classical_exponential_family} 
is actually known, which is stated below. 
Let $\cH$ be a finite-dimensional Hilbert space, 
$\sS = \sS(\cH)$ be the totality of strictly positive density operators on $\cH$, 
and $\sM=\{\rho_\xi\;|\;\xi\in\Xi\}$ be an 
arbitrary quantum statistical model consisting of  states
$\rho_\xi$ in $\sS$. 
It is well known that the variance matrix $V_\xi$ of 
an arbitrary unbiased estimator for the parameter $\xi$ satisfies the 
SLD Cram\'{e}r-Rao inequality $V_\xi \geq G_\xi^{-1}$
\cite{Helstrom:1967, Helstrom:1976}, 
where $G_\xi$  is the SLD Fisher information matrix 
(see Section~\ref{sec_efficient_estimator} for details).  
When an unbiased estimator satisfies $V_\xi = G_\xi^{-1}$ for 
every $\xi\in \Xi$, we call it an efficient estimator for 
the parameter $\xi$ as in the classical case.  
Then we have the following theorem (Theorem~7.6 of \cite{amanag}). 

\begin{theorem}
\label{thm:commutative}
For a quantum statistical model $\sM = \{\rho_\xi\}$, the following conditions are equivalent.
\begin{itemize}
\item[(i)] 
There exists an efficient estimator for the parameter $\xi$.
\item[(ii)] There exist mutually commuting Hermitian operators $F^1, \ldots , F^n$ 
and a strictly positive operator $P$ such that the elements of $\sM$ are 
represented as
\be
\rho_\xi 
=
\exp\Bigl[\, \frac{1}{2}
\Bigl\{\sum_{i=1}^n \theta_i(\xi) F^i - \psi(\xi)\Bigr\}
\Bigr]
\, P  \, 
\exp\Bigl[\, \frac{1}{2}
\Bigl\{\sum_{i=1}^n \theta_i(\xi) F^i - \psi(\xi)\Bigr\}
\Bigr]
\label{q_c:e-family}
\ee
and that 
\be
\forall \xi\in\Xi, \; \forall i, \quad  \xi^i = \Tr\left(\rho_\xi\, F^i\right). 
\label{q_c:expectation_coordinates}
\ee
(When a model $\sM = \{\rho_\xi\}$ is represented as 
\eqref{q_c:e-family},  we call it a \emph{quasi-classical exponential family}. )
\end{itemize}
\end{theorem}

Compared with Theorem~\!\ref{thm_classical_exponential_family}, this result is insufficient in several points 
except when $n=1$.

\red{
From an estimation-theoretical point of view, 
although the primary interest in the theory of multi-parameter quantum estimation lies in 
the mathematical difficulties arising from the noncommutativity of operators, which often limit the significance of 
the SLD Cram\'{e}r-Rao inequality itself, 
the above definition of efficient estimator considers only the highly special case in which these difficulties do not arise. 
} 

\red{
From a geometric point of view, on the other hand, condition (ii) is somewhat unsatisfactory in that it cannot be expressed as a simple geometric condition like condition (ii)' in the classical case.  More specifically, in terms of  the triple  $(g, \nablae, \nablam)$ of the Fisher metric and the e, m-connections defined by the SLD structure, which is among a family of information-geometric structures  introduced on $\sS$, 
conditions (ii) implies that $\sM$ is e-{\ap} in $\sS$ and that $\xi$ forms an m-affine coordinate system, but the converse does not hold 
unless $n=1$  (see Section~7.4 of \cite{amanag} and Section~\ref{sec_efficient_estimator} of this paper). 
Namely,  a quasi-classical exponential family is only a special case of an e-autoparallel submanifold, which means that  the existence of an efficient estimator is too strong as a characterization of the e-{\apty}. 
} 

\red{
Noting that Theorem~\ref{thm:commutative} may be extended in several ways, 
we formulate in this paper the problem of determining an estimation-theoretic condition, which should be a weaker version of condition (i), 
 equivalent to the geometric condition obtained by 
replacing (ii) with one of the same form as condition (ii)' in the classical case.  
In other words, the problem asks how the notion of efficient estimator should be extended in order for the e-{\apty} to be characterized by its existence. 
We give  two answers to the problem in Section~\!\ref{sec_efficient_filtration} and Section~\!\ref{sec_another_characterization}, which constitute the highlights of this paper. 
In addition, we provide a detailed discussion of the properties of {\ap} submanifolds and the conditions for their existence. This is because, unlike in the classical case where the e-connection is flat, 
the e-connection in the quantum case is not generally flat due to the non-vanishing torsion, making new insights into {\ap} submanifolds essential.
} 

\red{Let us outline the paper by following its organization.} 

In Section~\!\ref{sec_basic_autoparallelity}, we explain 
 the basic issues about the {\apty} for an affine connection $\nabla$ on a general differential manifold, 
 focusing in particular on 
 a curvature-free $\nabla$ with its relation to the dual connection $\nabla^*$ w.r.t.\ a Riemannian metric $g$. 
 Note that a connection is flat if and only if  it is curvature-free and torsion-free and that, 
even though both $\nabla$ and $\nabla^*$ are curvature-free in this case, they may have non-vanishing torsions and hence 
 $(g, \nabla, \nabla^*)$ may not define a dually flat structure. 
 In Section~\!\ref{sec_quantum_e-autoparallelity}, we introduce a family of information-geometric structures on 
 the space $\sS (\cH)$, where the e-connection has mostly a non-vanishing torsion, and derive the basic issues concerning the e-{\apty} by applying  
 the results of Section~\!\ref{sec_basic_autoparallelity}. 
 These two sections are preliminaries for later sections.  Although the results shown there are mostly known, 
 we present them together with their derivations so that the description is as self-contained as possible. 
 
In Section~\!\ref{sec_efficient_estimator},  preliminaries concerning quantum estimation are given. 
We give the definitions of (local-)unbiasedness and several kinds of (local-)efficiency with or without weight matrix, where the word ``local'' means the versions of these notions weakened by localizing them at a given point of the model. 
\red{
Essential for later arguments is Prop.~\!\ref{prop_inf_uVu}, which states that the equality in the scalar version of SLD Cram\'{e}r-Rao inequality 
$\transpose{\bu}V_\xi \bu 
\geq \transpose{\bu}G_\xi^{-1} \bu$ is attained as an infimum for  each point $\xi$ and each direction represented by a 
vector $\bu$. 
Although the result itself is not new (e.g. Eq.~\!(7.93) of \cite{amanag} and 
Theorem~5.2 of \cite{suzuki_nuisance_review}), considering its importance we provide it with 
a proof and a remark that the infimum is not generally replaced with the minimum even locally at each point. 
} 
We introduce a family of randomized estimators, which is  used for the proof of this proposition as well as for a main result of the paper presented in the next section. 
We revisit Theorem~\!\ref{thm:commutative} 
and clarify that the existence of efficient estimator only partially characterizes the e-{\apty} for the SLD structure. 

In Section~\!\ref{sec_efficient_filtration}, we treat a 
one-parameter family of estimators $\filtPi= (\Pi_\eps )_{\eps >0}$, which we call a \emph{filtration} of estimators, 
instead of a single estimator $\Pi$.  We show Theorem~\ref{thm:autoparallel_m-affine}, which gives the first answer to our problem, 
stating that a quantum statistical model is e-{\ap} in $\sS$ if and only if for every direction $\bu$, there exists a filtration $\filtPi= (\Pi_\eps )_{\eps >0}$ such that $\Pi_\eps$ is an unbiased estimator for every $\eps >0$ and that the equality in $\transpose{\bu}V_\xi \bu 
\geq \transpose{\bu}G_\xi^{-1} \bu$ is attained for every point $\xi$ as the limit of $\eps \downarrow 0$.  
(The condition imposed on the parametrization is omitted here for \rred{brevity.}) 
This theorem can be regarded as an extension of 
Theorem~\ref{thm:commutative} since an efficient estimator $\Pi$ defines such a filtration by  $\Pi_\eps = \Pi$ ($\any \eps >0$), independently of $\bu$. In Section~\!\ref{sec_gaussian}, it is shown that 
a quantum Gaussian shift model has such a filtration for every direction $\bu$ and that the model in fact exhibits an analogous property to the e-{\apty} in $\sS (\cH)$, although the Hilbert space $\cH$ is infinite-dimensional in this case, so that 
we cannot fully develop a differential geometric argument there. 

\red{While  our concern  in the previous sections lies in  estimation of a multi-dimensional parameter $\xi = (\xi^i)$ that injectively represents the states $\rho_\xi$ in a model even though the estimation error is measured only in one direction expressed by $\bu$, Section~\!\ref{sec_another_characterization} studies estimation of a scalar-valued function on the model, for which the efficiency of an estiator is defined in a similar way. 
This includes estimation of a single component $\xi^i$ of the parameter $\xi$ as a special case, which can be considered equivalent to a problem of estimation with nuisance parameters (e.g.\ \cite{suzuki_nuisance_review}). } 
 Theorems~\ref{thm_another_characterization_1} and \ref{thm_another_characterization_2} give the second answer to our question, where it is shown that an $n$-dimensional quantum statistical model is e-{\ap} in $\sS$ \rred{if and only if  
 every component} $\xi^i$ of the parameter $\xi$ has an efficient estimator.  
Since an efficient estimator $\Pi$ for $\xi$ induces an efficient estimator for each component $\xi^i$ in a trivial way, this result can also be regarded as an extension of Theorem~\ref{thm:commutative}, although 
 the set of efficient estimators for components induced from $\Pi$ is  not conceptually identified with $\Pi$ itself. 
In contrast to the discussion in Section~\!\ref{sec_efficient_filtration},  
the approach of this section does not involve the technical difficulties inherent in estimation of a multi-dimensional parameter of quantum states.  
We show that some of the estimation-theoretic results obtained in this section can be refined within a purely differential-geometric framework, 
where the use of differential forms proves to be particularly useful. 

In Section~\!\ref{sec_integrability}, we move away from estimation theory and address 
the condition for existence of e-{\ap} submanifolds from a purely geometrical point of view.
It should be noted that the e-connection in the SLD structure 
is curvature-free but is not flat due to the non-vanishing torsion. 
In the case of a flat connection, an {\ap} submanifold corresponds to an affine subspace in 
the coordinate space of an affine coordinate system, 
so that the existence condition for  {\ap} submanifolds is trivial. 
For a non-flat connection, on the other hand, we cannot see the whole picture of  
 {\ap} submanifolds  in general. 
 and are faced with the new problem of what kind of condition ensures the existence of {\ap} submanifold.  
 At present, our understanding of this problem remains quite limited even for a curvature-free connection. 
 We focus on the fact that an {\ap} submanifold is an integral manifold of a parallel distribution of tangent vectors, 
 and investigate the involutivity of parallel distributions in relation to the torsion tensor. 
 \red{
 We show that, even when $\nabla$ has a non-vanishing torsion, there is a case in which there exists a pair of dual foliations 
 consisting of mutually orthogonal $\nabla$-{\ap} submanifolds and $\nabla^*$-{\ap} submanifolds, as in the case of dually flat spaces. } 

In Section~\!\ref{sec_qubit}, we treat the case when  $\dim \cH =2$ 
and study the SLD structure of the space $\sS (\cH)$ of qubit states in detail. 
We show that  $\sS (\cH)$ in this case has a characteristic property that 
every e-parallel distribution is involutive, 
which means that the existence condition for {\ap} submanifolds is trivial as in the classical case.  
\red{
We also give several remarks on our randomized estimator for the qubit case, 
where its relation to some related results in quantum estimation theory is clarified. } 

Section~\!\ref{sec_conclusion} is devoted to concluding remarks.  Some proofs and additional results are included in Appendices for the sake of readability of the main text. 

\medskip

\begin{remark}
\label{remark_nomenclature}
We make some remarks on the nomenclature and the notation of the paper.
\begin{enumerate}
\item 
Throughout the paper, when we refer to a manifold, say $\mM$, it means that 
$\mM$ is a manifold with a trivial global structure, 
so that we need not worry about the difference between 
global properties and local properties of $\mM$. 
For instance, 
$\mM$ is always supposed to have a global coordinate system, 
and every closed differential form on $\mM$ is considered to be exact. 
\item 
When we say that $\xi= (\xi^1, \ldots , \xi^n)$ is a coordinate system of 
a manifold $\mM$ in the subsequent sections, 
it basically means that $\xi$ is a map $: \mM \rightarrow \bR^n$ 
(a global chart of $\mM$) which 
represents each point $p\in \mM$ by an $n$-dimensional vector 
$\xi(p)=  (\xi^1(p), \ldots , \xi^n(p))\in \bR^n$, 
although the same symbol $\xi$  has appeared above 
as a parameter to specify a point in the manifold. 
They are equivalent by $\xi (p) = \xi' \Leftrightarrow p = p_{\xi'}$.
A parametrization is often more convenient than a coordinate system when dealing with concrete examples. 
In fact, we will use parametrizations in Section~\ref{sec_gaussian} for quantum Gaussian states and 
in Section~\ref{sec_qubit} for qubit states. 
\item This paper contains both arguments on quantum statistical manifolds (manifolds consisting of density operators) 
and those on general manifolds. 
We denote quantum statistical manifolds by 
$\sS, \sM, \sN, \dots$, while general manifolds are denoted by 
$\mS, \mM, \mN, \dots$. 
\end{enumerate}
\end{remark}

\section{Basic issues about {\apty}}
\label{sec_basic_autoparallelity}

In this section we summarize basic issues related to {\apty} from the perspective of general differential geometry, which will be necessary for later discussions.

Let $\mS$ be an arbitrary manifold, and denote the totality of smooth functions and that of
smooth vector fields on $\mS$ by $\func(\mS)$ and $\vf (\mS)$, respectively. 
Suppose that $\mS$ is provided with an affine connection $\nabla$, which is 
a map $\vf(\mS) \times \vf(\mS) \rightarrow \vf (\mS), \; (X, Y) \mapsto \nabla_X Y$.
Given a submanifold $\mM$ of $\mS$, let $\vf (\mS / \mM)$ denote 
the totality of 
smooth mappings which map each point $p\in \mM$ to a 
tangent vector in $T_p (\mS)$, i.e., sections of the vector bundle $\bigsqcup_{p\in \mM} T_p (\mS)$. 
Then $\nabla$ naturally induces a map $\vf (\mM) \times\vf (\mS / \mM) \rightarrow \vf (\mS / \mM)$ 
so that for any $X\in \vf (\mM)$ and any $Y \in \vf (\mS / \mM)$, $\nabla_X Y$ is defined as 
an element of $\vf (\mS / \mM)$. Since $\vf (\mM) = \vf (\mM / \mM) \subset  \vf (\mS / \mM)$, 
$\nabla_XY \in  \vf (\mS / \mM)$ is defined for any $X, Y \in \vf (\mM)$, although it does not necessarily belong to $\vf (\mM)$.  

When $\nabla_X Y$ belongs to $\vf (\mM)$ for every $X, Y \in \vf (\mM)$, $\mM$ is said to be \emph{\ap} 
w.r.t.~\!$\nabla$ or \emph{$\nabla$-{\ap}} in $\mS$ (e.g., Sec.~\!8 in Chap.~\!VII of \cite{KobayashiNomizu}). 
In particular, $\mS$ itself is $\nabla$-{\ap} in $\mS$. An autoparallel curve is usually called a \emph{geodesic}  
(or \emph{pregeodesic} when we wish to clarify that our interest lies only in the image of the curve), 
so that the autoparallelity is a multi-dimensional extension of the notion of geodesic. 
When $\mM$ is $\nabla$-{\ap} in $\mS$, $\nabla$ defines an affine connection on $\mM$.  We denote this connection 
by $\nabla|_\mM$ when we wish to distinguish it from the original connection $\nabla$ on $\mS$. 
The autoparallelity is transitive in the sense that if $\mM$ is $\nabla$-{\ap} in $\mS$ and $\mN$ is $\nabla|_{\mM}$-{\ap} in 
$\mM$, then $\mN$ is $\nabla$-{\ap} in $\mS$. 

\begin{remark}
If $\mM$ is $\nabla$-{\ap} in $\mS$ and $\mN$ is a nonempty open set of $\mM$, then 
$\mN$ is also $\nabla$-{\ap} in $\mS$ having the same dimension as $\mM$. In this paper, 
we restrict ourselves to maximal {\ap} submanifolds to avoid this ambiguity. 
In particular, an {\ap} submanifold of $\mS$ having the same dimension as $\mS$ 
is considered to be only $\mS$. 
This restriction is solely for simplicity of exposition. 
\end{remark}

\begin{remark}
\label{remark_totally_geodesic}
\red{A notion similar to, but different from,  {\apty} is total geodesicity.} 
A submanifold $\mM$ is said to be \emph{totally geodesic} 
w.r.t.~\!$\nabla$ or \emph{$\nabla$-totally geodesic} in $\mS$ when 
for any point $p \in \mM$ and any tangent vector $X_p\in T_p (\mM)$ of 
$\mM$, the $\nabla$-geodesic passing through $p$ in direction 
$X_p$ lies in $\mM$.  It is obvious that the {\apty} implies the  total \rred{geodesicity,}  
but the converse is not true in general except when $\nabla$ is torsion-free 
(\cite{KobayashiNomizu}, Theorem~\!8.4 in Chap.~\!VII).
We will revisit this topic in Remark~\!\ref{remark_totally_geodesic_revisited}. 
\end{remark}

A vector field $X\in \vf (\mS)$ is said to be \emph{parallel w.r.t.\ $\nabla$}  
or \emph{$\nabla$-parallel} when $\any Y\in \vf (\mS), \,\nabla_Y X = 0$.
More generally, $X\in \vf (\mS/\mM)$ (including the case $X\in \vf (\mM)$) 
is said to be $\nabla$-parallel when $\any Y\in \vf (\mM), \,\nabla_Y X = 0$.

When there exist on $\mS$ the same number of  linearly independent $\nabla$-parallel vector fields as $\dim \mS$, 
we say that $(\mS, \nabla)$ is \emph{curvature-free}.  This condition is known to be equivalent to 
the curvature tensor of $\nabla$ vanishing on $\mS$ (: recall that $\mS$ is assumed to be globally trivial). 
When  $(\mS, \nabla)$ is curvature-free, 
the parallel transport $\para^{(\nabla)}_{p,q} : T_p (\mS) \rightarrow T_q (\mS)$ 
is defined for arbitrary two points $p, q\in \mS$ so that a vector field $X\in \vf (\mS)$ is $\nabla$-parallel 
iff 
$ \any p, q\in \mS, \; \para^{(\nabla)}_{p,q} (X_p) = X_q$. 

\medskip

The following two propositions are straightforward, where the curvature-freeness is essential.

\begin{proposition}
\label{prop_parallel_X(S/M)}
Let  $\mS$ be a manifold on which a curvature-free connection $\nabla$ is given and  $\mM$ 
be a submanifold of $\mS$. 
For $X\in \vf (\mS / \mM)$ (including the case when $X\in \vf (\mM)$), 
the following conditions are equivalent. 
\begin{itemize}
\item[(i)] $X$ is $\nabla$-parallel.
\item[(ii)] $\some \tilde{X} \in \vf (\mS)$, $\tilde{X}$ is $\nabla$-parallel and $X = \tilde{X}|_{\mM}$. 
\end{itemize}
\end{proposition}

\begin{proposition} 
\label{prop_equiv_autoparallel_general}
Let  $\mS$ be a manifold on which a curvature-free connection $\nabla$ is given and $\mM$ 
be an $n$-dimensional submanifold of $\mS$. Then 
the following conditions are equivalent. 
\begin{itemize}
\item[(i)] $\mM$ is $\nabla$-autoparallel in $\mS$.
\item[(ii)] $\any p, q\in\mM, \; \para_{p,q}^{(\nabla)} (T_p(\mM)) = T_q (\mM)$.
\item[(iii)] There exist $n$ linearly independent $\nabla$-parallel vector fields on $\mM$.
\item[(iv)] There exist $n$ linearly independent $\nabla$-parallel vector fields 
$\tilde{X}^1 , \ldots , \tilde{X}^n$ on $\mS$ such that 
$\any i$, $\tilde{X^i}|_{\mM} \in \vf (\sM)$. 
\end{itemize}
When these conditions hold, a vector field $X$ on $\mM$ is $\nabla|_{\mM}$-parallel iff it is 
$\nabla$-parallel, and $\nabla|_{\mM}$ is curvature-free. 
\end{proposition}

In the following, we consider the case when 
$(\mS, \nabla)$ is additionally provided with a Riemannian metric $g$ for which the dual connection $\nabla^*$ of $\nabla$ 
is defined. 
Namely, the triple $(g, \nabla, \nabla^*)$ satisfies the duality \cite{amanag} \cite{metr} : 
\be
\any X, Y, Z\in \vf (\mS), \; X g (Y, Z) = g(\nabla_X Y, Z) + g(Y, \nabla_X^* Z). 
\label{duality_on_S}
\ee
As is well known \red{(see Theorems~3.2 and 3.3 of \cite{amanag})}, 
 the curvature-freeness of $\nabla$ and that of $\nabla^*$ are equivalent, 
\red{
and when they are curvature-free, their parallel transports satisfy 
\be
\any X_p, Y_p\in T_p(\mS), \; g_p( X_p, Y_p) = g_q (\para^{(\nabla)}_{p,q} (X_p), \para^{(\nabla^*)}_{p,q} (Y_p)). 
\label{duality_parallel_transport}
\ee
} 

\begin{proposition}
\label{prop_parallel_nabla_nablastar}
When $\nabla$ and $\nabla^*$ are curvature-free in $(\mS, g, \nabla, \nabla^*)$, 
we have:
\begin{itemize}
\item[(1)] For a vector field $X\in \vf(\mS)$, $X$ is $\nabla$-parallel iff  $g(X, Y)$ is constant on $\mS$ 
for every $\nabla^*$-parallel vector field  $Y\in \vf(\mS)$. 
\item[(2)] 
For a vector field $X\in \vf(\mS)$, $X$ is $\nabla^*$-parallel iff  $g(X, Y)$ is constant on $\mS$ 
for every $\nabla$-parallel vector field  $Y\in \vf(\mS)$. 
\end{itemize}
\end{proposition}

\begin{proofs}
It suffices to show (1). 
For an $X\in \vf(\mS)$, we have
\begin{align}
\text{ $X$ is $\nabla$-parallel} 
\; &\Leftrightarrow \; \any Y, \any Z, \; g (\nabla_Z X , Y) =0 
\noret 
& \stackrel{\text{a}}{\Leftrightarrow} 
 \any Y: \text{$\nabla^*$-parallel}, \any Z, \;  g (\nabla_Z X , Y) =0
 \noret 
& \stackrel{\text{b}}{\Leftrightarrow}  \; \any Y: \text{$\nabla^*$-parallel}, \any Z, \; 
Z g (X, Y) = 0
\noret 
& \Leftrightarrow \; \any Y: \text{$\nabla^*$-parallel}, \; 
\text{$g (X, Y)$ is constant},
\end{align}
where $\Leftarrow$ in $\stackrel{\text{a}}{\Leftrightarrow}$  follows since 
the set of $\nabla^*$-parallel vector fields has the same dimension as $\dim \mS$ 
due to the curvature-freeness of $\nabla^*$, and $\stackrel{\text{b}}{\Leftrightarrow}$ 
follows since the duality \eqref{duality_on_S} and the $\nabla^*$-parallelity of $Y$ imply 
\be
Z g (X, Y) = g (\nabla_Z X , Y) + g (X , \nabla^*_Z Y) = g (\nabla_Z X , Y).
\ee 
\end{proofs}

In order to state the next proposition, which will be of fundamental importance for later arguments, 
we need some preliminaries. First, 
suppose that $\mM$ is a $\nabla$-autoparallel submanifold of $\mS$, so that $\mM$ is a Riemannian 
submanifold of $(\mS, g)$ equipped with the affine connection $\nabla|_\mM$. Then 
the dual connection of $\nabla|_\mM$ w.r.t.\ $g$ 
(more precisely, w.r.t.\ the induced metric $g|_{\mM}$ on $\mM$) 
is defined, 
which we denote by $\nabla^*_\mM := (\nabla|_\mM)^*$; i.e., 
\be
\any X, Y, Z\in \vf (\mM), \; X g(Y, Z) = g (\nabla_X Y, Z) + g( Y,  (\nabla^*_\mM)_X Z), 
\label{duality_on_M}
\ee
where we have applied $(\nabla|_\mM)_X Y = \nabla_X Y$.  From \eqref{duality_on_S} and \eqref{duality_on_M}, we have
\be
\any X, Y, Z \in \vf (\mM), \; g(Y,  (\nabla^*_\mM)_X Z) = g(Y, \nabla^*_X Z), 
\label{projection_nabla*}
\ee
which means that $ (\nabla^*_\mM)_X Z$ is the $g$-projection of $\nabla^*_X Z \in \vf (\mS / \mM)$ onto $\vf (\mM)$. 
\[
  \begin{CD}
     \nabla @>{\text{restriction}}>> \nabla|_{\mM} \\
  @V{\text{$g$-dual}}VV    @VV{\text{$g|_{\mM}$-dual}}V \\
     \nabla^*   @>{\text{$g$-projection}}>>  \nabla_\mM^*
  \end{CD}
\]
Next, we introduce the notions of flat connection and affine coordinate system.  An affine connection $\nabla$ is said to be \emph{flat} when it is curvature-free and torsion-free. The flatness is known to be 
equivalent to the existence of 
a (generally local) coordinate system $\xi = (\xi^i)$, which is called an \emph{affine coordinate system w.r.t.\ $\nabla$} or \emph{a $\nabla$-affine coordinate system},  
such that the vector fields $\partial_i = \frac{\partial}{\partial\xi^i}, i \in\{ 1, \ldots , \dim \mS\}$, are all $\nabla$-parallel. 

\begin{proposition}
\label{prop_autopara_affine_coordinate_general}
Suppose that $\nabla$ and $\nabla^*$ are curvature-free in $(\mS, g, \nabla, \nabla^*)$, and let $\mM$ 
be an $n$-dimensional submanifold of $\mS$ with a coordinate system $\xi = (\xi^i)$. Then the following conditions are equivalent.
\begin{itemize}
\item[(i)] $\mM$ is $\nabla$-autoparallel in $\mS$, and $\xi$ is a $\nabla^*_\mM$-affine coordinate system 
(so that $\nabla^*_\mM$ is flat). 
\item[(ii)] For every $i\in\{1, \ldots, n\}$, the vector field 
\be X^i :=
\sum_j g^{ij} \partial_j \in \vf (\mM) 
\label{X^i=g_partial}
\ee
is $\nabla$-parallel, where $\partial_i := \frac{\partial}{\partial\xi^i}$ and 
$[g^{ij}] :=[g_{ij} := g(\partial_i, \partial_j)]^{-1}$. 
\end{itemize}
\end{proposition}

\begin{proofs}
We first show  (i)  $\Rightarrow $ (ii).
Assume (i) and define $X^i\in \vf (\mM)$ by \eqref{X^i=g_partial}. 
Then we have $g(X^i, \partial_j) = \delta^i_j$ for every 
$i, j$, which is constant on $\mM$.
Noting that $n(=\dim \mM)$ vector fields 
$\{\partial_j\}$ are $\nabla^*_\mM$-parallel and that 
Prop.~\!\ref{prop_parallel_nabla_nablastar} can be applied to 
$(\mM, g|_{\mM}, \nabla|_{\mM}, \nabla^*_{\mM})$ 
due to the $\nabla$-{\apty} of $\mM$, 
it follows from item (1) of Prop.~\!\ref{prop_parallel_nabla_nablastar} that 
 $X^i$ is  $\nabla|_{\mM}$-parallel, and hence 
it is  $\nabla$-parallel (: recall the last sentence of Prop.~\!\ref{prop_equiv_autoparallel_general}). 

We next show  (ii)  $\Rightarrow $ (i).
Assume (ii). 
Then according to Prop.~\!\ref{prop_equiv_autoparallel_general}, $\mM$ is $\nabla$-autoparallel in $\mS$. 
Noting that $g(X^i, \partial_j) = \delta^i_j$ is constant on $\mM$  
 and applying item (2) of Prop.~\!\ref{prop_parallel_nabla_nablastar} to $(\mM, g|_{\mM}, \nabla|_{\mM}, \nabla^*_{\mM})$, 
we have that $\{\partial_j\}$ are $\nabla^*_\mM$-parallel, which means that $\xi$ is $\nabla^*_\mM$-affine. 
\end{proofs}

\begin{remark} 
\label{remark_nablastar_M_flat}
In applying this proposition to later arguments, $\nabla^*$ will appear as the flat connection $\nablam$. 
When $\nabla^*$ is flat, its torsion-freeness is inherited by $\nabla^*_\mM$, so that $\nabla^*_\mM$ 
turns out to be flat for every $\nabla$-{\ap} $\mM$. That is, the existence of $\nabla^*_\mM$-affine coordinate system $\xi$ 
follows from the $\nabla$-{\apty} of $\mM$ in condition~(i) in this case. 
%
\end{remark}

\section{\rred{Information-geometric} structures on 
quantum statistical manifolds
}
\label{sec_quantum_e-autoparallelity}

In this section, we introduce a family of information-geometric structures on quantum statistical manifolds and apply 
the results of the previous section to them. 
The geometric structures treated here are essentially the same as those studied in \S~\!7.3 of \cite{amanag}.

Let $\cL = \cL (\cH)$, $\cLh = \cLh (\cH)$ and $\sS = \sS (\cH)$ 
be the totality of linear operators on $\cH$, that of Hermitian operators on $\cH$ 
and that of strictly positive density operators on $\cH$, respectively. 
Then $\sS$ is an open subset of the affine space $\cLhone :=  \{ A\in \cLh\,|\, \Tr A = 1\}$, 
so that a flat affine connection is naturally introduced on $\sS$, which we call  
the {\em m-connection} and denote by $\nablam$. 
In order to express $\nablam$ more explicitly, we introduce the embedding map 
$\iota : \sS \rightarrow\cLhone$ so that $\rho\in\sS$ is denoted 
by $\iota (\rho)$ when treating it as an element of $\cLhone$. 
Since $\iota$ is a smooth map, it has 
the differential  at every point $\rho\in\sS$, which we denote by
$\iota_{*} = (d\iota)_\rho : T_\rho (\sS) \rightarrow \cLhzero :=   \{ A\in \cLh\,|\, \Tr A = 0\}$. 
For a vector field $X\in \vf (\sS)$, the map $\iota_* (X) : \sS \rightarrow \cLhzero$ is defined to be 
$\rho \mapsto \iota_* (X_\rho) = (d\iota)_\rho (X_\rho)$. Then the definition of the m-connection is represented as follows:
\be
\any X, Y\in \vf (\sS), \; 
\iota_* (\nabla^{(\m)}_X Y ) = X \iota_* (Y), 
\label{def_m-connection}
\ee
where 
$X \iota_* (Y) : \sS \rightarrow \cLhzero$ is the  derivative of $\iota_* (Y)$ w.r.t.~\!$X$.  
When a coordinate system 
$\xi = (\xi^i)$ is arbitrarily given and 
the elements of $\sS$ is parametrized by it as $\rho_\xi$, we have
\be
\iota_* \left( \partial_i \right) =  \partial_i \rho_\xi, 
\label{m-rep_partial_rho}
\ee
and 
\be
\iota_* \left( 
\nabla^{(\m)}_{\partial_i}\partial_j \right) 
= 
\partial_i \, \iota_* \left( \partial_j \right) =  
\partial_i \partial_j \rho_\xi, 
\ee
where $\partial_i := \frac{\partial}{\partial\xi^i}$. 

Suppose that we are given a family of inner products $\{\inprod{\cdot}{\cdot}_\rho \,|\, \rho \in \sS (\cH)\}$ 
on the $\bR$-linear space $\cLh$, where 
the correspondence $\rho\mapsto \inprod{\cdot}{\cdot}_\rho$ is smooth, and assume that 
\be
\any \rho\in \sS, \any A\in \cLh, \; 
\inprod{A}{I}_\rho = \expect{A}_\rho := \Tr ( \rho A),
\label{<A,I>=<A>_1}
\ee
where $I$ denotes the identity operator on $\cH$.  (As usual,  $cI$ for $c\in \bC$ is often denoted as $c$ throughout the paper.) 
The inner products are represented as
\be
\inprod{A}{B}_\rho = \inprod{A}{\Omega_\rho (B)}_{\rm HS} = \Tr ( A\,\Omega_\rho (B))
\label{inprod_Omega}
\ee
by a family of  super-operators $\{ \Omega_\rho 
: \cLh \rightarrow \cLh\}_{\rho\in \sS}$, where $\inprod{\cdot}{\cdot}_{\rm HS}$ denotes 
the Hilbert-Schmidt inner product.  Note that the assumption \eqref{<A,I>=<A>_1} is equivalent to
\be
\any \rho\in \sS , \; \Omega_\rho (I) = \rho.
\label{Omega(I)=rho}
\ee

For an arbitrary tangent vector $X_\rho\in T_\rho (\sS)$, a Hermitian operator 
$L_{X_\rho} \in \cLh$ is defined by the relation
\be
\any A\in \cLh, \; X_\rho \expect{A} = \inprod{L_{X_\rho}}{A}_\rho,
\label{def_LX_1}
\ee
where the LHS denotes the derivative of the function $\expect{A}: \sS \rightarrow \bR$, 
$\rho \mapsto \expect{A}_\rho$  w.r.t.\ $X_\rho$. 
Noting that  the LHS and the RHS are represented as 
$\Tr ( \iota_*(X_\rho) A)$ and $\Tr (\Omega_\rho (L_{X_\rho})\, A )$, respectively, 
we can rewrite \eqref{def_LX_1} into 
\be
 \iota_*(X_\rho) = \Omega_\rho  (L_{X_\rho}).
\label{def_LX_2}
\ee
From \eqref{<A,I>=<A>_1} and  \eqref{def_LX_1}, we have
\be
\expect{L_{X_\rho}}_\rho = \inprod{L_{X_\rho}}{I}_\rho = X_\rho \expect{I} = X_\rho 1 = 0.
\label{<LX>=0}
\ee
Since $X_\rho \leftrightarrow \iota_* (X_\rho) \leftrightarrow L_{X_\rho}$ are 
one-to-one correspondences, we obtain the following identity: 
\be
\{ L_{X_\rho}\, |\, X_\rho \in T_\rho (\sS) \} = \{ A\in \cLh\, |\, \expect{A}_\rho = 0\}.
\label{e-tangent_space}
\ee

In the following, we often express \eqref{<A,I>=<A>_1} as
\be
\any A\in \cLh, \; \inprod{A}{I} = \expect{A}
\label{<A,I>=<A>_2}
\ee
as an identity for functions on $\sS$.  Similarly, \eqref{def_LX_1} is expressed as
\be
\any A\in \cLh, \; X \expect{A} = \inprod{L_{X}}{A},
\label{def_LX_3}
\ee
where 
$X$ is a vector field on $\sS$,  $L_X$ denotes 
the map $\sS \rightarrow \cLh$, $\rho\mapsto L_{X_\rho}$, 
 and $\inprod{L_{X}}{A}$ denotes 
the function $\rho\mapsto  \inprod{L_{X_\rho}}{A}_\rho$.  

For a submanifold $\sM$ of $\sS$ and a vector field $X\in \vf (\sM)$ on it, the map 
$L_X : \sM \rightarrow \cLh, \rho\mapsto L_{X_\rho}$ is defined, for which 
\eqref{def_LX_3} holds as an identity for functions on $\sM$. 
We may write $X\expect{A}$ as $X\expect{A}\!|_{\sM}$ in this case.
In particular, given a coordinate system 
$\xi=(\xi^i)$ of $\sM$,  we have 
\be
\any A\in \cLh, \;  \partial_i \expect{A} = \partial_i \expect{A}\!|_{\sM} = \inprod{L_i}{A}, 
\label{def_Li_3}
\ee
where $\partial_i := \frac{\partial}{\partial\xi^i}$ and $L_i := L_{\partial_i}$.

\medskip

\begin{remark}
\label{remark_em_representation}
In the terminology of \cite{amanag}, $\iota_* (X_\rho)$ 
and $L_{X_\rho}$ are called the m-representation and e-representation,  
and are denoted by $\iota_* (X_\rho)= X_\rho^{(\m)}$ and $L_{X_\rho}
= X_\rho^{(\e)}$, respectively.  We do not use the symbols $X_\rho^{(\m)}$ and 
$X_\rho^{(\e)}$ in this paper, but 
may use the following notation when convenient: 
\begin{align}
T^{(\m)}_\rho (\sM) &:= \{ \iota_* (X_\rho)\,|\, X_\rho \in T_\rho (\sM)\} , 
\label{def_Tm(M)}
\\
T^{(\e)}_\rho (\sM) &:= \{ L_{X_\rho} \,|\, X_\rho \in T_\rho (\sM)\} .
\label{def_Te(M)}
\end{align}
Note that 
\begin{align}
T^{(\m)}_\rho (\sS) &= \cLhzero 
\quad\text{and} \quad 
T^{(\e)}_\rho (\sS) = \{A\in \cLh\,|\, \expect{A}_\rho =0 \}.
\end{align}
\end{remark} 

\medskip

Now, we define a Riemannian metric $g$ on $\sS$  by
\be
\any \rho\in \sS, \any X_\rho, Y_\rho \in T_\rho (\sS), \; 
g_\rho (X_\rho, Y_\rho) := \inprod{L_{X_\rho}}{L_{Y_\rho}}_\rho
= \Tr \left( \iota_* (X_\rho) L_{Y_\rho}\right), 
\label{def_g_1}
\ee
which is equivalently written as
\be
\any X, Y \in \vf (\sS), \; 
g (X, Y) := \inprod{L_X}{L_Y} = \Tr \left(  \iota_* (X) L_Y\right). 
\label{def_g_2}
\ee
We also have the expression
\be
\any X, Y \in \vf (\sS), \; g(X, Y) = - \expect{X L_Y} = - \expect{Y L_X}, 
\label{g(XY)=-<XLY>}
\ee
where $X L_Y : \sS \rightarrow \cLh$, $\rho \mapsto X_\rho L_Y$,
denotes the derivative 
of the map $L_Y : \sS \rightarrow  \cLh$, $\rho\mapsto L_{Y_\rho}$, w.r.t.\ $X$, and 
$\expect{X L_Y}$ denotes the function $ \rho\mapsto \expect{X_\rho L_Y}_\rho$. 
This expression is derived as
\begin{align}
0 = X_\rho \expect{L_Y} = X_\rho \expect{L_{Y_\bcdot}}_\bcdot 
& = X_\rho \expect{L_{Y_\rho}}_\bcdot + X_\rho \expect{L_{Y_\bcdot}}_\rho 
\noret 
& = 
\inprod{L_{X_\rho}}{L_{Y_\rho}}_\rho + 
\expect{X_\rho L_Y}_\rho, 
\end{align}
where 
the first equality follows from \eqref{<LX>=0} and the last from \eqref{def_LX_1}, 
with dots $\bcdot$  added to clarify the positions of variables of maps. 

An important class of such Riemannian metrics is that of {\em monotone metrics} \cite{petz_monotone} for which 
$\Omega_\rho$ is represented as
\be
\Omega_\rho (A) = f(\Delta_\rho) (A \rho) = (f(\Delta_\rho) (A)) \rho, 
\label{Omega_f_Delta}
\ee
where $f: (0, \infty) \rightarrow (0, \infty)$ is an operator monotone function satisfying $\any x>0,\,  x f(1/x) = f(x)$ and 
$f(1) = 1$, and $\Delta_\rho: \cL \rightarrow \cL$ is the modular operator defined by 
$\Delta_\rho (A) = \rho A \rho^{-1}$. 
The class contains the SLD metric, which plays the main role in this paper, defined by $f(x) = (x+1)/2$ and
\be
\inprod{A}{B}_\rho = {\rm Re}\, \Tr (\rho A B) =  \Tr ( \rho (A \circ B) ) = \Tr ((\rho\circ A) B), 
\;\; A, B\in \cLh, 
\label{symmetrized_inprod}
\ee
where $\circ$ denotes the symmetrized product: $A \circ B = \frac{1}{2} (A B + BA)$. 
In this case, $L_{X_\rho}$ ($L_X$, resp.) is called the {\em SLD 
(symmetric logarithmic derivative)} 
of a tangent vector $X_\rho$ 
(a vector field $X$, resp.).  In particular, $L_i := L_{\partial_i}$ obeys the equation
\be
\partial_i \rho_\xi =\rho_\xi \circ L_{i,\xi},
\ee
which is a popular expression for the SLD. 

Given a family of inner products $\{\inprod{\cdot}{\cdot}_\rho\,|\, \rho\in \sS\}$ which 
determines a  Riemannian metric $g$, let the {\em e-connection} $\nablae$ be defined 
as the dual connection of $\nablam$ w.r.t.~\!$g$; i.e., 
\be
\any X, Y, Z\in \vf (\sS), \; X g (Y, Z) = g (\nabla^{(\e)}_X Y, Z) + g(Y, \nabla^{(\m)}_X Z).
\label{duality_em_quantum}
\ee
We have thus obtained $(\sS, g, \nablae, \nablam)$  as an example of $(\mS, g, \nabla, \nabla^*)$ treated in 
the previous section, where $\nablae$ and $\nablam$ are dual w.r.t.\ $g$, 
$\nablae$ is curvature-free and $\nablam$ is flat (see Remark~\ref{remark_nablastar_M_flat}).
 The triple $(\sS, g, \nablae, \nablam)$ is called the {\em information-geometric structure} 
on $\sS$ 
induced from a family of inner products $\{\inprod{\cdot}{\cdot}_\rho\,|\, \rho\in \sS\}$. 
In particular, the information-geometric structure induced from the symmetrized inner product \eqref{symmetrized_inprod} is called 
the {\em SLD structure}.  It is the SLD structure that will play a leading role in subsequent sections in relation to estimation theory, 
but  this section will continue the discussion on general information-geometric structures. 

\medskip

\begin{remark} 
\label{remark_BKM}
As is shown in Theorem~\!7.1 of \cite{amanag}, there is only one information-geometric structure 
defined in the manner described above 
for which the e-connection is torsion-free (so that 
$(\sS, g, \nablae, \nablam)$ is dually flat). That is the structure induced from 
the BKM (Bogoliubov-Kubo-Mori) inner product 
\be
\inprod{A}{B}_\rho = \int_0^1 \Tr \left( \rho^s A\, \rho^{1-s} B\right) ds, \;\; A, B\in\cLh.
\ee
The induced Riemannian metric is a monotone metric corresponding to $f(x) =\frac{x-1}{\log x} = \int_0^1 x^s ds$. 
In the other cases, the torsion $
{\torsion}^{(\e)} (X, Y) =
\nabla^{(\e)}_X Y - \nabla^{(\e)}_Y X - 
[X, Y]$ does not vanish, where $[\cdot  , \cdot]$ denotes the Lie bracket for vector fields.
 \end{remark}
 
 \medskip
 
 \begin{remark} 
 For the SLD structure, it is known (\cite{amanag}, Eq.~\!(7.80)) that the torsion has the following representation
  : for each point $\rho\in\sS$ and each tangent vectors $X_\rho, Y_\rho\in T_\rho (\sS)$, we have
\be
\iota_* ({\torsion}^{(\e)} (X_\rho, Y_\rho) ) = \frac{1}{4} [ [L_{X_\rho}, L_{Y_\rho}], \rho], 
\label{SLD_torsion}
\ee
where  $[\cdot  , \cdot]$ in the RHS denotes the commutator for operators on $\cH$. 
Since this representation will be used in Sections~\ref{sec_integrability} and \ref{sec_qubit}, 
we show its proof in Appendix~\ref{sec_proof_SLD_torsion}  for the reader's convenience. 
\end{remark}

\medskip

Henceforth, we use the prefixes e- and m- for notions concerning the e-connection and m-connection; e.g., e-parallel, e-{\ap}, m-affine, etc.

\begin{proposition}
\label{prop_e-connection_operator_representation}
For any vector fields $X, Y, W\in \vf (\sS)$,
we have
\begin{align}
W  = \nabla^{(\e)}_X Y \; & \Leftrightarrow \; L_W = X L_Y - \expect{X L_Y} = X L_Y + g(X, Y).
\label{e-connection_operator_representation}
\end{align}
\end{proposition}

\begin{proofs} Differentiating $g(Y, Z) = \Tr (L_Y\, \iota_* (Z))$ (see  \eqref{def_g_2}) by $X$, we have
\begin{align}
X g (Y, Z) & = \Tr \left( (X L_Y)\, \iota_* (Z)\right) + \Tr \left( L_Y\,  (X \iota_* (Z))\right) 
\noret 
&=  \Tr \left( (X L_Y)\, \iota_* (Z)\right) + g (Y, \nabla^{(\m)}_X Z),
\nonumber
\end{align}
where the second equality follows from \eqref{def_m-connection}. 
Letting $W'\in \vf (\sS)$ be defined by $L_{W'} = X L_Y - \expect{X L_Y}$, 
whose existence is ensured by \eqref{e-tangent_space}, the above 
equation is represented as
\be
X g (Y, Z) 
=  g(W', Z) + g (Y, \nabla^{(\m)}_X Z).
\nonumber
\ee
This means that $W' = \nabla^{(\e)}_X Y$, and proves the proposition. 
\end{proofs}

\begin{proposition}
\label{prop_e-parallel_vactor_field}
For a vector field $X\in \vf (\sS)$, we have
\begin{align}
\text{$X$ is e-parallel}\; & \Leftrightarrow \; 
\some A\in \cLh, \; L_X = A - \expect{A}
\noret 
&  \Leftrightarrow \; 
\some A\in \cLh, \any \rho\in \sS, \; L_{X_\rho}  = A - \expect{A}_\rho. 
\label{e-parallel_vactor_field}
\end{align}
\end{proposition}

\begin{proofs}
We may use Prop.~\!\ref{prop_e-connection_operator_representation}  to prove this, but 
here we show an alternative proof. 
We first note that, according to   \eqref{def_m-connection}, 
 a vector field $Y\in \vf (\sS)$ is m-parallel if and only if 
$\iota_* (Y) : \sS \rightarrow \cLhzero$ is a constant map represented by 
an operator $B\in \cLhzero$ as  $\iota_* (Y) =B$. 
Invoking Prop.~\!\ref{prop_parallel_nabla_nablastar} and \eqref{def_g_2}, 
we have
\begin{align}
\text{$X$ is e-parallel}\; & \Leftrightarrow \; 
\any Y : \text{m-parallel},  \; \text{$g(X, Y)$ is constant on $\sS$}
\noret 
 & \Leftrightarrow \; \any B\in \cLhzero, \; 
  \text{$\Tr(L_X B )$ is constant on $\sS$} 
 \noret 
 & \Leftrightarrow \; \some A\in \cLh, \; L_X = A - \expect{A}.
 \nonumber
\end{align}
\end{proofs}

Recalling Prop.~\!\ref{prop_parallel_X(S/M)}, the following corollary is immediate.

\begin{corollary}
\label{cor_e-parallel_vactor_field_extended}
For a submanifold $\sM$ of $\sS$ and for $X\in \vf (\sS / \sM)$ (including the case when $X\in \vf (\sM)$), 
we have
\begin{align}
& \text{$X$ is e-parallel (i.e., parallel w.r.t.\ the e-connection on $\sS$)
}\; 
\noret 
& \Leftrightarrow \; 
\some A\in \cLh, \; L_X = A - \expect{A}|_{\sM}
\noret 
&  \Leftrightarrow \; 
\some A\in \cLh, \any \rho\in \sM, \; L_{X_\rho}  = A - \expect{A}_\rho.
\label{e-parallel_vactor_field_extended}
\end{align}
\end{corollary}

The following corollary is also immediate from Prop.~\!\ref{prop_e-parallel_vactor_field}.

\begin{corollary}
The e-parallel transport $\para^{(\e)}_{\rho, \sigma}$ $: 
T_\rho (\sS) \rightarrow T_\sigma (\sS)$ for arbitrary two points $\rho, \sigma\in \sS$ 
is represented as follows: $\any X_\rho\in T_\rho (\sS), \any X_\sigma\in T_\sigma (\sS)$, 
\be
X_\sigma = \para^{(\e)}_{\rho, \sigma} (X_\rho) \; \Leftrightarrow \; L_{X_\sigma} = L_{X_\rho} - \expect{L_{X_\rho}}_\sigma.
\label{e-parallel_transport}
\ee
\end{corollary}

\medskip

In the following, a pair  $(\sM, \xi)$ of a submanifold $\sM$ of $\sS$ and a coordinate system $\xi$ of $\sM$ 
is called a {\em model}.
\begin{proposition}
\label{prop_autopara_affine_coordinate_quantum}
For an $n$-dimensional model $(\sM, \xi)$, the following conditions are equivalent.
\begin{itemize}
\item[(i)] $\sM$ is e-autoparallel in $\sS$, and $\xi$ is an m-affine coordinate system.
\\
(Note: ``m-affine'' means ``affine w.r.t.\ the m-connection $\nabla^{(\m)}_{\sM}$ on $\sM$''.)
\item[(ii)] $\some \{F^1, \ldots, F^n\} \subset \cLh$ such that for every $i\in \{1, \ldots, n\}$, 
\be
\sum_j g^{ij} L_j = F^i - \expect{F^i}|_{\sM},
\label{gij_Lj_1}
\ee
where $\partial_i := \frac{\partial}{\partial\xi^i}$, 
$L_j := L_{\partial_j}$ and 
$[g^{ij}] :=[g_{ij} := g(\partial_i, \partial_j)]^{-1}$. 
\item[(iii)] $\some \{F^1, \ldots, F^n\} \subset \cLh$ such that for every $i\in \{1, \ldots, n\}$, 
\be
\sum_j g^{ij} L_j = F^i - \xi^i.
\label{gij_Lj_2}
\ee
(Note:  \eqref{gij_Lj_2} implies $\xi^i = \expect{F^i}|_{\sM}$.)
\item[(iv)] $\some \{F^1, \ldots, F^n\} \subset \cLh$ such that  for every $i\in \{1, \ldots, n\}$, 
\be
\any \rho \in \sM, \;\; L_{i, \rho} \in {\rm span}\, \{F^j \}_{j=1}^n \oplus \bR 
\quad\text{and}\quad \xi^i(\rho) = \expect{F^i}_\rho, 
\label{eq_(iv)_prop_autopara_affine_coordinate_quantum}
\ee
where $\bR$ is identified with $\{cI \, |\,c\in \bR\}$ (see Remark~\!\ref{remark_Fi_linear_indep} below).
\end{itemize}
\end{proposition}

\begin{proofs}
The equivalence 
(i) $\Leftrightarrow$ (ii) is immediate from Prop.~\!\ref{prop_autopara_affine_coordinate_general} and 
Cor.~\!\ref{cor_e-parallel_vactor_field_extended}, 
and (iii) $\Rightarrow$ (ii) is obvious since $\expect{L_i} =0$. 

To show (ii) $\Rightarrow$ (iii), assume (ii).  Then we have 
\[
\sum_j g^{ij} \inprod{L_j}{L_k} = \inprod{F^i}{L_k}. 
\]
Here, the LHS is $\sum_j g^{ij} g_{jk} = \delta^i_k = \partial_k \xi^i$ and  the RHS is $\partial_k \expect{F^i}$
due to \eqref{def_Li_3}. 
Hence,  there exists a constant vector 
$(c^i)\in \bR^n$ such that $\expect{F^i}|_{\sM} = \xi^i + c^i$. Redefining $F^i:= F^i - c^i$, 
 \eqref{gij_Lj_1} is rewritten as \eqref{gij_Lj_2}, and (iii) is verified. 

Since \eqref{gij_Lj_2} implies that 
\be
L_{i, \rho} = \sum_j g_{ij} (\rho) F^j - \Bigl(\sum_j g_{ij} (\rho) \xi^j(\rho)\Bigr) 
\in  {\rm span}\, \{F^j \}_{j=1}^n \oplus \bR,
\nonumber
\ee
 we have (iii) $\Rightarrow$ (iv). 
 To show the converse, we assume the existence of $\{F^1, \ldots , F^n\}$ in  (iv). 
 Then we have $\xi^i = \expect{F^i}|_{\sM}$, and 
 for each $\rho\in \sM$ there exist $[a_{ij}]\in \bR^{n\times n}$ and 
 $[b_i]\in \bR^n$ such that for any $i$
 \be
 L_{i, \rho} = \sum_j a_{ij} F^j + b_i.
 \nonumber
 \ee
 This implies that 
 \begin{align}
g_{ik} (\rho) & = \inprod{L_{k, \rho}}{L_{i, \rho}}_\rho = \sum_j a_{ij} \inprod{L_{k, \rho}}{F^j}_\rho
\noret 
 & =  \sum_j a_{ij} (\partial_k \expect{F^j})_\rho = \sum_j a_{ij}  (\partial_k \xi^j)_\rho = a_{ik}.
 \nonumber
 \end{align}
 Hence we have
 \be
 \sum_j g^{ij}(\rho) L_{j, \rho} = F^i  +  \sum_j g^{ij}(\rho) b_j.
  \nonumber
 \ee
 Here,  the constant $\sum_j g^{ij}(\rho) b_j$ should be equal to $-\expect{F^i}_\rho = -\xi^i (\rho)$ due to $\expect{L_{j, \rho} }_\rho =0$. 
 Thus (iii) is concluded. 
\end{proofs}

\begin{remark}
As noted in Remark~\ref{remark_nablastar_M_flat}, the flatness of the m-connection on $\sS$ ensures that every e-{\ap} submanifold is m-flat and has an m-affine coordinate system. 
This fact can be seen directly from Prop.~\ref{prop_autopara_affine_coordinate_quantum} as follows. 
The e-{\apty} of $\sM$ is equivalent to the existence of $\{F^1, \ldots, F^n\}\subset \cLh$ s.t. $\any \rho\in\sM$, 
$T^{(\e)}_\rho (\sM) = {\rm span} \{F^i - \expect{F^i}_\rho\}_{i=1}^n$ (see \eqref{def_Te(M)} for  $T^{(\e)}_\rho (\sM)$), 
which is equivalent to the first condition that $\any i, \any \rho \in \sM, \; L_{i, \rho} \in {\rm span}\, \{F^j \}_{j=1}^n \oplus \bR$ 
of (iv) in the proposition. When $\sM$ satisfies this condition, letting $\xi = (\xi^i)$ be defined by $\xi^i = \expect{F^i}|_{\sM}$, we have that $(\sM, \xi)$ satisfies the two conditions of (iv), so that $\xi$ is m-affine due to (iv) $\Rightarrow$ (i).
\end{remark}

\begin{remark}
\label{remark_Fi_linear_indep}
In \eqref{gij_Lj_1} and \eqref{gij_Lj_2}, 
the operators $\{F^i - \expect{F^i}_\rho\}_{i=1}^n $ turn out to be linearly independent for each $\rho\in\sM$, which implies that  
 $\{F^1 , \ldots , F^n, I\}$ are linearly independent, or equivalently that $\{F^1 , \ldots , F^n\}$ are linearly independent 
in the quotient space $\cLh / \bR$ with identification $\bR = \{cI \, |\,c\in \bR\}$. 
 \end{remark}

\medskip

At the end of this section, we present a proposition which claims that i.i.d.\ extensions of a model preserves 
the e-{\apty}.  For the proposition, we assume the following condition on the family of inner products from which the 
information-geometric structure is defined:
\be
\any \{A_t\}_{t=1}^{\iidn}, \, \any \{B_t\}_{t=1}^{\iidn} \subset \cLh (\cH), \; 
\bigl\langle\, \Motimes_{t=1}^{\iidn} A_t , \Motimes_{t=1}^{\iidn} B_t\, \bigr\rangle_{\rho^{\iidtensor}} 
=
\prod_{t=1}^{\iidn} \inprod{A_t}{B_t}_\rho, 
\label{inprod_tensor-1}
\ee
which is equivalent to
\be
\any \{A_t\}_{t=1}^{\iidn} \subset \cLh (\cH), \; 
\Omega_{\rho^{\iidtensor}} \bigl(  \Motimes_{t=1}^{\iidn} A_t \bigr) 
= \Motimes_{t=1}^{\iidn} \Omega_{\rho}  (A_t). 
\label{inprod_tensor-2}
\ee
The assumption is satisfied 
when the inner products are defined  from a function $f$ by \eqref{inprod_Omega} and \eqref{Omega_f_Delta} 
for which 
$\Omega_{\rho} = f (\Delta_\rho )$ and 
$\Omega_{\rho^{\iidtensor}} =f \bigl( \Delta_{\rho^{\iidtensor}}\bigr)$ hold. 
In particular, the proposition hols for the SLD structure.

\begin{proposition}
\label{prop_iid_extension}
Given a model $(\sM, \xi)$ in $\sS(\cH)$ and a natural number $\iidn\geq 2$, define 
the model $(\tilde{\sM}, \tilde{\xi})$ in $\sS (\cH^{\iidtensor})$ by 
$ \tilde{\sM}:= \{ \rho^{\iidtensor}\,|\, \rho\in \sM\}$ 
and $
\tilde{\xi}^i (\rho^{\iidtensor}) = 
\xi^i (\rho)$ for $\rho\in \sM$. 
Under the assumption \eqref{inprod_tensor-1}-\eqref{inprod_tensor-2}, 
the following conditions are equivalent. 
\begin{itemize}
\item[(i)] $\sM$ is e-{\ap} in $\sS$, and $\xi$ is m-affine.
\item[(ii)] $\tilde{\sM}$ is e-{\ap} in $\sS (\cH^{\iidtensor})$, and $\tilde{\xi}$ is m-affine.
\end{itemize}
\end{proposition}

\begin{proofs}
Let $\partial_i := \frac{\partial}{\partial\xi^i}$, $\tilde{\partial}_i := \frac{\partial}{\partial\tilde{\xi^i}}$, 
and $L_i = L_{\partial_i}$, $\tilde{L}_{i} = L_{\tilde{\partial}_i}$, which are determined by 
\be
\iota_* ((\partial_i)_\rho) = \Omega_\rho (L_{i, \rho})
 \nonumber
\ee
and
\be
\tilde{\iota}_* ((\tilde{\partial}_i)_{\rho^{\iidtensor}})  = \Omega_{\rho^{\iidtensor}} (\tilde{L}_{i, \rho^{\iidtensor}}), 
 \nonumber
\ee
where $\tilde{\iota}$ denotes the natural embedding $\sS (\cH^{\iidtensor}) \rightarrow \cLhone (\cH^{\iidtensor})$. 
With the aid of the parametric representation \eqref{m-rep_partial_rho}, we see that
\begin{align}
\tilde{\iota}_* ((\tilde{\partial}_i)_{\rho^{\iidtensor}}) 
& = \tilde{\partial}_i (\rho^{\iidtensor})_{\tilde{\xi}}
= \partial_i \rho_\xi^{\iidtensor} 
= \sum_{t=1}^{\iidn} \rho_\xi^{\otimes (t-1)} \otimes \partial_i \rho_\xi  \otimes \rho_\xi^{\otimes (\iidn - t)}
\noret
&=  \sum_{t=1}^{\iidn} \rho^{\otimes (t-1)}\otimes \iota_* ((\partial_i)_\rho) \otimes \rho^{\otimes (\iidn - t)}
\noret 
&= \sum_{t=1}^{\iidn} \rho^{\otimes (t-1)}\otimes \Omega_\rho(L_{i, \rho}) \otimes \rho^{\otimes (\iidn - t)}
\noret 
& \stackrel{\star}{=}
\sum_{t=1}^{\iidn}
\Omega_{\rho^{\iidtensor}} (
 I^{\otimes (t-1)} \otimes L_{i, \rho} \otimes I^{\otimes (\iidn - t)})
= \Omega_{\rho^{\iidtensor}} (L_{i, \rho}^{(\iidn)}), 
 \nonumber
\end{align}
where $\stackrel{\star}{=}$ follows from \eqref{Omega(I)=rho} and \eqref{inprod_tensor-2}, and we have used the notation 
\be
A^{(\iidn)} := \sum_{t=1}^{\iidn} I^{\otimes (t-1)} \otimes A \otimes I^{\otimes (\iidn - t)} 
\quad\text{for} \;\; A\in \cLh (\cH).
 \nonumber
\ee
This implies that  
$\tilde{L}_{i, \rho^{\iidtensor}} = (L_{i, \rho})^{(\iidn)}$ 
and that $\tilde{g}_{ij} = \iidn\, g_{ij}$ for $g_{ij} (\rho) = \inprod{L_{i, \rho}}{L_{j,\rho}}_\rho$ 
and $\tilde{g}_{ij} (\rho^{\iidtensor}) = 
\bigl\langle \tilde{L}_{i, \rho^{\iidtensor}}, \tilde{L}_{j, \rho^{\iidtensor}}\bigr\rangle_{\rho^{\iidtensor}}$, 
which leads to $\tilde{g}^{ij} = \frac{1}{\iidn}\, g^{ij}$. Hence, $L^i := \sum_j g^{ij} L_j$ and $\tilde{L}^i := \sum_j \tilde{g}^{ij} \tilde{L}_j$ are linked by 
$
\tilde{L}^i_{\rho^{\iidtensor}} = \frac{1}{\iidn} \, (L^i_{\rho})^{(\iidn)}. 
$
 Now, according to Prop.~\!\ref{prop_autopara_affine_coordinate_quantum}, 
 conditions (i), (ii) of the present proposition are respectively expressed as
 \begin{itemize}
 \item[(i)'] $\exists \{F^i\} \subset \cL (\cH), \; 
 \any i, \any \rho\in \sM, \; L^i_\rho = F^i - \xi^i (\rho)$.
 \item[(ii)']
 $\exists \{\tilde{F}^i\} \subset \cL (\cH^{\iidtensor}), \; 
 \any i, \any \rho\in \sM, \; 
  \frac{1}{\iidn} \, (L^i_{\rho})^{(\iidn)}
= \tilde{F}^i - \xi^i (\rho)$.
 \end{itemize}
 They are obviously equivalent with relation $\tilde{F^i} = \frac{1}{\iidn} (F^i)^{(\iidn)}$. 
\end{proofs}

\section{\red{Preliminaries from quantum estimation theory} 
}
\label{sec_efficient_estimator}

From this section, we investigate the relationship between estimation problems and geometric properties for 
quantum statistical models. 
Henceforth,  we will consider only the SLD structure as an \rred{information-geometric} structure unless otherwise stated. 
\red{As a first step, 
we introduce several notions of quantum estimation theory and discuss their basic properties in this section.
} 

Given a model $(\sM, \xi)$ in $\sS (\cH)$, 
an {\em estimator} for coordinates 
$\xi$ is generally represented by a POVM $\Pi = \Pi(d\xihat)$ 
on $\cH$, where $\xihat$ is a variable representing an estimate. 
A representative case is when 
$\Pi$ is expressed as $\Pi (d\xihat) = \sum_{\omega\in \Omega}\pi_\omega\,  \delta_{f(\omega)} (d\xihat) $ by 
a POVM $\pi = (\pi_\omega)_{\omega\in \Omega}$ on a discrete set $\Omega$ and a function 
$f : \Omega \rightarrow \bR^n$, where 
$\delta_{f(\omega)}$ denotes the $\delta$-measure concentrated on the point $f(\omega)\in \bR^n$. 
This estimator, which is  denoted by $\Pi = (\pi, f)$, represents the estimation procedure in which the estimate 
is determined as $\xi = f(\omega)$ from the outcome $\omega$ of the measurement $\pi$. 

The expectation $E_\rho (\Pi) \in \bR^n $ and 
the mean squared error (the variance in the unbiased case) $V_\rho (\Pi) \in \bR^{n\times n}$ 
of $\Pi$ for a state $\rho$ 
are defined by
\begin{align}
 E_\rho (\Pi) & :=
 \int \hat{\xi} \, \Tr \left( \rho\,  \Pi (d\hat{\xi} )\right), 
 \label{def_expectation}
\\
  V_\rho (\Pi) &:= 
 \int (\hat{\xi} - \xi(\rho)) \transpose{(\hat{\xi} - \xi(\rho))} 
\, \Tr \left(\rho\, \Pi (d\hat{\xi})\right), 
 \label{def_variance}
\end{align}
where $\bR^n$ is regarded as the space of column vectors $\bR^{n\times 1}$ and 
$\transpose{}$ denotes the transpose.
For $\Pi = (\pi, f)$, we have
\begin{align}
 E_\rho (\Pi) &=
 \sum_\omega f (\omega) \,
 \Tr ( \rho \pi_\omega), 
 \label{def_expectation_var}
 \\
  V_\rho (\Pi) &=
   \sum_\omega (f (\omega) - \xi(\rho))  \transpose{(f (\omega) - \xi(\rho))}\,  \Tr ( \rho \pi_\omega).
  \label{def_variance_var}
\end{align}

An estimator $\Pi$ is said to be {\em locally unbiased} 
for a coordinate system $\xi$ at $\rho\in \sM$ 
when the elements $E^i_\rho (\Pi)$, $i\in \{1, \ldots , n\}$, of $E_\rho (\Pi)$ satisfy 
\be
E^i_\rho (\Pi) = \xi^i(\rho) \quad\text{and} \quad 
\partial_j E^i_\rho (\Pi) = \delta^i_j, 
\ee
where $\partial_j E^i_\rho (\Pi)$ denotes 
the derivative of the function $E^i (\Pi): 
\sigma \mapsto E^i_\sigma (\Pi)$ by $\partial_i = \frac{\partial}{\partial\xi^i}$ evaluated at the point $\rho$. 
We denote by $\cU(\rho, \xi)$ 
the totality of locally unbiased estimators for $\xi$ at $\rho$. 
Using the symmetrized inner product $\inprod{\cdot}{\cdot}$ of \eqref{symmetrized_inprod} 
and 
 the SLDs $L_i = L_{\partial_i}$, $i\in \{1, \ldots , n\}$, we have 
\be
\Pi \in \cU(\rho, \xi) \; \Leftrightarrow\; \any i, j, \; \expect{A^i}_\rho = 
\xi^i(\rho) 
\;\;\text{and}\;\; \inprod{A^i}{L_{j,\rho}}_\rho = \delta_j^i, 
\label{locally_unbiased_Ai_Lj}
\ee
where 
\be
A^i := \int \xihat^i \Pi (d\xihat) \in \cLh(\cH).
\ee

The well-known SLD Cram\'er-Rao inequality 
\cite{Helstrom:1967, Helstrom:1976} 
states that every $\Pi\in \cU(\rho, \xi)$ 
obeys 
\begin{equation}\label{eqn:CR}
V_\rho(\Pi) \geq  G_\rho^{-1}, 
\end{equation}
 where $G_\rho = [g_{ij} (\rho)]$ denotes the SLD Fisher information matrix
defined by $g_{ij} = \inprod{L_i}{L_j} = 
g(\partial_i, \partial_j)$ with $g$ being the SLD metric. 
Furthermore,  we have the following proposition, whose proof will be given after Lemma~\ref{lemma_Pi=(pi_f)}.

\begin{proposition}
\label{prop_inf_uVu}
For every column vector  $\bu\in\bR^n=\bR^{n\times 1}$, we have
\be \inf_{\Pi\in\cU(\rho, \xi)}\,\tbu V_\rho(\Pi)\bu
 =\tbu G_\rho^{-1}\bu.
 \label{eq_inf_uVu}
 \ee
\end{proposition}

\red{
\begin{remark}
\label{remark_attainability_min_uVu}
The infimum on the LHS of \eqref{eq_inf_uVu} cannot, in general,  be attained as a minimum. 
Using the terminology that will be introduced later, this can be rephrased as saying that a locally \rred{$\bu \tbu$-efficient} estimator at a point $\rho$ for $\xi$  does not generally exist. 
A necessary condition for the attainability is given  in Appendix~\ref{sec_attainability_min_uVu}.
\end{remark}
} 

Let us introduce a class of randomized procedures for estimation that will be useful in the proofs of both Prop.~\!\ref{prop_inf_uVu}  above and Theorem~\!\ref{thm:autoparallel_m-affine} later.
Suppose that a point $\rho\in \sM$, a basis $\{\bu^1, \ldots, \bu^n\}$ of $\bR^n$, a 
positive probability vector $(p_1, \ldots , p_n)$ 
s.t.\ $\any i$, $p_i > 0, \sum_i p_i =1$ 
and $n^2$ real numbers $\{\gamma_k^i\}$ satisfying
\be
\sum_k p_k\, \gamma_k^i = \xi^i(\rho)
\label{constraint_gammaki}
\ee
are 
arbitrarily given.  Let 
\be
X^k := \sum_i u^k_i\, L^i_\rho 
\in \cLh (\cH), 
\label{def_Xk}
\ee
where $\bu^k = (u^k_i)$, $L^i_\rho := \sum_j g^{ij}(\rho) L_{j, \rho}$, $G_\rho^{-1} = [g^{ij}(\rho)]$, 
and consider 
the random measurement such that 
$k\in \{1, \ldots , n\}$ is randomly chosen according to 
the probability distribution $(p_1, \ldots , p_n)$ and then the observable $X^k$ is measured. 
This measurement is represented by the POVM 
$\pi= \{\pi_{k, r}\} = \{ p_k \pi^k_r\}$, where $\{\pi^k_r\}$ are the projectors 
in the spectral decomposition 
$X^k = \sum_r x^k_r\, \pi^k_r$.  (Do not confuse $X^k$,  $x^k_r$ and $\pi^k_r$ with the $n$th powers of $X$, $x_r$
and $\pi_r$.) 
When an eigenvalue $x^k_r$ is observed by measuring $X^k$, 
we estimate $\xi$ 
by $\xihat = f (k,x^k_r)$, where $f = \transpose{(f^1, \ldots , f^n)}$ is defined by
\be
 f^i (k,x) := \gamma_k^i +\frac{w^i_k}{p_k}\, x, 
 \label{def_fi}
\ee
and $[w^i_k] = [u^k_i]^{-1}$, i.e., 
$\sum_k u^k_i w^j_k = \delta_i^j$ and $\sum_i u^k_i w^i_l = \delta_l^k$. 
This estimation procedure defines the estimator $\Pi:=(\pi, f)$, 
which is characterized by the following property: 
for any polynomial function $\varphi : \bR^n \rightarrow \bR$, 
\be
\int \varphi (\xihat) \Pi (d \xihat) = \sum_k p_k\, \varphi (f(k, X^k)).
\label{charactrize_Pi_fpi}
\ee
\red{
The  procedure for obtaining the estimate $\hat{\xi}$ is illustrated below:
\be
\randest{(p_k)}{k}{ \fbox{$\sum_i u^k_i\, L^i_\rho$}}{x}{\displaystyle \hat{\xi} = 
{\mbox{\boldmath$\gamma$}}_k +\frac{x}{p_k}\, \bw_k, }
\ee
where 
${\mbox{\boldmath$\gamma$}}_k := (\gamma_k^i),  
\bw_k := (w_k^i) \in \bR^n$ and ``meas.'' stands for ``measure''.
} 

In this situation we have the following lemma, whose proof is given in Appendix~\ref{sec_proof_lemma_Pi=(pi_f)}.

\begin{lemma} The estimator  $\Pi =(\pi, f)$ satisfies:
\label{lemma_Pi=(pi_f)}
\begin{enumerate}
\item[(1)]
$\Pi \in \cU (\rho, \xi)$. 
\red{\item[(2)] $\disp V_\rho(\Pi) = \sum_k \frac{1}{p_k} \,
\Bigl( \transpose{(\bu^k)} G_\rho^{-1}\bu^k \Bigr)  \bw_k \transpose{(\bw_k)}
+ \sum_k p_k (\gamma_k - \xi(\rho)) \transpose{(\gamma_k - \xi(\rho))}
 $, \\
where 
$\bw_k := (w_k^i), \gamma_k:= (\gamma_k^i) 
\in \bR^n=\bR^{n\times 1}$. 
} 
\item[\red{(3)}] 
$\disp \any k, \; 
\transpose{(\bu^k)} V_\rho(\Pi)\bu^k=\frac{1}{p_k}\,\transpose{(\bu^k)} G_\rho^{-1}\bu^k
+ \sum_l p_l (a_l^k)^2
$, \\ where 
$\disp a_l^k := \sum_i u^k_i (\gamma_l^i - \xi^i(\rho))$.
\end{enumerate}
\end{lemma}

\begin{proofsvar}{of Prop.~\!\ref{prop_inf_uVu}} \ 
Let $\gamma_k^i := \xi^i (\rho)$, which satisfies \eqref{constraint_gammaki} so that  
Lemma~\!\ref{lemma_Pi=(pi_f)} is applicable\red{:
\be
\randest{(p_k)}{k}{ \fbox{$\sum_i u^k_i\, L^i_\rho$}}{x}{\displaystyle \hat{\xi} = 
\xi (\rho)  +\frac{x}{p_k}\, \bw_k. }
 \label{estimation_procedure_gamma_xi}
\ee
} 
Since $a_l^k =0$ in this case, we have
\be 
\transpose{(\bu^k)} V_\rho(\Pi) \bu^k = \frac{1}{p_k} \transpose{(\bu^k)} G_\rho^{-1} \bu^k.
\label{uVu_1/p}
\ee
The proposition is obvious when $\bu =0$, so we assume  $\bu \neq 0$ and 
choose $p\in (0, 1)$ arbitrarily. Taking $(\bu^1, \ldots , \bu^n)$ and 
$(p_1, \ldots, p_n)$ in the above construction such that $\bu^1 = \bu$ and $p_1 =p$, 
the resulting $\Pi$ satisfies $\transpose{\bu} V_\rho(\Pi) \bu = \frac{1}{p} \transpose{\bu} G_\rho^{-1} \bu$.
Since $p$ can be arbitrarily close to $1$, we have the proposition. 
\end{proofsvar}

\medskip

\red{
In the following,  
we introduce several notions of efficiency for estimators for a model $(\sM, \xi)$. 
We begin with local  notions which are 
properties of estimators  
at each point $\rho$ of the model. 
\begin{itemize}
\item A locally unbiased estimator $\Pi\in\cU(\rho, \xi)$ 
is said to be  {\em locally efficient} for $\xi$ at $\rho$ if 
\be \any \Pi'\in \cU (\rho, \xi), \; 
V_\rho (\Pi) \leq V_\rho (\Pi').
\ee
\item Given a positive-semidefinite matrix $\wgt \in \bR^{n\times n}$ as 
a weight,
an estimator $\Pi\in\cU(\rho, \xi)$ is said to be  {\em locally $\wgt$-efficient} for $\xi$ at $\rho$ if
\be
\any \Pi'\in\cU(\rho, \xi), \; 
\tr (\wgt V_\rho (\Pi)) \leq \tr (\wgt V_\rho (\Pi')), 
\ee
 or equivalently if 
\begin{equation}\label{eqn:G-efficient}
  \tr (\wgt V_\rho(\Pi)) =\min_{\Pi'\in\cU(\rho, \xi)}\tr (\wgt V_\rho(\Pi')).
\end{equation}
\end{itemize}
} 

\begin{proposition}
\label{prop_locally_efficient_estimator}
Given a model $(\sM, \xi)$, a point $\rho\in\sM$ and an estimator $\Pi\in\cU (\rho, \xi)$, the following conditions are equivalent. 
\begin{itemize}
\item[(i)] $\Pi$ is locally efficient for $\xi$ at $\rho$.
\item[(ii)] $V_\rho(\Pi) = G_\rho^{-1}$.
\item[(iii)] $\Pi$ is  locally $\bu\tbu$-efficient for $\xi$ at $\rho$ for every column vector $\bu\in \bR^n$.
\item[(iv)] $\Pi$ is  locally $\wgt$-efficient for $\xi$ at $\rho$ for every positive weight $\wgt  >0$.
\end{itemize}
\end{proposition}

\begin{proofs}
The equivalence (i) $\Leftrightarrow$ (iii) $\Leftrightarrow$ (iv) is obvious since 
\begin{align}
V_\rho (\Pi) \leq V_\rho (\Pi') \; & \Leftrightarrow \; 
\any \bu\in \bR^n, \; \tbu V_\rho (\Pi)\bu \leq  \tbu V_\rho (\Pi')\bu \noret 
& \Leftrightarrow \; 
\any \wgt >0, \; \tr(\wgt V_\rho (\Pi)) \leq \tr(\wgt V_\rho (\Pi')) ,
\nonumber 
\end{align}
and (ii) $\Leftrightarrow$ (iii) follows from Prop.~\!\ref{prop_inf_uVu}.
\end{proofs}

\begin{remark}
\label{remark_locally_efficient_commutative_SLDS}
As is well known, there exists a locally efficient estimator at $\rho$ iff  the SLDs $L_{1, \rho}, \ldots , L_{n, \rho}$ 
mutually commute (e.g.\ \S~\!7.4 of \cite{amanag}). 
\end{remark}

\red{
We next extend the notions of local efficiency introduced above to those that pertain to the model as a whole. 
\begin{itemize}
\item An estimator $\Pi$ is said to be  {\em efficient} for a coordinate system $\xi$ 
if $\Pi$ is locally efficient for $\xi$ at every point $\rho\in \sM$. 
\medskip
\item 
Given a weight field
${\mathcal \wgt} = \{\wgt_\rho \; |\; \rho \in \sM\}$, 
$\Pi$ is said to be  {\em ${\mathcal \wgt}$-efficient} for $\xi$ if 
$\Pi$ is locally $\wgt_\rho$-efficient for $\xi$ at every $\rho\in \sM$.
\medskip
\item 
When $\wgt_\rho = \wgt$ for all $\rho$, we simply call it
{\em $\wgt$-efficient} for $\xi$.  
\end{itemize}
} 

According to Prop.~\!\ref{prop_locally_efficient_estimator}, 
$\Pi$ is efficient $\Leftrightarrow$ $\Pi$ is $\bu\tbu$-efficient for every $\bu\in \bR^n$ 
$\Leftrightarrow$  $\Pi$ is $\wgt$-efficient for every $\wgt >0$. 

The condition for existence of efficient estimator was mentioned in the introduction as 
Theorem~\!\ref{thm:commutative}.  Suppose that  $(\sM, \xi)$ is represented as 
\eqref{q_c:e-family} and \eqref{q_c:expectation_coordinates} in the theorem; namely, 
every $\rho\in \sM$ is represented in the form
\be
\rho = 
\exp\Bigl[\, \frac{1}{2}
\Bigl\{\sum_{i=1}^n \theta_i(\rho) F^i - \psi(\rho)\Bigr\}
\Bigr]
\, P 
\exp\Bigl[\, \frac{1}{2}
\Bigl\{\sum_{i=1}^n \theta_i(\rho) F^i - \psi(\rho)\Bigr\}
\Bigr]
\label{q_c:e-family_var}
\ee
and satisfies $\xi^i (\rho) = \expect{F^i}_\rho$, where 
$F^1, \ldots , F^n$ are mutually commuting observables. Note that $P$ can be chosen to be 
an arbitrary element of $\sM$ if we wish. 
The SLDs of $\sM$ w.r.t.\ $\xi$ are represented as
\be
L_i =  \partial_i \Bigl( \sum_j \theta_j F^j - \psi \Bigr) 
= \sum_j (\partial_i \theta_j) (F^j - \partial^j \psi), 
\ee
where  $\partial_i := \frac{\partial}{\partial\xi^i}$ and $\partial^i:= \frac{\partial}{\partial\theta_i}$. 
Noting that the positions of upper/lower indices (superscripts and subscripts) are reversed from the standard notation 
of information geometry as in \cite{amanag} (see Remark~\!\ref{remark_index_positions}), 
we have
$\partial_i \theta_j = g_{ij}$, $\partial^i \psi = \xi^i$ and 
\be
L^i = \sum_j g^{ij} L_j = F^i - \xi^i.
\ee
According to Prop.~\!\ref{prop_autopara_affine_coordinate_quantum}, 
this means that $\sM$ is e-autoparallel in $\sS$ and that $\xi$ is an m-affine coordinate system 
w.r.t.~\!the SLD structure. 
Furthermore, the induced e-connection $\nabla^{(\e)}|_{\sM}$ on $\mM$ turns out to be torsion-free and hence flat, 
for which $\theta = (\theta_i)$ forms an affine coordinate system.  We have thus seen that $\sM$ is dually flat just 
as  classical exponential families. 

When $n=1$, \eqref{q_c:e-family_var} is written as 
\be
\rho = 
\exp\left[\frac{1}{2}
\bigl\{ \theta (\rho) F - \psi(\rho)\bigr\}
\right]
\, P
\exp\left[\frac{1}{2}
\bigl\{ \theta (\rho) F - \psi(\rho)\bigr\}
\right] .
\label{q_c:e-family_var_1dim}
\ee
This is a general form of e-geodesic in the sense that every e-geodesic is 
represented in this form by some $P$ and $F$. 
In the multi-dimensional case $n\geq 2$, on the other hand, 
\eqref{q_c:e-family_var} provides merely a special case of e-autoparallel submanifolds.  
In order to characterize the e-autoparallelity by an estimation-theoretical notion, 
the existence of efficient estimator is too strong, and 
we need a new variant of the notion of efficient estimators, which will be introduced in the next section. 

 \section{\red{Characterization of e-autoparallelity  in terms of filtrations of estimators} 
 }
\label{sec_efficient_filtration}

We now consider a one-parameter family of estimators $\filtPi= (\Pi_\eps )_{\eps\in (0, \eps_1)}$ 
instead of a single estimator, and call it a {\em filtration of estimators} 
or simply a {\em filtration}.  The upper limit $\eps_1$ can be an arbitrary positive number or $\infty$, 
but our interest lies only in the  limiting property of $\eps\downarrow 0$ and the value of $\eps_1$ 
plays no role.  So we simply write $\filtPi= (\Pi_\eps )_{\eps >0}$. 
\red{
The notions of efficiency for estimators introduced in the previous section are extended to those for filtrations as follows:
\begin{itemize}
\item 
Given a nonnegative (i.e., positive-semidefinite)
real matrix $\wgt\in \bR^{n\times n}$ as a weight, 
a filtration $\filtPi= (\Pi_\eps )_{\eps >0}$ is said to be  
{\em locally $\wgt$-efficient}  for $\xi$ at $\rho$ if $\Pi_\eps\in \cU (\rho, \xi)$ for 
every $\eps >0$ and 
$\lim_{\eps\downarrow 0} \tr (\wgt V_\rho (\Pi_\eps )) \leq 
\tr( \wgt V_\rho (\Pi'))$ for every $\Pi'\in\cU(\rho, \xi)$, which is equivalent to 
\be \label{cond_W-efficient_filtration}
  \lim_{\eps\downarrow 0}\tr (\wgt V_\rho(\Pi_\eps))
 =\inf_{\Pi'\in\cU(\rho, \xi)} \tr (\wgt V_\rho(\Pi')).
\ee 
When $\wgt = \bu \tbu$ with $\bu\in \bR^n$, in particular, 
this is represented as 
\be 
\label{cond_uuT-efficient_filtration}
  \lim_{\eps\downarrow 0} \tbu V_\rho(\Pi_\eps) \bu
 = \tbu G_\rho^{-1} \bu
\ee 
due to Prop.~\!\ref{prop_inf_uVu}. 
\medskip
\item Given a weight field 
${\mathcal \wgt} = \{\wgt_\rho \,|\, \rho \in \sM\}$, 
a  filtration $\filtPi= (\Pi_\eps )_{\eps >0}$ is said to be  
{\em ${\mathcal \wgt}$-efficient} for $\xi$ if 
it is  $\wgt_\rho$-locally efficient 
 for $\xi$ at every $\rho\in \sM$. 
 \medskip
 \item 
When $\wgt_\rho = \wgt$ for all $\rho$, we simply call it 
{\em $\wgt$-efficient} for $\xi$. 
\end{itemize}
}
Compare  \eqref{cond_W-efficient_filtration} with (\ref{eqn:G-efficient}), and note that 
a locally $\wgt$-efficient filtration at $\rho$ always exists, 
even when a locally $\wgt$-efficient estimator does not exist\red{; see Remark~\!\ref{remark_attainability_min_uVu}} 
.
\medskip

Now, we have the following theorem, which gives an estimation-theoretical characterization 
of the e-{\apty} w.r.t.\ the SLD structure. 

\begin{theorem}
\label{thm:autoparallel_m-affine} 
For a model $(\sM, \xi)$, 
the following conditions are equivalent. 
\begin{itemize}
\item[(i)] $\sM$ is e-autoparallel in $\sS$, and $\xi$ is an m-affine coordinate system.
\item[(ii)]
For every $\bu\in\bR^{n}$, there exists a 
$\bu\tbu$-efficient filtration for $\xi$. 
\end{itemize}
\end{theorem}

\begin{proofs}
According to Prop.~\!\ref{prop_autopara_affine_coordinate_quantum}, 
it suffices to show the equivalence of  (ii) and 
the condition that 
\be
\some \{F^i \},\;  
\any i, \;\; 
L^i = F^i - \xi^i, 
\label{gij_Lj_2_var}
\ee
where $L^i := \sum_j g^{ij} L_i$. 

We first show (ii) $\Ra$ \eqref{gij_Lj_2_var}. 
Fix $i\in \{1, \ldots , n\}$ arbitrarily, and  let $\filtPi=(\Pi_\eps)_{\eps>0}$ be an 
${\bbe}^i \transpose{(\bbe^i)}$-efficient 
filtration for $\xi$, where $\bbe^i = (\delta_j^i)$ 
denotes the $i$th vector of the natural basis of $\bR^n$. For each $\rho\in \sM$, we have 
\begin{align}
 \transpose{(\bbe^i)} \rred{V_\rho (\Pi_{\eps})} \bbe^i & = \int (\xihat^i - \xi^i (\rho))^2 \Tr (\rho \Pi_{\eps} (d\xihat)) 
 \noret 
&  \geq
 \Tr (\rho (F_\eps^i - \xi^i (\rho))^2) 
 =
  \| F_\eps^i - \xi^i (\rho)\|_\rho^2, 
 \label{eiVei_geq_|F-xi|^2}
\end{align}
where 
\be
F_\eps^i := \int \xihat^i \Pi_\eps (d \xihat) \in \cLh (\cH), 
 \nonumber
\ee
and $\|\cdot \|_\rho$ denotes the norm for the symmetrized inner product $\inprod{\cdot}{\cdot}_\rho$. 
(We also denote the norm for the metric $g_\rho$ by the same symbol.)  
Note that the inequality in \eqref{eiVei_geq_|F-xi|^2} follows from 
\[
\int (\xihat^i - \xi^i (\rho))^2  \Pi_{\eps} (d\xihat)
- 
 (F_\eps^i - \xi^i (\rho))^2 
 = \int (\xihat^i - F_\eps^i ) \Pi_{\eps} (d\xihat)  (\xihat^i - F_\eps^i ) \geq 0.
\]
From the local unbiasedness condition \eqref{locally_unbiased_Ai_Lj} applied to $\Pi_\eps$, we have
\be
\inprod{F_\eps^i - \xi^i (\rho)}{L_{j, \rho}}_\rho 
= \inprod{F_\eps^i}{L_{j, \rho}}_\rho = 
\delta_j^i = \inprod{L_\rho^i}{L_{j, \rho}}_\rho,
 \nonumber
\ee
where the first equality follows from \eqref{<LX>=0}. 
This means that $L_\rho^i$ is the orthogonal projection of $F_\eps^i - \xi^i (\rho)$ onto 
${\rm span}\{L_{j, \rho}\}_{j=1}^n$. Hence we have
\be
\| F_\eps^i - \xi^i (\rho)\|_\rho^2 = \| L_\rho^i\|_\rho^2 + \| F_\eps^i - \xi^i (\rho) - L_\rho^i\|_\rho^2. 
\label{F-xi_pythagorean}
\ee
The ${\bbe}^i \transpose{(\bbe^i)}$-efficiency of $\filtPi$ is represented as
\be
\lim_{\eps\downarrow 0} \transpose{(\bbe^i)} \rred{V_\rho (\Pi_{\eps})} \bbe^i = \transpose{(\bbe^i)} G_\rho^{-1} \bbe^i 
= g^{ii}(\rho) = \| L^i\|_\rho^2, 
 \nonumber
\ee
which, combined with \eqref{eiVei_geq_|F-xi|^2} and \eqref{F-xi_pythagorean}, yields 
\be
\lim_{\eps\downarrow 0} \| F_\eps^i - \xi^i (\rho) - L_\rho^i\|_\rho^2 =0.
 \nonumber
\ee
This implies that a $\rho$-independent Hermitian operator $F^i := \lim_{\eps\downarrow 0} F_\eps^i$ 
exists and satisfies $  L_\rho^i= F^i - \xi^i (\rho) $ for every $\rho\in\sM$, which concludes \eqref{gij_Lj_2_var}. 

We next show \eqref{gij_Lj_2_var} $\Ra$ (ii).  Let $\bu = (u_i) \in \bR^n$ 
be arbitrarily given, for which we will construct $\bu \tbu$-efficient filtration by assuming 
the existence of $\{F^i\}$ of \eqref{gij_Lj_2_var}. 
We can assume $\bu\neq 0$, and take a basis $\{\bu^1, \ldots , \bu^n\}$, $\bu^k = (u_i^k)$, 
of $\bR^n$ such that $\bu^1 = \bu$, whereby for each $k$ we define
\be Y^k := \sum_i u^k_i F^i. 
\label{def_Yk}
\ee
Let $\eps\in (0, 1)$, and take a positive probability vector $(p_1, \ldots , p_n)$ such that $p_1 = 1-\eps$. 
We define the estimator $\Pi_\eps$ by the following estimation procedure: randomly choose $k\in \{1, \ldots , n\}$ 
according to the probability distribution $(p_1, \ldots , p_n)$, measure the observable $Y^k$ to 
get an outcome $y$, and then estimate $\xi$ by $\xihat = g(k, y)$ using 
the function $g = \transpose{(g^1, \ldots , g^n)}$ defined by
\be
g^i (k, y) := \frac{w_k^i}{p_k} y, 
\label{def_gi}
\ee
where $[w^i_k] = [u^k_i]^{-1}$\red{:
\be
\randest{(p_k)}{k}{ \fbox{$\sum_i u^k_i\, F^i$}}{y}{\displaystyle \hat{\xi} = 
\frac{y}{p_k}\, \bw_k. }
 \label{estimation_procedure_filtration}
\ee
} 
 Invoking \eqref{gij_Lj_2_var} evaluated at an 
arbitrary point $\rho\in\sM$, we have
\begin{align}
g^i (k, Y^k) & =  \frac{w_k^i}{p_k} \sum_j u^k_j F^j 
= \frac{w_k^i}{p_k} \sum_j u^k_j (L_\rho^j + \xi^j (\rho)) 
\noret 
& = \gamma_k^i + \frac{w_k^i}{p_k} X^k = f^i (k, X^k),
 \nonumber
\end{align}
where $X^k$ and $f^i$ are those defined by \eqref{def_Xk} and \eqref{def_fi} with
\be
\gamma_k^i := \frac{w_k^i}{p_k} \sum_j u^k_j \xi^j (\rho) \red{= \frac{w_k^i}{p_k}\, \bu^k\cdot \xi(\rho).} 
\label{def_gammki}
\ee
\red{
Namely, \eqref{estimation_procedure_filtration} is equivalent to the following:
\be
\randest{(p_k)}{k}{ \fbox{$\sum_i u^k_i\, L^i_\rho$}}{x}{\displaystyle \hat{\xi} = 
\frac{\bu^k\cdot \xi(\rho)+x}{p_k} \, \bw_k.}
\ee
} 
Since this $\gamma_k^i$ satisfies \eqref{constraint_gammaki}, Lemma~\!\ref{lemma_Pi=(pi_f)} 
applies to conclude that $\Pi_\eps$ is locally unbiased at $\rho$ and satisfies,  for every $k$ and every $\rho\in \sM$, 
 \be
\transpose{(\bu^k)} V_\rho(\Pi_\eps)\bu^k=\frac{1}{p_k}\,\transpose{(\bu^k)} G_\rho^{-1}\bu^k
+ \sum_l p_l (a_l^k)^2, 
\label{uVu_proof_thm:autoparallel_m-affine}
 \ee
 where $ a_l^k  := \sum_i u^k_i (\gamma_l^i - \xi^i(\rho))$.
From \eqref{def_gammki}, we have
\[
 a_l^k  = \frac{1}{p_l} \sum_{i,j} u_i^k w_l^i u_j^l \xi^j(\rho) - 
 \sum_i u_i^k \xi^i (\rho) 
 = \Bigl( \frac{\delta_l^k}{p_k} -1 \Bigr) \sum_i u_i^k \xi^i (\rho)
\]
 and hence 
 \be
 \sum_l p_l (a_l^k)^2 = \sum_l p_l  \Bigl( \frac{\delta_l^k}{p_k} -1 \Bigr)^2 \Bigl( \sum_i u_i^k \xi^i (\rho) \Bigr)^2
 = \frac{1-p_k}{p_k}\Bigl( \sum_i u_i^k \xi^i (\rho) \Bigr)^2. 
 \label{uVu_proof_thm:autoparallel_m-affine_aux}
 \ee
 Thus, 
 letting $k=1$ in \eqref{uVu_proof_thm:autoparallel_m-affine} with \eqref{uVu_proof_thm:autoparallel_m-affine_aux}, 
  we obtain
 \be
 \tbu V_\rho (\Pi_\eps) \bu = \frac{1}{1-\eps} \tbu G_\rho^{-1} \bu + \frac{\eps}{1-\eps} \Bigl( \sum_i u_i^k \xi^i (\rho) \Bigr)^2.
 \nonumber
\ee
This implies that $\lim_{\eps\downarrow 0} \tbu V_\rho (\Pi_\eps) \bu
=  \tbu G_\rho^{-1} \bu$ for every $\rho$ 
and that $\filtPi := (\Pi_\eps)_{\eps\in (0, 1)}$ is a $\bu \tbu$-efficient filtration. 
\end{proofs}

\red{
\begin{remark}
\label{rem_variance_additional}
The term $ \frac{1-p_k}{p_k}\Bigl( \sum_i u_i^k \xi^i (\rho) \Bigr)^2$ in  \eqref{uVu_proof_thm:autoparallel_m-affine} with  \eqref{uVu_proof_thm:autoparallel_m-affine_aux}, which does not appear in \eqref{uVu_1/p}, is an additional estimation error 
incurred  
to ensure that the estimation procedure does not depend on the state $\rho \in \sM$. 
\end{remark}
} 

\medskip

The following proposition follows  immediately from Theorem~\!\ref{thm:autoparallel_m-affine} and Prop.~\!\ref{prop_iid_extension}.

\begin{proposition}
In the situation of Prop.~\!\ref{prop_iid_extension} where $(\tilde{\sM}, \tilde{\xi})$  
is the $\iidn$th i.i.d.\ extension of $(\sM, \xi)$, 
$(\tilde{\sM}, \tilde{\xi})$ has an efficient filtration if and only if $(\sM, \xi)$ 
has an efficient filtration. 
\end{proposition}

\section{A sufficient condition for the e-autoparallelity and its relation to the Gaussian states}
\label{sec_gaussian}

\begin{proposition}
\label{prop_cor_maintheorem}
For a model $(\sM, \xi)$, the following condition is sufficient 
for (i) and (ii) of Theorem~\!\ref{thm:autoparallel_m-affine}:
\begin{itemize}
\item For every positive weight $\wgt>0$, there exists a $\wgt$-efficient estimator for $\xi$. 
\end{itemize}
\end{proposition}

\begin{proofs}
Given $\bu\in \bR^n$ and $\eps >0$, arbitrarily, let $\Pi_\eps$ be a $(\bu \tbu+\eps I)$-efficient estimator. 
Then, for an arbitrary $\rho\in \sM$ and an arbitrary $\Pi'\in \cU (\rho, \xi)$, we have
\begin{align}
\tbu V_\rho (\Pi_\eps) \bu & \leq 
\tr ( (\bu \tbu + \eps I)  V_\rho (\Pi_\eps) ) 
\noret 
& \leq \tr ( (\bu \tbu + \eps I)  V_\rho (\Pi') ) 
= \tbu V_\rho (\Pi') \bu + \eps \tr (V_\rho (\Pi') ) .
 \nonumber
\end{align}
This implies that $\lim_{\eps\downarrow 0} \tbu V_\rho (\Pi_\eps) \bu \leq \tbu V_\rho (\Pi') \bu$ 
for every $\Pi'\in \cU (\rho, \xi)$, so that the filtration $\filtPi = (\Pi_\eps)_{\eps >0}$ is $\bu \tbu$-efficient. 
\end{proofs}

An important example of a model satisfying the condition of Prop.~\!\ref{prop_cor_maintheorem} 
is given by a model consisting of quantum Gaussian states. 
Strictly speaking, the model is mathematically out of our scope in that the underling Hilbert space is infinite-dimensional. 
Nevertheless, it is worthwhile to consider the relationship between this important model and the e-{\apty}, even at the expense of some rigor.

We begin by quickly reviewing the definition of quantum Gaussian states based on the description in Chapter~\!5 of  \cite{Holevo}. 
 Let $Z$ be an even-dimensional real linear space on which a symplectic form $\Delta (\cdot , \cdot)$ is given, and $U (\cdot)$ be an irreducible projective representation of 
$(Z, \Delta)$ on a separable Hilbert space $\cH$; i.e., $\{U (z)\,|\, z\in Z\}$ is a family of unitary operators on $\cH$ 
satisfying $\any z, z'\in Z$, $U(z) U(z') = e^{\sqrt{-1} \Delta (z, z')/2} U (z+ z')$ and having no nontrivial invariant subspace. 
For each $z\in Z$, a self-adjoint operator $R(z)$ 
is defined by $U(tz) = e^{\sqrt{-1}\, t R(z)}, t\in \bR$, and satisfies
\be
\any z, z'\in Z, \; 
[R (z), R(z') ] = - \sqrt{-1}\, \Delta (z, z') I.
\ee
Given a symmetric bilinear form $\alpha (\cdot, \cdot)$ on $Z$ satisfying $\any z, z'\in Z$, $\alpha (z, z) + \alpha (z', z') \geq \Delta (z, z')$ 
and a linear functional $\mu (\cdot)$ on $Z$,  
there uniquely exists 
a density operator $\rho$ on $\cH$ such that 
\be
\any z \in Z, \; 
\Tr ( \rho \, U (z) ) = e^{\sqrt{-1}\, \mu (z) - \frac{1}{2}\alpha (z, z)}. 
\ee
This $\rho$ is called the {\em Gaussian state} determined by $(\mu, \alpha)$, and satisfies
\be
\mu (z) = \inprod{I}{R(z)}_\rho = \expect{R(z)}_\rho, 
\ee
\be
\alpha (z, z') = \inprod{R(z) - \mu (z)}{R(z') - \mu (z')}_\rho, 
\ee
where $\inprod{\cdot}{\cdot}_\rho$ denotes the symmetrized inner product \eqref{symmetrized_inprod}.

Assuming that $\alpha$ and linearly independent $\mu_1, \ldots, \mu_n$ are 
arbitrarily given, consider the model $\sM = \{\rho_\xi\,|\, \xi = (\xi^1, \ldots, \xi^n)\in \bR^n\}$, 
where $\rho_\xi$ is the Gaussian state determined by $(\mu_\xi, \alpha)$ with 
$\mu_\xi := \sum_i \xi^i \mu_i$. 
We call  $\sM = \{\rho_\xi\}$ a {\em quantum Gaussian shift model}. 
Holevo showed (\S~\!6.9 of \cite{Holevo}) that the model has $\wgt$-efficient estimator for every positive weight $\wgt$. 
Namely, the sufficient condition presented in Prop.~\!\ref{prop_cor_maintheorem} is fulfilled. 
Hence, if $\cH$ were finite-dimensional, it would have been concluded that $\sM$ is e-{\ap} in $\sS$ and that 
$\xi$ is m-affine as a coordinate system of $\sM$. In reality, however, $\cH$ is infinite-dimensional, and 
the e-{\apty} for a model in $\sS$ is not given a mathematical definition in the framework of the present paper. 
Nevertheless, there is no essential difference from the finite-dimensional case.  In fact,  
we can verify that the model $(\sM, \xi)$ satisfies conditions (ii)-(iv) of Prop.~\!\ref{prop_autopara_affine_coordinate_quantum} 
as follows. According to \cite{Holevo}, the $i$th SLD $L_{i, \xi}$ is represented as
\be
L_{i, \xi} = R(z_i ) - \mu_\xi (z_i), 
\ee
where $z_i$ is the element of $Z$ determined by the condition $\any z\in Z$, $\mu_i (z) = \alpha (z_i, z)$.
The SLD Fisher information matrix $G = [g_{ij}]$ is given by 
\be
g_{ij} = \inprod{L_{i, \xi}}{L_{j, \xi}}_\xi = \alpha (z_i, z_j), 
\ee
which does not depend on the parameter $\xi$. Letting $F^i:= \sum_j g^{ij} R(z_j)$, where $G^{-1} = [g^{ij}]$, we have
\be
 \sum_j g^{ij} L_{j, \xi} = 
F^i - \sum_j g^{ij} \mu_\xi (z_j) = F^i - \xi^i,
\ee
where the second equality follows from $\mu_\xi (z_j) = \sum_{k}  \xi^k \mu_k (z_j)$ and 
$\mu_k (z_j) = \alpha (z_k, z_j) = g_{kj}$.  
We thus have verified that $(\sM, \xi)$ satisfies condition (iii) of Prop.~\!\ref{prop_autopara_affine_coordinate_quantum}, which is 
evidently equivalent to (ii) and (iv) even in this case. 
Hence, we may consider that the model also satisfies condition (i) at least in a naive sense. In order to 
mathematically justify this consideration, we need a rigorous treatment of $\sS (\cH)$ as an infinite-dimensional 
manifold equipped with an \rred{information-geometric} structure, which is out of scope of the present paper. 

The fact that $G = [g_{ij}]$ is constant on $\sM$ means that $(\sM, g)$ is a Euclidean manifold and 
the m-affine coordinate system $\xi$ is also affine w.r.t.\ the flat Levi-Civita connection of $g$. 
This implies that $\sM$ is dually flat and that the e, m-connections on $\sM$ coincide with the Levi-Civita connection.  
Note also that the SLDs $\{L_{i, \xi}\}$ do not commute and hence $\sM$ has no efficient estimator. 

\medskip

\begin{remark}
Let $\alpha$ be arbitrarily fixed, and let $\sN := \{ \rho_\mu \,|\, \mu\in Z^*\}$, where $Z^*$ denotes the 
dual linear space of $Z$ and  $\rho_\mu$ denotes the Gaussian state determined by $(\mu, \alpha)$. 
This is a special (maximal) case of $\sM$ treated above, so that $\sN$ is ``e-{\ap} in $\sS$'' 
in the same naive sense.  The SLD structure of $\sN$ is Euclidean, and  the model $\sM = \{\rho_\xi\,|\, \xi \in \bR^n\}$ 
treated above, where $\mu_\xi = \sum_i \xi^i \mu_i$, forms an e,m-{\ap} submanifold of $\sN$. 
Generally, a submanifold $\sM$ of $\sN$ is e,m-{\ap} in $\sN$ iff there exists an affine subspace 
${\mathcal A}$ of $Z^*$ such that $\sM = \{ \rho_\mu\, |\, \mu \in {\mathcal A}\}$, which is 
represented as $\sM = \{\rho_\xi\,|\, \xi \in \bR^n\}$ 
with $\mu_\xi = \mu_0 + \sum_{i=1}^n \xi^i \mu_i$. 
Note that the construction of $\wgt$-efficient estimator for the case $\mu_0 =0$ which was established by Holevo 
is 
immediately applied to this extended model, so that it satisfies the sufficient condition of Prop.~\!\ref{prop_cor_maintheorem}. 
\end{remark}

\section{\red{Characterization of e-autoparallelity in terms of estimation of scalar functions}
} 
\label{sec_another_characterization}

In this section we give another characterization to the e-autoparallelity 
w.r.t.\ the SLD structure by considering a different type of estimation problem. 
Before we get into the main discussion, some preliminaries on geometrical language are in order.

On a general Riemannian manifold $(\mM, g)$, a one-to-one correspondence between a tangent vector $X_p\in T_p (\mM)$ and a cotangent vector $\omega_p\in T^*_p (\mM)$ at a point $p\in \mM$ 
is naturally defined; denoting the correspondence by $\stackrel{g_p}{\longleftrightarrow}$, we have
\be
X_p \stackrel{g_p}{\longleftrightarrow} \omega_p \; \Leftrightarrow \; \any Y_p\in T_p(\mM), \; \omega_p (Y_p) = g_p (X_p, Y_p).
\ee 
\red{
In terms of the Dirac notation, $X_p \stackrel{g_p}{\longleftrightarrow} \omega_p$ means that 
$X_p = \ket{X_p}$ and $\omega_p = \bra{X_p}$ for the inner product $\langle \cdot |\cdot \rangle = g_p$. 
} 
This  is extended to the correspondence $\stackrel{g}{\longleftrightarrow}$ between a 
vector field $X\in \vf (\mM)$ and a differential 1-form $\omega\in \cD (\mM)$, 
where $\cD (\mM)$ denotes the totality of 1-forms on $\mM$, 
such that 
\begin{align}
X \stackrel{g}{\longleftrightarrow} \omega\; & \Leftrightarrow \; 
\any p\in \mM, \; X_p  \stackrel{g_p}{\longleftrightarrow} \omega_p 
\noret &  \Leftrightarrow \; 
\any Y\in \vf (\mM), \; \omega (Y) = g(X, Y).
\end{align}
When a coordinate system $\xi = (\xi^i)$ is given on $\mM$, and $X\in \vf (\mM)$ and $\omega \in \cD (\mM)$ are represented as $X= \sum_i X^i \partial_i$ and  $\omega = \sum_j \omega_j \,d\xi^j$ by 
functions $\{X^i\},  \{\omega_j\} \subset \func (\mM)$, we have 
\be
X \stackrel{g}{\longleftrightarrow} \omega 
\; \Leftrightarrow \; \any j, \; \omega_j = \sum_i X^i g_{ij} 
\; \Leftrightarrow \; \any i, \; X^i  = \sum_j \omega_j g^{ij}, 
\ee
where $g_{ij} = g(\partial_i, \partial_j)$ and  \rred{$[g^{ij}] = [g_{ij}]^{-1}$}. 

For a function $f\in \func (\mM)$, its {\em gradient} w.r.t.\ $g$ is defined as the vector field $X\in \vf (\mM)$ such that  
$X \stackrel{g}{\longleftrightarrow} df$, which we denote by $X={\rm grad}~\!f$.  This is represented as
\be
{\rm grad}~\!f= \sum_{i, j}  (\partial_i f) g^{ij} \partial_j. 
\label{grad_f_coordinate}
\ee

The correspondence $\stackrel{g_p}{\longleftrightarrow}$ induces an inner product and a norm on the cotangent space 
$T^*_p (\mM)$ such that $\stackrel{g_p}{\longleftrightarrow}$ is an isometry; i.e., 
$X_p\stackrel{g_p}{\longleftrightarrow} \omega_p$ $\Rightarrow$ $\|X_p\|_p = \|\omega_p\|_p$. 
In particular, we have 
\be
\| ({\rm grad}~\!f)_p\|^2_p = \| (df)_p\|^2_p = \sum_{i, j} g^{ij} (p)\, \partial_i f (p)\, \partial_j f (p) . 
\label{|grad_f|=|df|}
\ee

Now, we are ready to start the main discussion of this section. 
Let $\sM$ be an $n$-dimensional submanifold of $\sS = \sS (\cH)$, and $f \in \func (\sM)$ be a smooth function on it. 
We consider the problem of estimating the scalar value $f(\rho)$ for unknown $\rho\in \sM$. 
An estimator is generally represented by a POVM $\Lambda = \Lambda (d \hat{t}\,)$, 
where $\hat{t}$ is a scalar variable representing an estimate for $t =f(\rho)$. 
The expectation $E_\rho (\Lambda)$ and 
the mean squared error (the variance in the unbiased case) $V_\rho (\Lambda)$ 
of $\Lambda$ for a state $\rho$ 
are defined by 
\begin{align}
 E_\rho (\Lambda) & :=
 \int \hat{t} \, \Tr \left( \rho\,  \Lambda (d\hat{t} \,)\right), 
 \label{def_expectation_Lambda}
\\
  V_\rho (\Lambda) &:= 
 \int (\hat{t} - f (\rho))^2  
\, \Tr \left(\rho\, \Lambda (d\hat{t}\,)\right).
 \label{def_variance_Lambla}
\end{align}

\red{
\begin{remark} 
\label{remark_nuisance}
When $f$ is a component, say $\xi^1$, of a coordinate system $\xi = (\xi^i)$ of $\sM$, our problem treated here can be regarded as an example  of 
estimation with nuisance parameters (see e.g.\ \cite{suzuki_nuisance_review}), with $\xi^1$ being the parameter of interest and $\xi^2, \ldots , \xi^n$  being the nuisance parameters. 
Note, however, that the problem itself depends only on $f = \xi^1$, not on the nuisance parameters, 
and that in our coordinate-free description, the nuisance parameters as well as their parameter orthogonalization \cite{suzuki_nuisance_review} do not appear explicitly. 
See Remark~\ref{remark_local_orthogonalization} below. 
\end{remark}
} 

Localizing the unbiasedness condition $E (\Lambda) = f$, where 
the LHS denotes the function $\sM \rightarrow \bR, \; \rho\mapsto E_{\rho} (\Lambda)$, 
we say that $\Lambda$  is \emph{locally unbiased} for $f$ at $\rho\in \sM$ when 
\be
E_\rho (\Lambda) = f(\rho) \quad \text{and} \quad (d E (\Lambda))_\rho = (df)_\rho.
\label{locally_unbiased_Lambda}
\ee
When a coordinate system $\xi = (\xi^i)$  is arbitrarily given on $\sM$, the 
second condition in \eqref{locally_unbiased_Lambda} is expressed as
\be
\any i \in \{1, \ldots , n\}, \;\; 
\partial_i E_\rho  (\Lambda) = \partial_i f (\rho), 
\label{locally_unbiased_Lambda_var}
\ee
where $\partial_i E_\rho  (\Lambda)$ and $\partial_i f (\rho)$ denote the 
derivatives of the functions $E (\Lambda)$ and $f$ by 
$\partial_i = \frac{\partial}{\partial\xi^i}$ evaluated at $\rho$. 
We denote by $\cU(\rho, f)$ 
the totality of locally unbiased estimators for $f$ at $\rho$. 

\begin{proposition} 
\label{prop_min_V(Lambda)}
For any $f\in \func (\sM)$ and any $\rho\in \sM$, we have
\be
\min_{\Lambda\in \cU(\rho, f)} V_\rho (\Lambda) = \| (d f)_\rho \|_\rho^2, 
\label{min_V(Lambda)}
\ee
where $\|\cdot\|_\rho$ denotes the norm for cotangent vectors w.r.t.\ the SLD metric on $\sM$.
The minimum of \eqref{min_V(Lambda)} is achieved by the spectral measure of 
the observable 
\be F_\rho := f(\rho) + \sum_{i} \partial_i f (\rho) \, L^i_\rho, 
\label{def_Frho}
\ee
where $L^i_\rho := \sum_j g^{ij} (\rho) L_{i, \rho}$ and $L_i := L_{\partial_i}$. 
\end{proposition}

\begin{proofs}  Given an estimator $\Lambda$, let 
\be
A := \int \hat{t}\, \Lambda (d\hat{t}) \in \cLh. 
 \nonumber
\ee
Then the local unbiasedness 
of $\Lambda$ at $\rho$ is represented as
\be
\expect{A}_\rho = f(\rho) \quad\text{and}\quad \any i, \; \partial_i \expect{A}_\rho = \partial_i f (\rho), 
\label{locally_unbiased_A}
\ee
and we have
\be
V_\rho (\Lambda) \geq \expect{(A - f(\rho))^2}_\rho = \| A - f(\rho) \|^2_\rho,  
\label{V(Lambda)_geq_|A-f|^2}
\ee
where $\| \cdot \|_\rho$ denotes the norm w.r.t.\ the symmetrized inner product $\inprod{\cdot}{\cdot}_\rho$ on $\cLh$. 
Noting that the second condition of \eqref{locally_unbiased_A} 
is equivalent to 
\be
\any i, \; \inprod{L_{i, \rho}}{A - f(\rho)}_\rho 
= \inprod{L_{i, \rho}}{A}_\rho
= \Bigl\langle L_{i, \rho}, \sum_j \partial_j f (\rho) L^j_\rho \Bigr\rangle_\rho, 
 \nonumber
\ee
where the first equality follows from \eqref{<LX>=0}, 
we see that $\sum_j \partial_j f (\rho) L^j_\rho$ is the orthogonal projection of $A - f(\rho)$ onto 
the space ${\rm span}~\!\{L_{j, \rho}\}_{j=1}^n$. 
Hence we have 
\begin{align}
 \| A - f(\rho) \|^2_\rho &=  \|\sum_j \partial_j f (\rho) L^j_\rho \|_\rho^2  + \| A - f(\rho) - \sum_j \partial_j f (\rho) L^j_\rho \|_\rho^2 
\noret 
&\geq \|\sum_j \partial_j f (\rho) L^j_\rho \|_\rho^2 
\noret 
& = 
 \sum_{i, j} g^{ij} (\rho)\, \partial_i f(\rho)\,  \partial_j f(\rho) =\| (d f)_\rho \|_\rho^2.
 \label{|A-f|^2_geq_|df|^2}
\end{align}
The inequality in \eqref{V(Lambda)_geq_|A-f|^2} holds with equality when $\Lambda$ is the spectral measure of 
$A$, and the inequality in \eqref{|A-f|^2_geq_|df|^2} holds with equality 
when $A = f(\rho) +  \sum_j \partial_j f (\rho) L^j_\rho = F_\rho$. 
These observations prove the proposition. 
\end{proofs}

\red{
\begin{remark}
Prop.~\!\ref{prop_min_V(Lambda)} is essentially equivalent to Theorem~7.2 of \cite{amanag} when 
the symmetrized inner product \eqref{symmetrized_inprod} is taken as what is called a generalized inner product in \cite{amanag} 
(which is the same as a family of inner products $\{\inprod{\cdot}{\cdot}_\rho \,|\, \rho \in \sS (\cH)\}$ introduced in 
Section~\ref{sec_quantum_e-autoparallelity} of the present paper). 
\end{remark}
} 

\red{
\begin{remark}
\label{remark_local_orthogonalization}
Letting $f = \xi^1$ in \eqref{min_V(Lambda)}, 
we have
\be
\min_{\Lambda\in \cU(\rho, \xi^1)} V_\rho (\Lambda) = \| (d \xi^1)_\rho \|_\rho^2 = g^{11} (\rho). 
\label{min_V(Lambda)_xi1}
\ee
This formula is equivalent to Theorem~5.3 of \cite{suzuki_nuisance_review}, where $g^{11} (\rho)$ 
is expressed as the inverse of the Schur complement $\tilde{g}_{11} (\rho)$ of the matrix $G_\rho = [g_{ij} (\rho)]$. 
Letting $(\tilde{\partial}_1)_\rho$ be the $g_\rho$-projection of $(\partial_1)_\rho$ onto the orthogonal complement of 
${\rm span} \{(\partial_2)_\rho, \ldots (\partial_n)_\rho\}$, the Schur complement is represented as 
$\tilde{g}_{11} (\rho) = g_\rho ((\tilde{\partial}_1)_\rho, (\tilde{\partial}_1)_\rho) = 
 \inprod{\tilde{L}_{1, \rho}}{\tilde{L}_{1, \rho}}_\rho$, where $\tilde{L}_{1, \rho} :=
  L_{(\tilde{\partial}_1)_\rho} $, and the conversion from 
  $(\partial_1)_\rho$ to $(\tilde{\partial_1})_\rho$, or from $L_{1, \rho}$ to $\tilde{L}_{1, \rho}$, is called  the local parameter orthogonalization in \cite{suzuki_nuisance_review}. 
\end{remark}
} 

Based on Proposition~\!\ref{prop_min_V(Lambda)}, we call an estimator $\Lambda$ \emph{locally efficient} for $f$ at $\rho$ when 
$\Lambda \in \cU (\rho, f)$ and $V_\rho (\Lambda) = \|d f\|_\rho^2$, and call it \emph{efficient} for $f$ 
when it is  locally efficient for $f$ at every $\rho\in \sM$.  Note that, unlike the case of estimation for 
multi-dimensional coordinates $(\xi^i)$ where the infimum in \eqref{eq_inf_uVu} cannot be replaced with minimum in general, 
there always exists a locally efficient estimator for a scalar function $f$ 
at each $\rho$. 
Furthermore, since a locally efficient estimator is obtained as the spectral measure 
of an observable as is shown in the proof of Prop.~\!\ref{prop_min_V(Lambda)}, it suffices to treat only 
estimators represented by Hermitian operators. Note that an estimator $F \in \cLh$ is 
efficient for $f$ iff 
\be
\any \rho\in \sM, \; \expect{F}_\rho = f(\rho) \quad 
\text{and}\quad
V_\rho (F) := \bigl\langle (F- \expect{F}_\rho)^2\bigr\rangle_\rho = \| (df)_\rho\|_\rho^2.
\label{efficiency_F}
\ee

We define 
\be
\cE (\sM) := \{ f\in \func (\sM)\,|\, \text{there exists an efficient estimator for $f$} \}.
\label{def_E(M)}
\ee

\begin{proposition}
\label{prop_efficient_f_1}
For a function $f\in \func (\sM)$, the following conditions are equivalent. 
\begin{itemize}
\item[(i)] $f\in \cE (\sM)$.
\item[(ii)] $\exists F\in \cLh, \; F - f = \sum_i (\partial_i f)\, L^i$.
\item[(iii)] $\exists F\in \cLh, \; F - \expect{F}|_{\sM} = \sum_i (\partial_i f)\, L^i$.
\item[(iv)] ${\rm grad}~\!f$ is e-parallel (i.e.\ parallel w.r.t.\ the e-connection on $\sS$). 
\end{itemize}
In (ii), the observable $F$ gives an efficient estimator for $f$. 
\end{proposition}

\begin{proofs}
From Prop.~\!\ref{prop_min_V(Lambda)}, it immediately follows that 
 (i) $\Leftrightarrow$ (ii) and that $F$ in (ii) gives an efficient estimator for $f$. 

It is obvious that (ii) $\Rightarrow$ (iii). 
To show the converse, 
assume that an operator $F\in \cLh$ satisfies $F - \expect{F}|_{\sM} = \sum_i (\partial_i f)\, L^i$. Then we have
\begin{align}
\partial_i \expect{F} = \inprod{L_i}{F} = \inprod{L_i}{F - \expect{F}|_{\sM}} = 
\bigl\langle L_i , \sum_j (\partial_j f)\, L^i \bigr\rangle
=
\partial_i f, 
 \nonumber
\end{align}
which implies that $\some c\in \bR$, $f = \expect{F}|_{\sM} + c$. 
Redefining $F:= F + c$, we have $F - f = \sum_i (\partial_i f) L^i$. This proves  (iii) $\Rightarrow$ (ii).

Let $X:= {\rm grad}~\!f$. Then \eqref{grad_f_coordinate} yields 
\be
L_X  = \sum_{i, j} (\partial_i f) g^{ij} L_j 
= \sum_i (\partial_i f)  L^i,
 \nonumber
\ee
and Cor.~\!\ref{cor_e-parallel_vactor_field_extended} yields
\be
\text{$X$ is e-parallel} \; \Leftrightarrow \; \some F\in \cLh, \; L_X = F - \expect{F}|_{\sM}.
 \nonumber
\ee
Thus we obtain  (iii) $\Leftrightarrow$ (iv). 
\end{proofs}

\begin{corollary}
$\cE (\sM)$ is an $\bR$-linear space.
\end{corollary}

\begin{proofs}
Obvious from Prop.~\!\ref{prop_efficient_f_1}. 
\end{proofs}

\begin{proposition}
\label{prop_X_e-parallel_grad}
For a vector field $X\in \vf (\sM)$, we have
\be
\text{$X$ is e-parallel}\;\Leftrightarrow \; \some f\in \cE (\sM), \; X = {\rm grad}\, f.
\label{eq_prop_X_e-parallel_grad}
\ee
\end{proposition}

\begin{proofs}
The implication $\Leftarrow$ follows from (i) $\Rightarrow$ (iv) in Prop.~\!\ref{prop_efficient_f_1}. 
To show the converse, assume that $X$ is e-parallel. Then, according to 
Cor.~\!\ref{cor_e-parallel_vactor_field_extended}, 
 there exists $F\in \cLh$ such that 
$L_X = F -f$, where  $f:= \expect{F}|_{\sM}$. For any $Y\in \vf (\sM)$ we have
\be
g(X, Y) = \inprod{L_Y}{F-f} =  \inprod{L_Y}{F} \stackrel{\star}{=} Y\expect{F}|_{\sM} = Y f
= df (Y), 
\label{g(XY)=df(Y)}
\ee
where $\stackrel{\star}{=}$ follows from \eqref{def_LX_1}. 
This implies that $X = {\rm grad}~\! f$. Since $X$ is e-parallel, 
it follows from (iv) $\Rightarrow$ (i) in Prop.~\ref{prop_efficient_f_1} that $f\in \cE (\sM)$.  Thus, $\Rightarrow$ in \eqref{eq_prop_X_e-parallel_grad} 
has been verified. 
\end{proofs}

\medskip

Define 
\be
d \cE (\sM) := \{ df \, |\, f \in \cE (\sM) \} \; \subset \cD (\sM).
\ee
Since $df = d f' \; \Leftrightarrow \; f - f' =$ const., we have the natural identification 
$d\cE (\sM)  \simeq \cE (\sM) / \bR$. 
We also define 
\be
\vfparae (\sM) := \{ X \in \vf (\sM) \,|\, \text{$X$ is e-parallel}\}.
\ee
Then we have the following proposition.

\begin{proposition}
\label{prop_dim_E_X_epara}
The correspondence $\stackrel{g|_{\sM}}{\longleftrightarrow}$ establishes a linear isomorphism between 
$\vfparae (\sM)$ and $d \cE (\sM)$. As a consequence, we have
\be 
\dim d \cE (\sM) = \dim \cE  (\sM) -1 = \dim \vfparae (\sM) \leq \dim \sM.
\label{eq_prop_dim_E_X_epar}
\ee
\end{proposition}

\begin{proofs} It suffices to show that for an arbitrary pair $(X, \omega) \in \vf (\sM) \times \cD (\sM)$ satisfying 
$X \stackrel{g|_{\sM}}{\longleftrightarrow} \omega$, the following equivalence holds:
\be
 X \in \vfparae (\sM) \; \Leftrightarrow \; \some f\in \cE (\sM), \; \omega = df. 
\label{eq_proof_prop_dim_E_X_epara}
\ee
Since $\omega = df \; \Leftrightarrow \; X = {\rm grad}~\!f$, this is just Prop.~\!\ref{prop_X_e-parallel_grad}. 
\end{proofs}

Now, we present two theorems for characterization of the e-{\apty} in terms of $\cE (\sM)$. 

\begin{theorem} 
\label{thm_another_characterization_1}
For an $n$-dimensional submanifold $\sM$ of $\sS$, 
the following conditions are equivalent.
\begin{itemize}
\item[(i)] $\sM$ is e-{\ap} in $\sS$. 
\item[(ii)] $\dim \cE (\sM) = n+1$. 
\end{itemize}
\end{theorem}

\begin{proofs}
We have  (i) $\Leftrightarrow$ $\dim \vfparae (\sM) =n$ by Prop.~\!\ref{prop_equiv_autoparallel_general}, 
and $\dim \vfparae (\sM) =n$ $\Leftrightarrow$ (ii)  by Prop.~\!\ref{prop_dim_E_X_epara}. 
\end{proofs}

\begin{theorem}
\label{thm_another_characterization_2}
For an $n$-dimensional model $(\sM, \xi)$, the following conditions are equivalent.
\begin{itemize}
\item[(i)] $\sM$ is e-{\ap} in $\sS$, and $\xi$ is an m-affine coordinate system. 
\item[(ii)]  $\any i\in \{1, \ldots , n\},  \; \xi^i \in \cE (\sM)$. 
\item[(iii)]  $\disp \cE (\sM) = \bigl\{ c + \sum_{i=1}^n u_i \xi^i \,\big|\, (c, u_1 , \ldots , u_n)\in \bR^{n+1} \bigr\}$. 
\end{itemize}
\end{theorem}

\begin{proofs}
Let $X^i :=  {\rm grad}~\!\xi^i = \sum_j g^{ij} \partial_j$.
Then we have 
\begin{align}
\text{(i)} \; &\Leftrightarrow \; \any i, \; X^i \in  \vfparae (\sM) 
\; \Leftrightarrow \; 
 \text{(ii)}, 
  \nonumber
\end{align}
where the first $\Leftrightarrow$ follows from Prop.~\!\ref{prop_autopara_affine_coordinate_general} 
and the second $\Leftrightarrow$ follows from Prop.~\!\ref{prop_efficient_f_1}. 

Next, noting that  constant functions on $\sM$ belong to $\cE (\sM)$, we have 
\begin{align}
\text{(ii)} \; &
\Leftrightarrow \; \bigl\{ c + \sum_{i=1}^n u_i \xi^i \,\big|\, (c, u_1 , \ldots , u_n)\in \bR^{n+1} \bigr\} 
\subset  \cE (\sM) 
\noret &  \Leftrightarrow \; \text{(iii)}, 
 \nonumber
\end{align}
where the second $\Leftrightarrow$ follows since $\dim \cE (\sM) \leq n+1$ by \eqref{eq_prop_dim_E_X_epar} 
and $\{1, \xi^1, \ldots , \xi^n\}$ are linearly independent. 
\end{proofs}

\red{
\begin{remark}
Suppose that condition (ii) in Theorem~\ref{thm_another_characterization_2} holds, and 
let $F^1, \ldots, F^n$ be the efficient estimators of $\xi^1, \ldots , \xi^n$, respectively, which means that 
$\expect{F^i}|_{\sM} = \xi^i$ and 
$\expect{(F^i - \xi^i)^2}|_{\sM} = g^{ii}$ hold for every $i$. 
If, in addition,  the operators $F^1, \ldots, F^n$ commute, their simultaneous measurement $\Pi (d\xihat)$ defines an efficient estimator 
of the coordinate system $\xi$. 
Conversely, if $\Pi (d\xihat)$ is an efficient estimator of $\xi$, using a similar argument to the proof of Theorem~\ref{thm:autoparallel_m-affine}, we can show that the operator $F^i := \int \xihat^i \Pi (d\xihat)$ 
satisfies $F^i - \xi^i(\rho) = L_\rho^i$ for every $\rho\in\sM$ and every $i$,  which implies that $F^i$ is an efficient estimator of $\xi^i$ and that $F^1 , \ldots , F^n$ commute. In this sense, Theorem~\ref{thm_another_characterization_2} can be regarded as 
an extension of Theorem~\ref{thm:commutative}. 
\end{remark}
} 

\begin{remark}
If we replace $\sS (\cH)$ by  $\cP (\Omega)$ in Theorems \ref{thm_another_characterization_1} and \ref{thm_another_characterization_2}, 
these theorems hold as they are in the classical case. 
When the coordinate functions $\xi^1, \ldots , \xi^n$ satisfy condition (ii) in 
Theorem~\!\ref{thm_another_characterization_2}, they have their efficient estimators 
$F^1 , \ldots , F^n$, which are functions $\Omega \rightarrow \bR$ in this case,  
and the map $(F^1 , \ldots , F^n) : \Omega \rightarrow \bR^n$ becomes an efficient estimator 
for $\xi = (\xi^1, \ldots , \xi^n)$. 
Thus, we see that the equivalence (i) $\Leftrightarrow$ (ii) in the theorem becomes just Theorem~\!\ref{thm_classical_exponential_family}. 
\end{remark}

\medskip

Finally, we present three propositions that will aid in understanding the above results  in a 
purely geometric context, whose proofs are given in Appendix~\ref{sec_proofs_prop_E(S)_prop_E(M)}. 

\begin{proposition}
\label{prop_V(F)_|d<F>|^2}
 $\any F\in \cLh, \any\rho\in\sS, 
 \; V_\rho (F) := \bigl\langle (F - \expect{F}_\rho)^2\bigr\rangle_\rho =  \| (d\expect{F})_\rho\|_\rho^2$. 
\end{proposition}

\begin{proposition}
\label{prop_E(S)}
We have
\begin{align}
\cE (\sS) & = \{\expect{F}\,|\, F \in \cLh\} 
\noret 
&= \{ f\in \func (\sS)\,|\, \text{${\rm grad}~\!f$ is e-parallel} \} 
\noret 
& = \{ f\in \func (\sS)\,|\, \text{$d f$ is m-parallel} \}, 
\label{eq_prop_E(S)}
\end{align}
where a 1-form $\omega \in \cD (\sS)$ is said to be m-parallel when
\be
\any X, Y\in \vf (\sS), \; (\nabla^{(\m)}_X \omega) (Y) := X \omega (Y) - \omega \bigl(\nabla^{(\m)}_X Y\bigr) = 0.
\label{omega_m-parallel}
\ee
\end{proposition}

\begin{proposition}
\label{prop_E(M)}
For an arbitrary submanifold $\sM$ of $\sS$, we have
\begin{align}
\cE (\sM) = \bigl\{ f\in \func (\sM)\, \big| \, 
& \some \tilde{f}\in \cE (\sS), \noret 
& \; f = \tilde{f}|_{\sM} \;\;\text{and}\;\; 
\any \rho\in \sM , \; \| (df)_\rho\|_\rho = \| (d\tilde{f})_\rho\|_\rho\, \bigr\}.
\label{eq_prop_E(M)}
\end{align}
\end{proposition}

\medskip

As these propositions suggest, the discussion for $\cE (\sM)$ given in this section 
can be extended to a more general geometrical setting. Let us recall  the situation treated 
in section~\ref{sec_basic_autoparallelity} where 
a manifold $\mS$ is provided with a Riemannian metric $g$ together with  
mutually dual \red{curvature-free connections $\nabla$ and $\nabla^*$,  
and  now assume the flatness of $\nabla^*$ additionally.} 
We define 
\begin{align}
\cE (\mS) := &
\{ f \in \func (\mS)\,|\, \text{${\rm grad}~\!f$ is $\nabla$-parallel} \} 
\noret 
 = & \{ f \in \func (\mS)\,|\, \text{$d f$ is $\nabla^*$-parallel} \}, 
\end{align}
where we have invoked the fact that the correspondence 
$\vf (\mS) \ni X \stackrel{g}{\longleftrightarrow} \omega \in \cD (\mS)$ 
implies (cf.\ \eqref{X_e-parallel_omega_m-parallel} in Appendix~\ref{sec_proofs_prop_E(S)_prop_E(M)}) 
\be
\text{$X$ is $\nabla$-parallel} \; \Leftrightarrow \; \text{$\omega$ is $\nabla^*$-parallel}. 
\label{X_nabla-parallel_omega_nabla*-parallel}
\ee
Given a submanifold $\mM$ of $\mS$, let 
\begin{align}
\cE (\mM) := \bigl\{ f\in \func (\mM)\, \big| \, 
\some \tilde{f}\in \cE (\mS),& \; f = \tilde{f}|_{\mM} \;\;\text{and}\;\; 
\noret 
& \any \rho\in \mM , \; \| (df)_\rho\|_\rho = \| (d\tilde{f})_\rho\|_\rho\, \bigr\}. 
\end{align}
Then it is not difficult to verify that Theorems \ref{thm_another_characterization_1} and \ref{thm_another_characterization_2} 
as well as their proofs are extended to this general situation almost as they are. 

It should be noted that the flatness of $\nabla^*$ is essential in that it ensures $\dim \cE (\mS) = \dim \mS +1$.  
To clarify the role of the flatness, 
let us consider a more general situation by  removing the 
assumption that $\nabla$ is curvature-free and $\nabla^*$ is flat, assuming only that 
they are dual w.r.t.\ $g$.  We start from the following general identity: 
for any 1-form $\omega\in \cD (\mS)$ and any vector fields $X, Y \in \vf (\mS)$, we have
\begin{align}
(d \omega) (X, Y) & := X \omega (Y) - Y \omega (X) - \omega ([X, Y]) 
\noret & =
(\nabla^*_X \omega) (Y) + \omega (\nabla^*_X Y) - 
(\nabla^*_Y \omega) (X) - \omega (\nabla^*_Y X) - \omega ([X, Y]) 
\noret & =
(\nabla^*_X \omega) (Y) - (\nabla^*_Y \omega) (X) 
+ \omega (\torsion^{(\nabla^*)} (X, Y)),
\end{align}
where $\torsion^{(\nabla^*)}$ denotes the torsion of $\nabla^*$: 
$\torsion^{(\nabla^*)} (X, Y) = \nabla^*_XY - \nabla^*_Y X - [X, Y]$. 
When $\nabla^*$ is torsion-free, this implies that for any $\omega\in \cD (\mS)$
\be
\text{$\omega$ is $\nabla^*$-parallel} \; \Rightarrow \; d \omega = 0 
\; \Leftrightarrow \; \some f\in \func (\mS), \; \omega = df
\label{omega_nabla*-parallel_df}
\ee
(see Remark \ref{remark_nomenclature}) and that for any $X\in \vf (\mS)$ 
\be
\text{$X$ is $\nabla$-parallel} \; \Rightarrow \; \some f\in \func (\mS), \; X = {\rm grad}~\!f.
\ee
This leads to
\be
d \cE (\mS) := \{d f\,|\,f \in \cE (\mS)\} = 
\{ \omega \,|\, \text{$\omega$ is $\nabla^*$-parallel}\}
\ee
and hence 
\be
\dim \cE (\mS) = \dim \{ \omega \,|\, \text{$\omega$ is $\nabla^*$-parallel}\} + 1 
= \dim \vf _{\text{$\nabla$-par}} (\mS) +1,
\ee
where $ \vf _{\text{$\nabla$-par}} (\mS)$ denotes the totality of $\nabla$-parallel vector fields on $\mS$. 
If, in addition, $\nabla$ is curvature-free, which is equivalent to the flatness of $\nabla^*$, then 
$ \dim \vf _{\text{$\nabla$-par}} (\mS) = \dim \mS$, and we obtain $\dim \cE (\mS) = \dim \mS +1$.

\section{Integrability conditions}
\label{sec_integrability}

We begin by considering the general situation where  an affine connection $\nabla$ is given on a manifold $\mS$. 
For an arbitrary point $p\in \mS$ and an arbitrary $1$-dimensional subspace $V$ of 
the tangent space $T_p (\mS)$, there always exists a $\nabla$-{\ap} curve, i.e. a $\nabla$-geodesic, that passes through 
$p$ in direction $V$.  
For the existence of multi-dimensional {\ap} submanifolds,  the situation differs greatly depending on whether $\nabla$ is flat or not.  
When $\nabla$ is flat (curvature-free and torsion-free), 
the {\ap} submanifolds are those determined by arbitrary affine constraints on $\nabla$-affine coordinates.  
This ensures that, for an arbitrary point $p\in \mS$ and an arbitrary linear subspace $V$ of the tangent space 
$T_p (\mS)$, there uniquely exists a $\nabla$-{\ap} submanifold $\mM$ satisfying $p\in \mM$ and 
$T_p (\mM) =V$.  This is the case with the e-connection on the space $\cP$ of probability distributions, 
for which the {\ap} submanifolds are the exponential families. 
We meet a similar situation in the quantum case if we consider the BKM structure:  
see Remark~\ref{remark_BKM}. 
When $\nabla$ is not flat, on the other hand, the existence of multi-dimensional {\ap} submanifolds is not ensured in general.  In this section we investigate conditions for existence of {\ap} submanifolds. 

Let us consider the case when $\nabla$ is curvature-free as in the e-connection on $\sS (\cH)$. 
According to (i) $\Leftrightarrow$ (iv) of Prop.~\!\ref{prop_equiv_autoparallel_general}, an $n$-dimensional submanifold $\mM$ of $\mS$ is $\nabla$-{\ap} iff 
there exist $n$ linearly independent $\nabla$-parallel vector fields 
$X^1, \ldots , X^n \in \vf (\mS)$ such that their restrictions $X^1|_{\mM}, \ldots , X^n|_{\mM}$ belong to $\vf (\mM)$.  This means that 
$\mM$ is an integral manifold of $\{X^1, \ldots , X^n\}$, or equivalently that $\mM$ is an integral manifold 
of the $n$-dimensional distribution 
\be \cV : \mS \ni p \mapsto V_p := {\rm span} \{X^1_p, \ldots , X^n_p\} \subset T_p (\mS), 
\ee
which is 
$\nabla$-parallel in the sense that $\any p, q\in \mS, \; \para^{(\nabla)}_{p,q} (V_p) = V_q$, 
where  $\para^{(\nabla)}_{p,q}$ denotes the parallel transport w.r.t.\ $\nabla$. 

\begin{proposition}
\label{prop_integrability}
Suppose that we are given a manifold $\mS$ with a curvature-free connection $\nabla$ and 
an $n$-dimensional $\nabla$-parallel distribution $\cV : \mS \ni p \mapsto V_p$. 
Define 
\be
\vf (\mS\colon\cV) := \{ X\in \vf (\mS)\,|\, \any p\in \mS, \, X_p \in V_p \}
\ee
and 
\be \vfparanabla (\mS\colon\cV) :=  \{ X\in\vf (\mS\colon\cV)\,|\, \text{$X$ is $\nabla$-parallel} \}.
\ee
Then the following conditions are equivalent.
\begin{itemize}
\item[(i)] For every $p\in \mS$, there exists a $\nabla$-{\ap} submanifold $\mM$ of $\mS$ satisfying $p\in \mM$ and $T_p(\mM) = V_p$. 
\item[(ii)] The distribution $\cV$ is involutive in the sense that $\any X, Y\in \vf (\mS\colon\cV) , \; [X, Y] \in \vf (\mS\colon\cV)$.
\item[(iii)] $\any X, Y\in \vfparanabla (\mS\colon\cV) , \; [X, Y] \in \vf (\mS\colon\cV)$.
\item[(iv)] The torsion  $\torsion^{(\nabla)}$ of $\nabla$ satisfies $\any p\in \mS, \; \torsion^{(\nabla)}_p (V_p\times V_p) \subset V_p$. 
\end{itemize}
\end{proposition}

\begin{proofs}
(i) is equivalent to the condition that for any point $p\in \mS$, there exists an 
integral manifold of $\cV$ containing $p$, and is equivalent to (ii) by the famous 
Frobenius theorem for integrability. 

(ii) $\Rightarrow$ (iii) is obvious, and (iii) $\Rightarrow$ (ii) follows since 
there exist $n$ linearly independent $\nabla$-parallel vector fields $\{X_1, \ldots , X_n\} 
\subset \vfparanabla (\mS\colon\cV)$, whereby every element of  $ \vf (\mS\colon\cV)$ 
is expressed as $\sum_i f_i X_i$ by some functions $\{f_1 , \ldots  , f_n\} \subset \func (\mM)$. 

For any $\nabla$-parallel vector fields $X$ and $Y$, we have
\be
\torsion^{(\nabla)} (X, Y) := \nabla_X Y - \nabla_Y X - [X, Y] = - [X, Y].
\label{torsion_parallel}
\ee
Hence (iii) is equivalent to 
\be \any X, Y\in \vfparanabla (\mS\colon\cV) , \;\; \torsion^{(\nabla)}(X, Y) \in \vf (\mS\colon\cV).
\label{involutive_torsion_parallel}
\ee
Since $\torsion^{(\nabla)}$ is a tensor field so that  
$(\torsion^{(\nabla)}(X, Y))_p = \torsion^{(\nabla)}_p (X_p, Y_p))$ holds 
at each point $p$, \eqref{involutive_torsion_parallel} is equivalent to (iv). 
\end{proofs}

\begin{remark}
\label{remark_foliation}
Condition (i) in Prop.~\!\ref{prop_integrability} 
(and hence (ii)-(iv) also) means that there exists a foliation 
$\mS = \bigsqcup_{\alpha} \mM_\alpha$ such that each leaf $M_\alpha$ is 
$\nabla$-{\ap} in $\mS$ and satisfies 
$T_p (M_\alpha) = V_p$ for every $p\in M_\alpha$.
\end{remark}

%

The following proposition is an immediate consequence of (i) $\Leftrightarrow$ (iv) in 
Prop.~\!\ref{prop_integrability}. 

\begin{proposition}
\label{cor_prop_integrability}
For a manifold $\mS$ with a curvature-free connection $\nabla$, the following conditions are equivalent.
\begin{itemize}
\item[(i)] For every point $p\in \mS$ and every linear subspace $V$ of $T_p (\mS)$, 
there exists a $\nabla$-{\ap} submanifold $\mM$ satisfying $p\in \mM$ and $T_p (\mM) =V$. 
\item[(ii)] $\any p\in \mS, \; \any X_p, Y_p\in T_p (\mS), \; \torsion^{(\nabla)}_p (X_p, Y_p) \in 
{\rm span} \{X_p, Y_p\}$, 
or equivalently, $\any X, Y\in \vf (\mS), \; \torsion^{(\nabla)} (X, Y) \in 
{\rm span}_{\func(\mS)} \{ X, Y\} := \{ f X + g Y\,|\, f, g\in \func (\mS)\}$. 
\end{itemize}
\end{proposition}

Condition (i) in the above proposition means that the condition for existence of autoparallel submanifolds 
is trivial as in the flat case, which corresponds to the vanishing torsion  $\torsion^{(\nabla)} =0$ implying condition (ii) obviously. 

\begin{remark}
\label{remark_totally_geodesic_revisited}
Let us revisit the relationship between autoparallelity and total \red{geodesicity} 
 described in 
Remark~\!\ref{remark_totally_geodesic} in the context of Prop.~\!\ref{cor_prop_integrability}. 
Suppose that $(\mS, \nabla)$ satisfies  conditions (i)-(ii) of  Prop.~\!\ref{cor_prop_integrability} and that 
a submanifold $\mM$ of $\mS$ is $\nabla$-totally geodesic. 
Given a point $p\in \mM$ arbitrarily, 
there exists a $\nabla$-{\ap} submanifold $\mN$ which satisfies 
$p\in \mN$ and $T_p (\mN) = T_p (\mM)$ by condition (i). 
Since $\mN$ is also $\nabla$-totally geodesic, 
we have $\mM = \mN$, so that $\mM$ is $\nabla$-{\ap}. Namely, condition (ii) together with the curvature-freeness 
of $\nabla$ implies the equivalence between $\nabla$-{\apty} and $\nabla$-total \red{geodesicity}. 
In fact, the curvature-freeness is unnecessary. 
\red{We have the following proposition, whose proof is given in Appendix~\ref{sec_proof_prop_totally_geodesic_autoparallel}.} 
\end{remark} 

\red{
\begin{proposition}
\label{prop_totally_geodesic_autoparallel}
Suppose that an affine connection $\nabla$ is given on a manifold $\mS$ whose 
torsion satisfies 
\be
\any X, Y\in \vf (\mS), \; \torsion^{(\nabla)} (X, Y) \in 
{\rm span}_{\func(\mS)} \{ X, Y\}.
\label{torsionXY_spanXY}
\ee
Then every $\nabla$-totally geodesic submanifold of $\mS$ is $\nabla$-autoparallel. 
\end{proposition}
} 

Let us apply Prop.~\!\ref{prop_integrability} to $\sS = \sS(\cH)$ with the SLD structure and its submanifolds. 
Let $\cA$ be an arbitrary linear subspace of $\cLh$, and define 
for each point $\rho\in \sS$ 
\begin{align}
V_{\cA, \rho} & := \{ X_\rho\in T_\rho (\sS)\, |\, \some A\in \cA, \; L_{X_\rho} = A - \expect{A}_\rho\}
\noret 
&= 
 \{ X_\rho\in T_\rho (\sS)\, |\,  L_{X_\rho} \in \cA + \bR \},
\label{def_V_A_rho}
\end{align}
where $\bR$ is identified with $\{ c I \, | \, c\in \bR\}$. 
Then $\cV_{\cA} : \sS \ni \rho \mapsto V_{\cA, \rho}$ defines an e-parallel distribution on $\sS$, 
whose dimension $\dim V_{\cA, \rho}$ is equal to $\dim \cA$ when $I\notin \cA$ and 
$\dim \cA -1$ otherwise. 
Every e-parallel distribution on $\sS$ is represented as $\cV_{\cA}$ by some $\cA$, 
and $\cV_{\cA} = \cV_{\cA'}$ iff 
$\cA + \bR = \cA' + \bR$.
This means that $\cA \mapsto \cV_{\cA}$ establishes 
 a one-to-one correspondence between linear subspaces of the quotient space $\cLh /\bR$ 
 and e-parallel distributions on $\sS$.

\begin{theorem}
\label{thm_SLD_integrability}
Given a subspace $\cA \subset \cLh$, the following conditions are 
equivalent. 
\begin{itemize}
\item[(i)]  For every $\rho\in \sS$, there exists an e-{\ap} 
(w.r.t.\ the SLD structure)
submanifold 
$\sM$ of $\sS$ satisfying $\rho\in \sM$ and 
$T_\rho (\sM) = V_{\cA, \rho}$. 
\item[(ii)] For every $\rho\in \sS$, 
\be 
\{ [ [A, B], \rho ] \,|\, A, B\in \cA \} \subset \{ \rho \circ C  \,|\, C\in \cA+\bR \}.
\ee
\end{itemize}
\end{theorem}

\begin{proofs}
From \eqref{SLD_torsion} it follows that for any $X_\rho, Y_\rho, Z_\rho \in T_\rho (\sS)$ 
\be
\torsion_\rho^{(\e)} (X_\rho, Y_\rho) = Z_\rho \; \Leftrightarrow \; 
\frac{1}{4} [ [L_{X_\rho}, L_{Y_\rho}], \rho] = \iota_* (Z_\rho) = \rho \circ L_{Z_\rho}.
 \nonumber
\ee
Hence, noting that $[A-\expect{A}_\rho, B-\expect{B}_\rho] = [A, B]$, 
we obtain (i) $\Leftrightarrow$ (ii) from (i) $\Leftrightarrow$ (iv) in Prop.~\!\ref{prop_integrability}. 
\end{proofs}

\red{
We have mentioned in Remark~\ref{remark_foliation} that 
the conditions in Prop.~\!\ref{prop_integrability} imply the existence of a foliation of $\mS$ 
with $\nabla$-{\ap} leaves. If, in addition, the dual connection of $\nabla$ w.r.t.\ a Riemannian metric $g$ is flat 
as in the case where $\nabla = \nablae$ and $\nabla^* = \nablam$ for an information-geometric structure $(g, \nablae, \nablam)$, then 
$\mS$ has also a foliation with $\nabla^*$-{\ap} leaves, which forms a pair of dual foliations 
as in the case of dually flat spaces (cf. \S~\!3.7 of \cite{amanag}).  Namely, we have 
the following proposition, whose proof is given in Appendix~\ref{sec_proof_prop_foliation}. 
} 

\red{
\begin{proposition}
\label{prop_foliation}
Given $(\mS, g, \nabla, \nabla^*)$, suppose that $\nabla$ is curvature-free and satisfies 
conditions (i)-(iv) in Prop.~\!\ref{prop_integrability} for a $\nabla$-parallel distribution $\cV : \mS \ni p \mapsto V_p$ 
and that $\nabla^*$ is flat. Then: 
\begin{enumerate}
\item[(1)] $\mS$ has two foliations $\mS = \bigsqcup_{\alpha} \mM_\alpha =  \bigsqcup_{\beta} \mN_\beta$ such that: 
\begin{enumerate}
\item $\any \alpha$, $\mM_\alpha$ is $\nabla$-{\ap} in $\mS$. 
\item  $\any \beta$, $\mN_\beta$ is $\nabla^*$-{\ap} in $\mS$. 
\item The leaves of the two foliations intersects orthogonally; i.e., if $p\in \mM_\alpha \cap \mN_\beta$, 
then  $T_p (\mM_\alpha) \perp T_p (\mN_\beta)$ w.r.t.\ the inner product $g_p$. 
\end{enumerate}
\item[(2)] Given $\nabla$-parallel vector fields $\{X^1 , \ldots , X^n\}$ spanning $\vfparanabla (\mS\colon\cV)$, 
there exist functions $\{\zeta^1, \ldots , \zeta^n\} \subset \func (\mS)$ such that $\any i$, $X^i \stackrel{g}{\longleftrightarrow} d\zeta^i$.  Letting $\zeta := (\zeta^i)$, we have: 
\begin{enumerate}
\item $\any \alpha$, \, $\zeta|_{\mM_\alpha}$ forms a $\nabla_{\mM_\alpha}^*$-affine coordinate system of $\mM_\alpha$, where  $\nabla_{\mM_\alpha}^*$ denotes the dual connection of $\nabla|_{\mM_\alpha}$ w.r.t.\ 
$g|_{\mM_\alpha}$ as defined in Section~\ref{sec_basic_autoparallelity} for \eqref{duality_on_M}. 
\item $\any \beta, \; \some c \in \bR^n,  \; \mN_\beta = \{ p\in \mS \,|\, \zeta (p) = c\}$. 
\end{enumerate}
\end{enumerate}
\end{proposition}
} 

\red{
\begin{corollary}
\label{cor_SLD_integrability_F}
Let $\{ F^1 , \ldots , F^n\}$ be Hermitian operators on $\cH$ such that $\{F^1 , \ldots , F^n, I\}$ are linearly 
independent, and suppose that $\cA := {\rm span}\, \{F^i\}$ satisfies condition (ii) in Theorem~\ref{thm_SLD_integrability}, 
which means that there exist $n^3$ functions $\{c^{ij}_k\} \subset \func (\sS)$ such that
\be
\any i, j, \; \any \rho\in  \sS, \;\; 
[[F^i, F^j], \rho ] = \sum_k c^{ij}_k (\rho) \, \rho \circ \bigl(F^k - \expect{F^k}_\rho\bigr).
\label{F_F_c_F}
\ee
\begin{itemize}
\item[(1)] The functions $\{c^{ij}_k\}$ satisfy
\be
\torsion^{(\e)} (X^i, X^j) = \frac{1}{4} \sum_k c^{ij}_k X^k, 
\ee
where $\{X^i\}$ denote the e-parallel vector fields on $\sS$ defined by $L_{X^i} = F^i - \expect{F^i}$. 
\item[(2)] Given a stete $\rho\in \sS$ arbitrarily, there exists an e-{\ap} submanifold $\sM$ of $\sS$ satisfying  $\rho\in \sM$ and 
$\any i$, $X^i |_{\sM} \in \vf (\sM)$. Letting $\xi^i := \expect{F^i}|_{\sM}$, $\xi = (\xi^i)$ forms an m-affine coordinate system of $\sM$. 
\item[(3)] In the situation of (2), the manifold 
\be
\sN := \{\rho' \in \sS \, |\, \any i, \; \expect{F^i}_{\rho'} = \xi^i (\rho)\}
\ee
is m-{\ap} in $\sS$ and intersects $\sM$  orthogonally at $\rho$. 
\item[(4)] Letting the state $\rho$  range over $\sS$ in (2) and (3),   a pair of 
dual foliations $\sS = \bigsqcup_{\alpha} \sM_\alpha =  \bigsqcup_{\beta} \sN_\beta$ 
 is generated as in Prop.~\!\ref{prop_foliation}. 
\end{itemize}
\end{corollary}
} 

\red{
\begin{proof}
(1), (2) and (4) are immediate from the discussion so far. 
It is not difficult to show (3) directly, while  it is also derived from  Prop.~\!\ref{prop_foliation} as follows. 
Let $\zeta^i := \expect{F^i}$. 
Invoking \eqref{<LX>=0}, \eqref{def_LX_3} and \eqref{def_g_2}  and noting that 
$L_{X^i}  = F^i - \zeta^i$, we have 
\[
\any Y\in \vf (\sS), \; 
g(X^i , Y) = \inprod{F^i-\zeta^i}{L_Y} =  \inprod{F^i}{L_Y}  = Y \zeta^i = d \zeta^i (Y), 
\]
which means that $X^i \stackrel{g}{\longleftrightarrow} d\zeta^i$. 
Now (3) follows from Prop.~\!\ref{prop_foliation}. 
\end{proof}
} 

\red{
\begin{example}
When  the operators  $F^1, \ldots , F^n$ commute in Cor.~\!\ref{cor_SLD_integrability_F}, 
which corresponds to the case  when \eqref{F_F_c_F} holds with vanishing  $\{c^{ij}_k\}$, 
the manifolds $\sM$ and $\sM_\alpha$ are 
\rred{quasi-classical exponential} 
families of the form \eqref{q_c:e-family_var}. 
In particular, for an arbitrary Hermitian operator $F$ such that $\{F, I\}$ are linearly independent, 
there always exists a pair of dual foliations $\sS = \bigsqcup_{\alpha} \sM_\alpha =  \bigsqcup_{\beta} \sN_\beta$ such 
that  
$\sM_\alpha$ is an e-geodesic of the form \eqref{q_c:e-family_var_1dim} and $\sN_\beta$ is 
an m-{\ap} submanifold of codimension $1$ represented by a constraint 
of the form $\expect{F}_\rho =$ const.
\end{example}
} 

\red{
A more interesting example of a pair of dual foliations will be given in Section~\ref{sec_qubit}. 
} 

\begin{remark}
Recall the situation when we defined the quantum Gaussian shift model 
$\sM =$ $\{\rho_\xi\,|\, \xi = (\xi^1, \ldots, \xi^n)\in \bR^n\}$ in Section~\ref{sec_gaussian}, and 
let $\cA :=$  ${\rm span}~\!\{R (z_1), \ldots , R(z_n)\}$.
Then, for any $A, B\in \cA$, we have $[A, B] =  c I$ with a purely imaginary constant $c$ 
 and hence $[[A, B], \rho] =0$. So, (ii) in Theorem~\!\ref{thm_SLD_integrability}  holds at least formally, 
 and the Gaussian model may be regarded as an integral manifold of $\cV_{\cA}$. Note, however, that 
 Theorem~\!\ref{thm_SLD_integrability} is not valid  in the infinite-dimensional case so that 
 (ii) does not imply (i), because 
 various mathematical problems arise that were not present in the finite dimensional case, such as 
 the fact that a positive operator does not always have finite trace and hence is not always normalizable. 
 \end{remark}

At the end of this section, we give an example of e-{\ap} submanifold 
that does not fall within the scope of Theorem~\!\ref{thm_SLD_integrability}. 
Let $\cB = \{\ket{1}, \ldots , \ket{d}\}$ with $d=\dim \cH$ 
be an arbitrary orthonormal basis of $\cH$, and let $\cL^{\cB} := 
\{ \sum_{i, j} a_{ij} \ket{i} \bra{j}\,|\, [a_{ij}] \in \bR^{d\times d}\}$, 
$\cLh^{\cB} := \cLh \cap \cL^{\cB}$  and 
$\sS^{\cB} := \sS \cap \cL^{\cB}$. 
That is, $\sS^{\cB}$ is the set of density operators whose matrix representations w.r.t.\ the basis $\cB$ are real matrices. 

\begin{proposition} 
\label{prop_real_e-ap}
$\sS^{\cB}$ is e-{\ap} in $\sS$ w.r.t.\ the SLD structure. 
(See Remark~\ref{remark_prop_real_e-ap} for a generalization.)
\end{proposition}

\begin{proofs}
It is easy to see that for each $\rho\in \sSB$ 
\be
T^{(\m)}_\rho (\sSB)  := \{ \iota_* (X_\rho)\,|\, X_\rho \in T_\rho (\sSB )\} 
= \{ A\in  \cLh^{\cB}\,|\, \Tr A = 0\} 
\label{Tm_SB}
\ee
and  that 
\begin{align}
T^{(\e)}_\rho (\sSB) & := \{L_{X_\rho}\,|\, X_\rho \in T_\rho (\sSB )\} 
\noret & = 
\{A \in  \cLh\,|\, \some B \in T^{(\m)}_\rho (\sSB), \; B = \rho \circ A \} 
\noret & = 
\{A \in  \cLh^{\cB}\,|\, \expect{A}_\rho =0\}.
\label{Te_SB}
\end{align}
(See Remark~\!\ref{remark_em_representation} for the symbols $T^{(\m)}$ and $T^{(\e)}$.) 
It follows from \eqref{Te_SB} that $A\in T^{(\e)}_\rho (\sSB) \; \Leftrightarrow \; A - \expect{A}_\sigma\in T^{(\e)}_\sigma (\sSB) $ 
for any $\rho, \sigma\in \sSB$, 
and hence we have from \eqref{e-parallel_transport} 
that $X_\rho \in T_\rho (\sSB) \; \Leftrightarrow \; \para_{\rho, \sigma}^{(\e)} (X_\rho )\in T_\sigma (\sSB) $. 
This proves the proposition by Prop.~\!\ref{prop_equiv_autoparallel_general}. 
\end{proofs}

Let us examine whether the e-autoparallelity of $\sSB$ can be understood as an example of Theorem~\!\ref{thm_SLD_integrability}. Namely, the problem is whether $\sSB$ is 
an integral manifold of an e-parallel distribution $\cV_{\cA}$ for some $\cA$ satisfying 
condition (ii) in Theorem~\!\ref{thm_SLD_integrability}. 
For each $\rho \in \sSB$, we have 
$T^{(\e)}_\rho  (\sSB) = \{ A - \expect{A}_\rho\,|\, A\in \cLh^{\cB}\}$ (see  \eqref{Te_SB}), 
which means that $T_\rho (\sSB) = V_{\cLh^{\cB}, \rho}$ and that $\sSB$ is an integral manifold of the distribution $\cV_{\cLh^{\cB}}$.  Noting that $\cLh^{\cB} + \bR = \cLh^{\cB}$, 
the problem comes down to whether 
\be
\{ [ [A, B], \rho ] \,|\, A, B\in  \cLh^{\cB} \} \subset \{ \rho \circ C \,|\, C\in  \cLh^{\cB} \}
\ee
holds for every $\rho\in \sS$.  The answer is no, except when $\dim \cH = 2$. 

\begin{proposition}
\label{prop_integrability_LBh_dim>=3}
When $\dim \cH \geq 3$, 
\be
\red{\any \rho\in \sS\setminus \sSB}, 
\; \{ [ [A, B], \rho ] \,|\, A, B\in  \cLh^{\cB} \} \not\subset \{ \rho \circ C \,|\, C\in  \cLh^{\cB} \}.
\ee
As a consequence, the distribution $\cV_{\cLh^{\cB}}$ is not involutive. 
\end{proposition}

\begin{proofs}
We represent operators on a $d$-dimensional Hilbert space by $d\times d$ matrices, and show that 
\red{
if a strictly positive density matrix $\rho$ satisfies 
\be
\{ [ [A, B], \rho ] \,|\, A, B : \text{real symmetric}
 \} \subset \{ \rho \circ C \,|\, C : \text{real symmetric} \}, 
 \nonumber
\ee
which is equivalent to 
\be
\{ [\rho, F ] \,|\, F : \text{real skew-symmetric}
 \} \subset \{ \rho \circ C \,|\, C : \text{real symmetric} \},
 \label{rhoF_subset_rhoC}
\ee
then $\rho$ is a real matrix. 
Denoting $\rho = Q + \sqrt{-1}\, R$ with $Q$: real symmetric and $R$: real skew-symmetric, 
\eqref{rhoF_subset_rhoC} is rewritten as
\begin{align}
\any F &:\text{real skew-symmetric}, \; \exists C :\text{real symmetric}, \; 
\noret
& [Q, F ] = Q\circ C \;\;\text{and} \;\; [R, F] = R \circ C.
\label{QFQC_RFRC}
\end{align}
Now it suffices to show the following lemma, whose proof will be given in Appendix~\ref{sec_proof_lemma_QFQC_RFRC_R=0}. 
} 
\end{proofs}

\red{
\begin{lemma}
\label{lemma_QFQC_RFRC_R=0}
Let $Q$ be a $d\times d$ real symmetric positive-definite matrix and $R$ be a $d\times d$ real skew-symmetric matrix. 
If they satisfy \eqref{QFQC_RFRC} and $d\geq 3$, then $R=0$. 
\end{lemma}
} 

The above result implies that 
the e-parallel distribution $\cV_{\cLh^{\cB}}$ 
does not induce a foliation with e-{\ap} leaves 
and that $\sS^{\cB}$ is an isolated integral manifold of 
$\cV_{\cLh^{\cB}}$ when $\dim \cH \geq 3$. 
The exceptional case $\dim \cH =2$ will be discussed in the next section. 

\begin{remark}
\label{remark_prop_real_e-ap} 
Prop.~\!\ref{prop_real_e-ap} holds for a wide class of \rred{information-geometric} structures, not limited to the SLD structure. 
In fact, the proof of Prop.~\!\ref{prop_real_e-ap} given above relies only upon the fact that if $\rho\in \sSB$ and $X_\rho \in T_\rho (\sSB)$, then 
$L_{X_\rho} \in \cLh^{\cB}$. Due to \eqref{def_LX_2} stating that 
$\iota_*(X_\rho) = \Omega_\rho  (L_{X_\rho})$, this fact is shared by the e-connection defined from 
an arbitrary family of inner products $\inprod{\cdot}{\cdot}_\rho = \inprod{\cdot}{\Omega_\rho (\cdot)}_{\rm HS}$, 
$\rho\in \sS$,  such that 
\be 
\any \rho \in \sSB, \; 
\Omega_\rho (\cLh^{\cB}) = \cLh^{\cB}.
\label{Omega_real}
\ee 
This means that Prop.~\!\ref{prop_real_e-ap} holds under this mild condition on $\{\Omega_\rho\}_{\rho\in\sS}$. 
In particular, 
if $\Omega_\rho$ is represented in the form \eqref{Omega_f_Delta} by a function $f: (0, \infty) \rightarrow (0, \infty)$ 
such that $\any x>0,\,  x f(1/x) = f(x)$ and $f(1) =1$ 
as in the case of monotone metrics,  condition \eqref{Omega_real} is satisfied. 
To verify this, we represent \eqref{Omega_f_Delta} as $\Omega_\rho = f(\Delta_\rho) {\mathcal R}_\rho$, 
where  ${\mathcal R}_\rho : A \mapsto A \rho$, and consider $\Omega_\rho$ as 
a $\bC$-linear map  $\cL \rightarrow \cL$. Then it is easy to see that 
if $\rho\in \sSB$, then 
${\mathcal R}_\rho (\cL^{\cB}) = \cL^{\cB}$ and 
$\Delta_\rho  (\cL^{\cB}) = \cL^{\cB}$, which yields $f(\Delta_\rho)  (\cL^{\cB}) = \cL^{\cB}$, 
and hence  we have
$ \Omega_\rho (\cL^{\cB}) = \cL^{\cB}$.
Combined with  $\Omega_\rho (\cLh) = \cLh$, 
this proves  \eqref{Omega_real}.   
\end{remark}

\begin{remark}
Since \eqref{Tm_SB} shows that the space $T^{(\m)}_\rho (\sSB) = \iota_* (T_\rho  (\sSB))$ does not depend on 
$\rho$, $\sSB$ is m-{\ap} in $\sS$, so that $\sSB$ is doubly {\ap} (e.g, \cite{ohara2004}) w.r.t.\ the e,~\!m-connections. 
In addition, the set $ \sS^{\cB}_{\text{diag}}:= \sS \cap {\mathcal L}_{{\rm h}, \text{diag}}^{\cB}$, where 
${\mathcal L}_{{\rm h}, \text{diag}}^{\cB} := 
\{ \sum_{i} a_{i} \ket{i} \bra{i}\,|\, (a_{i}) \in \bR^{d}\}$, is doubly {\ap}  in $\sSB$ 
w.r.t.\ the e,~\!m-connections of $\sSB$, and therefore is doubly {\ap}  in $\sS$ w.r.t.\ the e,~\!m-connections of $\sS$. 
These examples exhibit 
a remarkable contrast to the following fact for the classical case  \cite{nag_isit2017}:  if a submanifold 
$\sM$ of 
$\cP (\Omega)$, where $\Omega$ is an arbitrary finite set, is doubly {\ap} in $\cP (\Omega)$ w.r.t.\ the e, m-connections, then $\sM$ is statistically isomorphic to $\cP (\Omega')$ for some finite set $\Omega'$. 
\end{remark}

\section{Qubit manifolds}
\label{sec_qubit}

Throughout this section, we assume $\cH$ to be 2-dimensional. Our aim is to study the e-autoparallelity 
in $\sS (\cH)$ 
w.r.t.\ the SLD structure. 
To begin with, we make some preparations. 
Let $\{\sigma_1, \sigma_2, \sigma_3\} \subset \cLh$ be a triple of Pauli operators such that
\be
\Tr\, \sigma_i =0, \quad \sigma_i^2 = I, \quad\text{and}\quad \sigma_i \sigma_{i+1} = \sqrt{-1}\, \sigma_{i+2} 
\quad (i : {\rm mod}\ 3).
\label{pauli_operators}
\ee
Then $\{\sigma_1, \sigma_2, \sigma_3\}$ form a basis of $\cLhzero$.  For any $\avec = (a_i) \in \bR^3$, we write
$
\avec\cdot \sigmavec := \sum_i a_i \sigma_i$, 
so that we have
\be
\cLhzero = \{ \avec\cdot \sigmavec \,|\, \avec \in \bR^3\}.  
\ee
It follows that
\begin{align}
(\avec\cdot \sigmavec) (\bvec\cdot \sigmavec) & = (\avec\cdot \bvec) I + \sqrt{-1}\, (\avec \times \bvec) \cdot \sigmavec, 
\label{qubit_AB}
\\
(\avec\cdot \sigmavec) \circ (\bvec\cdot \sigmavec) &=  (\avec\cdot \bvec) I, 
\label{qubit_AcircB}
\\
[ \avec\cdot \sigmavec, \bvec\cdot \sigmavec ] &= 2 \sqrt{-1} \,  (\avec \times \bvec) \cdot \sigmavec, 
\label{qubit_[A,B]}
\end{align}
where $\avec \cdot \bvec = \sum_i a_i b_i$, and 
$\avec \times \bvec = \cvec$ $\Leftrightarrow$ $ \any i : {\rm mod}\ 3$, 
$a_{i} b_{i+1} - a_{i+1} b_{i} = c_{i+2}$. 
The manifold $\sS = \sS (\cH)$ is represented as
\be
\sS = \{\rho_{\rvec}\,|\, \rvec \in \cR\},
\ee
where 
\be
\rho_{\rvec} := \frac{1}{2} (I + \rvec\cdot \sigmavec), \quad 
\cR := \{\rvec\in \bR^3\,|\, \|\rvec\| := \sqrt{\rvec\cdot\rvec} < 1\}.
\ee
For $\rho = \rho_{\rvec}$ and $A = a_0 I + \avec\cdot \sigmavec$, we have $\expect{A}_\rho = a_0 + \rvec\cdot \avec$.

A tangent vector $X_\rho \in T_\rho (\sS)$ at $\rho = \rho_{\rvec}$ is represented by 
a 3-dimensional vector 
$\xvec  \in \bR^3$ such that
\be
\iota_* (X_\rho) = \frac{1}{2} \xvec\cdot \sigmavec.
\ee
The SLD of $X_\rho$ is then represented as
\be
L_{X_\rho} = \ellvec_{\rvec} (\xvec) \cdot \sigmavec - \lambda_{\rvec} (\xvec) I, 
\label{LX_qubit}
\ee
where
\be
\lambda_{\rvec} (\xvec) :=  \frac{\xvec\cdot\rvec}{1-\|\rvec\|^2} 
\quad\text{and}\quad
\ellvec_{\rvec} (\xvec) := \xvec +\lambda_{\rvec} (\xvec)\, \rvec. 
\label{def_ellvec_c}
\ee
In fact, \eqref{LX_qubit} is verified as follows:  
noting that \eqref{def_ellvec_c} yields $\rvec\cdot \ellvec_{\rvec} (\xvec) = \lambda_{\rvec} (\xvec)$, we have 
\begin{align}
\rho \circ L_{X_\rho} & = \frac{1}{2} ( I + \rvec\cdot \sigmavec) \circ (\ellvec_{\rvec} (\xvec) \cdot \sigmavec -\lambda_{\rvec} (\xvec) I)
\noret &= \frac{1}{2} (\ellvec_{\rvec} (\xvec)  - \lambda_{\rvec} (\xvec)\, \rvec )\cdot \sigmavec 
+ \frac{1}{2} ( (\rvec\cdot \ellvec_{\rvec} (\xvec) ) -\lambda_{\rvec} (\xvec)) I
\noret 
&= \frac{1}{2} \xvec \cdot \sigmavec = \iota_* (X_r).
\end{align}

Let us investigate the e-{\ap} submanifolds of $\sS$.  
We first consider the 1-dimensional  case, i.e., the e-geodesics. 
We recall that the general form of  e-geodesic is given by \eqref{q_c:e-family_var_1dim}. 
Treating the coordinate $\theta$ as a parameter to specify states and choosing $P$ in \eqref{q_c:e-family_var_1dim} 
to be a state $\rho_0$, an arbitrary e-geodesic $\sM$ is represented as the trajectory $\sM = \{\rho_\theta \,|\, \theta \in \bR\}$
of 
\be
\rho_\theta = \frac{1}{Z_\theta} \exp \Bigl(\frac{\theta}{2} F\Bigr) \, \rho_0 \exp \Bigl(\frac{\theta}{2} F\Bigr), 
\quad Z_\theta := \Tr( \rho_0 \exp (\theta F)), 
\label{e-geodesic_parameter}
\ee
where $F$ is a Hermitian operator such that $\{ F, I\}$ are linearly independent. 
Since the transformation 
$F \rightarrow a F + b$ by 
$a, b\in \bR$, $a\neq 0$, together with $\theta \rightarrow \frac{1}{a}\theta$ and 
$\psi \rightarrow \psi + \frac{b}{a} \theta$,  
keeps $\mM$ invariant, 
we can assume that $F$ is represented as $F = \uvec \cdot \sigmavec$ by a unit vector $\uvec$. 

\begin{proposition}
\label{prop_qubit_e-geodesic}
Let $\rho_0 = \rho_{\rvec_0}$ and $F = \uvec \cdot \sigmavec$ with $\| \uvec\| =1$ 
in \eqref{e-geodesic_parameter}. 
Letting $\vvec$ be a unit vector such that $\uvec\cdot \vvec =0$ and that 
$\rvec_0 \in {\rm span} \{\uvec, \vvec\}$, 
the e-geodesic $\sM = \{\rho_\theta \,|\, \theta \in \bR\}$ is 
represented as
\be
\sM = \{\rho_{\rvec}\,|\, \rvec \in \cQ\} 
\quad\text{with}\quad \cQ := \{ \rvec (\xi) \,|\, -1 < \xi < 1\},
\label{M_Q_prop_qubit_e-geodesic}
\ee
where
\begin{gather}
\rvec (\xi)  := \xi\, \uvec +  c \sqrt{1 -  \xi^2}\,\rred{\vvec} , 
\label{rvec_a_prop_qubit_e-geodesic} 
\\
\quad c  := \frac{b}{\sqrt{1- a^2}} , \quad a:= \rvec_0 \cdot \uvec, \quad b:= \rvec_0\cdot \rred{\vvec}.
\label{c_a_prop_qubit_e-geodesic} 
\end{gather}
(Here $\xi^2$ denotes the square of $\xi$, while the same symbol will appear as 
the second component of $\xi = (\xi^i)$ later.) 
The parameter $\xi$ is m-affine as a coordinate system of $\sM$ and 
in one-to-one correspondence with the e-affine parameter $\theta$ by 
\begin{align}
\xi & 
= \frac{(1+a) e^{2\theta} - (1-a)}{(1+a) e^{2\theta} + (1-a)} 
\quad\text{and}\quad 
\theta = \frac{1}{2} \log \frac{(1-a) (1+\xi)}{(1+a) (1-\xi)}.
\label{qubit_xi_theta_transformation}
\end{align}
\end{proposition}

\begin{proofs}
Noting that $F = \uvec\cdot \sigmavec$ is represented as
\be
F =  \rho_{\uvec} - \rho_{-\uvec} = 1  \rho_{\uvec} + (-1)  \rho_{-\uvec} 
 \nonumber
\ee
and that this is the spectral decomposition of $F$ with projectors $\{ \rho_{\uvec},  \rho_{-\uvec}\}$, we have
\be
\exp \Bigl(\frac{\theta}{2} F\Bigr) = e^{\theta/2}  \rho_{\uvec}  + e^{- \theta/2}  \rho_{-\uvec} 
= \cosh (\theta/2) I + \sinh (\theta/2) \uvec\cdot\sigmavec. 
 \nonumber
\ee
Using this expression and representing $\rvec_0$ as $\rvec_0 = a \uvec + b \vvec$ by $a := \rvec_0\cdot \uvec$ and 
$b := \rvec_0\cdot \vvec$, a direct calculation shows that
\begin{gather}
\exp \Bigl(\frac{\theta}{2} F\Bigr) \, \rho_{\rvec_0} \exp \Bigl(\frac{\theta}{2} F\Bigr) = 
\frac{Z_\theta}{2}\, I + \frac{1}{2} \bigl\{ (a \cosh \theta + \sinh \theta)\, \uvec + b\, \vvec \bigr\} \cdot \sigmavec
 \nonumber
\end{gather}
and
$
Z_\theta = \cosh\theta + a\sinh \theta, 
$
which yields
\be
\rho_\theta = \frac{1}{2} ( I + \svec(\theta) \cdot \sigmavec),
 \nonumber
\ee
where
\begin{align}
\svec(\theta) & := \frac{ a \cosh \theta + \sinh \theta}{\cosh\theta + a\sinh \theta}\,  \uvec 
+ \frac{b}{\cosh\theta + a\sinh \theta}\, \vvec.
 \nonumber
\end{align}
If we define $\xi$ from $\theta$ by \eqref{qubit_xi_theta_transformation}, 
we have
\be
\frac{ a \cosh \theta + \sinh \theta}{\cosh\theta + a\sinh \theta} = \xi
\quad
\text{and}
\quad
 \frac{b}{\cosh\theta + a\sinh \theta} = c \sqrt{1-\xi^2}, 
  \nonumber
\ee
so that $\svec (\theta) = \rvec (\xi)$. It is easy to see that the range of $\xi$ is $(-1, 1)$, and 
we obtain \eqref{M_Q_prop_qubit_e-geodesic}.  In addition, since 
\be
\expect{F}_{\rho_{\rvec (\xi)}}  = \rvec (\xi) \cdot \uvec = \xi, 
 \nonumber
\ee
the parameter $\xi$ is m-affine. 
\end{proofs}

Note that $\cQ$ in the above proposition forms a semi-ellipse in the open unit ball $\cR$ 
obtained by cutting an ellipse in half 
on the major axis: see Fig.1.  In the special case of $c=0$, the semi-ellipse becomes a straight line. 
\begin{figure}
\centering
\includegraphics[keepaspectratio, scale=1]{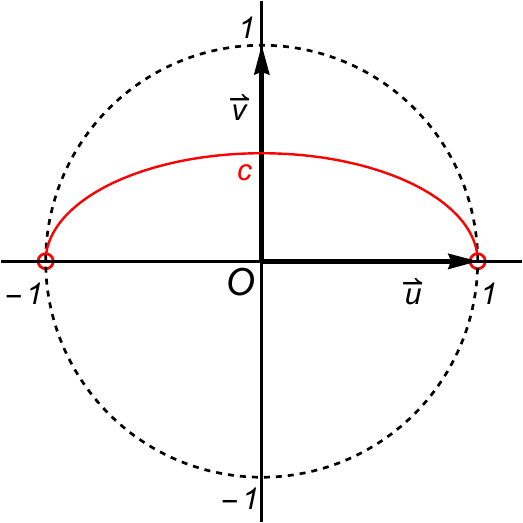}
\caption{The semi-ellipse representing an e-geodesic}
\label{fig:e-geodesic}
\end{figure}

Next, let us proceed to considering the 2-dimensional case. 
In searching for 2-dimensional e-{\ap} submanifolds, 
the previously obtained knowledge of e-geodesics provides an important clue. 
If a 2-dimensional submanifold $\sM = \{ \rho_{\rvec}\, |\, \rvec\in \cQ\}$ is e-{\ap}, 
it must be e-totally geodesic, and hence the surface $\cQ$ should be a union of semi-ellipses.  
The following proposition claims that a 2-dimensional e-{\ap} submanifold is obtained as 
a semi-ellipsoid formed by rotating a semi-ellipse representing 
an e-geodesic around its minor axis. 

\begin{proposition}
\label{prop_qubit_e-autoparallel}
Given an orthonormal basis $\{\uvec_1, \uvec_2, \vvec\}$ of $\bR^3$ and a real constant $c$ satisfying 
$|c| <1$, let 
\be
\cQ := \{ \rvec (\xi) \,|\, \xi = (\xi^1,\xi^2) \in \bR^2, \; (\xi^1)^2 + (\xi^2)^2 < 1\}, 
\label{prop_qubit_e-autoparallel_Q}
\ee
where
\be
\rvec (\xi) := \xi^1 \uvec_1 + \xi^2 \uvec_2 + c \sqrt{1 -  (\xi^1)^2 - (\xi^2)^2}\,\vvec .
\label{prop_qubit_e-autoparallel_r}
\ee
Then $\sM := \{\rho_{\rvec}\,|\, \rvec\in \cQ\}$ is 
 e-{\ap} in $\sS$, and the parameter $\xi=(\xi^1, \xi^2)$ is m-affine as a coordinate system of $\sM$.  
More specifically, letting $F^i := \uvec_i \cdot \sigmavec$, $\cA := {\rm span} \{ F^1, F^2\}$, 
 and $\cV_{\cA} : \sS \ni \rho \mapsto V_{\cA, \rho}$ be  the e-parallel distribution defined 
 from $\cA$ by \eqref{def_V_A_rho}, $\sM$ is an integral manifold of $\cV_{\cA}$ and 
 $\xi^i = \expect{F^i}_{\rho_{\rvec (\xi)}}$. 
\end{proposition}

\begin{proofs}
For $i\in \{1, 2\}$, let 
\be
\xvec_i := \partial_i \rvec (\xi) = \uvec_i - \frac{c\, \xi^i}{\alpha (\xi)}\, \vvec,
 \nonumber
\ee
where $\partial_i := \frac{\partial}{\partial\xi^i}$ and $\alpha (\xi) := \sqrt{1 - (\xi^1)^2 - (\xi^2)^2}$. 
Noting that 
\be 
\|\rvec (\xi)\|^2 =1- (1-c^2) \alpha (\xi)^2 \quad \text{and} \quad  \xvec_i \cdot \rvec(\xi) = (1-c^2) \xi^i, 
 \nonumber
\ee
we have
\begin{align}
\ellvec_{\rvec (\xi)} (\xvec_i) & = \xvec_i + \frac{\xi^i}{\alpha (\xi)^2}\, \rvec (\xi)
\noret 
&= \uvec_i + \frac{\xi^i \xi^1}{\alpha (\xi)^2}\, \uvec_1 +  \frac{\xi^i \xi^2}{\alpha (\xi)^2}\, \uvec_2
\; \in {\rm span}\,\{ \uvec_1, \uvec_2\}, 
 \nonumber
\end{align}
where the terms
proportional to $\vvec$ included in $\xvec_i$ and $ \rvec (\xi)$ cancel, yielding the last line. 
Owing to \eqref{LX_qubit} this implies that the SLDs satisfy $L_{i, \xi} \in  {\rm span}\,\{F^1, F^2\} \oplus \bR$ 
for $i\in\{1, 2\}$, which means that the first condition in \eqref{eq_(iv)_prop_autopara_affine_coordinate_quantum} 
is satisfied.  The second condition is also satisfied since $\expect{F^i}_{\rho_{\rvec (\xi)}} = \uvec_i \cdot \rvec (\xi) = \xi^i$. 
Thus the claim of the proposition follows from Prop.~\!\ref{prop_autopara_affine_coordinate_quantum}. 
\end{proofs}

As can be seen from naive geometric intuition, for any point $\rvec$ in $\cR$ and any plane $P=\rvec + V$ containing $\rvec$,  where $V$ is a 2-dimensional linear subspace of $\bR^3$,  
there always exist an orthonormal basis $\{\uvec_1, \uvec_2, \vvec\}$ and a constant $c \in (-1, 1)$ such that 
the semi-ellipsoid $\cQ$ defined from them by \eqref{prop_qubit_e-autoparallel_Q} and \eqref{prop_qubit_e-autoparallel_r} contains 
$\rvec$ and has $P$ as the tangent plane at $\rvec$. In fact, such  $\{\uvec_1, \uvec_2, \vvec\}$ and $c$ are obtained  
as follows: take an 
orthonormal basis $\{\uvec_1, \uvec_2, \vvec\}$ so that $\{\uvec_1, \uvec_2\} \subset 
\ellvec_{\rvec} (V) = 
\{ \ellvec_{\rvec} (\xvec) \,|\, \xvec\in V\}$, 
and then let $\beta^2:= (\rvec\cdot \uvec_1)^2 +  (\rvec\cdot \uvec_2)^2$ (i.e. the squared norm of the orthogonal projection of 
$\rvec$ onto $\ellvec_{\rvec} (V)$), $\gamma:=  \rvec\cdot \vvec$, and $c :=\gamma / \sqrt{1-\beta^2}$. 
Since $\dim \sS =3$, 
this fact means that $(\sS, \nabla^{(e)})$ satisfies condition (i) of Prop.~\!\ref{cor_prop_integrability}, and necessarily satisfies 
condition (ii) as well.  Invoking \eqref{SLD_torsion},  condition (ii) is expressed as follows.

\begin{proposition}
\label{prop_qubit_integrability}
When $\dim\cH =2$, for any $\rho\in \sS$ and any $A, B\in\cLh$ satisfying 
$\expect{A}_\rho = \expect{B}_\rho =0$ we have
\be
[[A, B], \rho] \in {\rm span}\,\{\rho\circ A, \rho\circ B\}.
\ee
\end{proposition}

This proposition can also be proved directly by the use of the following lemma, whose proof is given in 
Appendix~\ref{sec_proof_lemma_qubit_torsion}.

\begin{lemma}
\label{lemma_qubit_torsion}
When $\dim\cH =2$, 
for any $\rho\in \sS$ and any $A, B\in \cLh$, we have
\begin{align}
\frac{1}{2} [[A, B], \rho] = & 
(\Tr A - 2 \expect{A}_\rho ) (\rho\circ B) - 
(\Tr B - 2 \expect{B}_\rho ) (\rho \circ A) 
\noret & 
+ \bigl\{
(\Tr B) \expect{A}_\rho - (\Tr A) \expect{B}_\rho\bigr\} \,\rho.
\label{eq_lemma_qubit_torsion}
\end{align}
\end{lemma}

\medskip

Letting $\expect{A}_\rho = \expect{B}_\rho =0$ in \eqref{eq_lemma_qubit_torsion}, we obtain
\begin{align}
[[A, B], \rho] = & 
2 (\Tr A )\, (\rho\circ B) - 
2 (\Tr B )\, (\rho \circ A), 
\end{align}
which proves Prop.~\!\ref{prop_qubit_integrability}. 

\medskip

The following proposition immediately follows from Prop.~\!\ref{prop_qubit_integrability}, 
which  presents a remarkable contrast to Prop.~\!\ref{prop_integrability_LBh_dim>=3} 
for the case $\dim \cH \geq 3$. 

\begin{proposition}
\label{prop_integrability_LBh_dim=2}
When $\dim \cH = 2$, for any orthonormal basis $\cB$ of $\cH$ it holds that
\be
\any \rho\in \sS, \; \{ [ [A, B], \rho ] \,|\, A, B\in  \cLh^{\cB} \} \subset \{ \rho \circ C \,|\, C\in  \cLh^{\cB} \}.
\ee
\end{proposition}

\begin{proofs}
Obvious from Prop.~\!\ref{prop_qubit_integrability} since $\cLh^{\cB}$ is an $\bR$-linear space with $I \in \cLh^{\cB}$. 
\end{proofs}

Thus, the distribution $\cV_{ \cLh^{\cB}}$ is involutive, and induces a foliation 
$\sS = \bigsqcup_{\alpha} \sM_\alpha$ whose leaves $\{\sM_\alpha\}$ are 2-dimensional e-{\ap} submanifolds 
that are integral manifolds of $\cV_{ \cLh^{\cB}}$. 
Furthermore, we can see from the following lemma 
that every $2$-dimensional e-{\ap} submanifold of $\sS$ is an integral manifold of $\cV_{ \cLh^{\cB}}$ 
for some $\cB$. 

\begin{lemma}
When $\dim \cH =2$, 
for any $A, B\in \cLh$ there exists an orthonormal basis $\cB$ such that 
$\{A, B\} \subset \cLh^{\cB}$. 
\end{lemma}
\begin{proofs}
Let $\{\ket{1}, \ket{2}\}$ be an orthonormal basis that diagonalizes $A$, 
and choose $\beta\in \bC$ so that $|\beta|=1$ and  $\beta \langle 1 | B | 2\rangle \in \bR$. 
Then $\cB := \{ \ket{1}, \beta  \ket{2}\}$ satisfies the desired condition. 
\end{proofs}

\medskip

\red{
According to Prop.~\!\ref{prop_qubit_integrability}, every e-parallel distribusion is involutive in the qubit case, and should generate a pair of dual foliations due to Cor.~\!\ref{cor_SLD_integrability_F}.  
Let us observe these foliations based on the representations of e-{\ap} manifolds given in Prop.~\!\ref{prop_qubit_e-geodesic} and Prop.~\!\ref{prop_qubit_e-autoparallel}. 
} 

\red{
First, fix a unit vector $\vec{u} \in \bR^3$ arbitrarily and let $F := \vec{u}\cdot \vec{\sigma}$ as in 
Prop.~\!\ref{prop_qubit_e-geodesic}.  The distribution $\cV_{\cA}$ with $\cA = {\rm span}\, \{F\}$ is involutive, 
and an integral manifold $\sM$, which is an e-geodesic, is generally represented as 
\eqref{M_Q_prop_qubit_e-geodesic} and  \eqref{rvec_a_prop_qubit_e-geodesic}.  The representation has two parameters, a unit vector $\vec{v}$ orthogonal to $\vec{u}$ and a real number $c$ lying in the interval $(-1, 1)$, but the true parameter that determines $\sM$ injectively is $\vec{v}\,' := c \vec{v}$ lying 
in the open unit disc orthogonal to $\vec{u}$. Denoting the manifold $\sM$ corresponding to $\vec{v}\,'$ 
by $\sM_{\vec{v}\,'}$, we obtain the foliation $\sS = \bigsqcup_{\vec{v}\,'}\sM_{\vec{v}\,'}$ consisting of $1$-dimensional e-{\ap} leaves. On the other hand, letting 
\be
\sN_b := \{\rho\in\sS \,|\, \expect{F}_\rho = b\} 
\ee
for $b\in (-1, 1)$, we obtain the dual foliation $\sS = \bigsqcup_{b}\sN_b$ consisting of $2$-dimensional m-{\ap} leaves which intersect $\sM_{\vec{v}\,'}$ orthogonally for any $\vec{v}\,'$. 
} 

\red{
Next, fix a pair of orthonormal vectors $\{\vec{u}_1, \vec{u}_2\}$ in $\bR^3$ arbitrarily and let $F^i := \vec{u}_i \cdot \vec{\sigma}$ as in Prop.~\!\ref{prop_qubit_e-autoparallel}. 
The distribution $\cV_{\cA}$ with $\cA = {\rm span}\, \{F^1, F^2\}$ is involutive, 
and an integral manifold $\sM$ is generally represented as 
\eqref{prop_qubit_e-autoparallel_Q} and \eqref{prop_qubit_e-autoparallel_r}. 
Here the true parameter is $\vec{v}\,' := c \vec{v}$,  whose range is the open line segment 
given by the intersection of the open unit ball and the line through the origin orthogonal to both $\vec{u}_1$ and $\vec{u}_2$. Denoting the manifold $\sM$ corresponding to $\vec{v}\,'$ 
by $\sM_{\vec{v}\,'}$, we obtain the foliation $\sS = \bigsqcup_{\vec{v}\,'}\sM_{\vec{v}\,'}$ consisting of $2$-dimensional e-{\ap} leaves. The same foliation is also expressed as  $\sS = \bigsqcup_{c\in (-1, 1)} \sM_c$ when $\vec{v}$ is chosen 
as one of the two unit vectors orthogonal to $\vec{u}_1$ and $\vec{u}_2$. 
On the other hand, letting 
\be
\sN_{b} := \{\rho\in\sS \,|\, \expect{F^1}_\rho = b^1 \;\;\text{and}\;\;  \expect{F^2}_\rho = b^2\} 
\ee
for $b = (b^1, b^2) \in \bR^2$ satisfying $(b^1)^2 + (b^2)^2 <1$, we obtain the dual foliation $\sS = \bigsqcup_{b}\sN_{b}$ consisting of $1$-dimensional m-{\ap} leaves which intersect $\sM_{\vec{v}\,'}$ orthogonally for any $\vec{v}\,'$. 
} 

\medskip

\red{
\begin{remark}
In a context of quantum parameter estimation with nuisance parameters, Suzuki et al.\ introduced 
in Section~6.3 of \cite{suzuki_nuisance_review} 
the coordinate system $\tilde{\xi} = (\tilde{\xi}^i)$ 
on $\sS$ defined by
\be
\tilde{\xi}^i  = \expect{\sigma_i}, \;\; i \in\{1, 2\}, \;\; 
\text{and} \;\; 
\tilde{\xi}^3 =  \expect{\sigma_3} / \sqrt{1 -(\tilde{\xi}^1)^2   -(\tilde{\xi}^2)^2}. 
\label{def_tilde_xi_qubit}
\ee
It is shown there that 
\be
g( \tilde{\partial}_1, \tilde{\partial_3}) = g( \tilde{\partial}_2, \tilde{\partial_3}) =0, \quad 
\tilde{\partial}_i := \frac{\partial}{\partial\tilde{\xi}^i}, 
\label{global_orthogonal_qubit}
\ee
while the usual  coordinate system $\xi = (\xi^i)$ with $\xi^i = \expect{\sigma_i}$, $i \in \{1, 2, 3\}$, 
has non-vanishing $g( \partial_i, \partial_j)$ for any $i, j$. In the situation where 
$\xi^1$ and $\xi^2$ are parameters of interest to be estimated and $\xi^3$ is a nuisance parameter, 
the parameter change from $\xi$ to $\tilde{\xi}$, which is called the global orthogonalization in  \cite{suzuki_nuisance_review}, 
is considered to remove the influence  of nuisance parameter in some sense, while the parameters of interest are kept invariant as $\tilde{\xi}^i = \xi^i$, $i \in\{1, 2\}$.  The orthogonality in \eqref{global_orthogonal_qubit} can be viewed as a reflection of the orthogonality of the pair of dual foliations 
$\sS = \bigsqcup_{c} \sM_c =  \bigsqcup_{b} \sN_b$ given above. 
To see this, let $\vec{u}_1, \vec{u}_2, \vec{v}$ in 
Prop.~\!\ref{prop_qubit_e-autoparallel} be the natural basis $\vec{e}_1, \vec{e}_2, \vec{e}_3$ of $\bR^3$.  
Then the $2$-dimensional e-{\ap} leaf $\sM_c$ is represented by the constraint $\xi^3 = c \sqrt{1 - (\xi^1)^2 - (\xi^2)^2}$ on the 
coordinates $\xi^i = \expect{\sigma_i}$, $i \in \{1, 2, 3\}$, which means that $\tilde{\xi}^3$ is fixed to $c$ for the coordinates $(\tilde{\xi}^i)$ defined by \eqref{def_tilde_xi_qubit}.  
Therefore, $(\tilde{\partial}_1)_\rho$ and $(\tilde{\partial}_2)_\rho$ belong to $T_\rho (\sM_c)$ when $\rho\in \sM_c$. On the other hand, the $1$-dimensional m-{\ap} leaf $\sN_b$ is represented by the constraint 
$(\xi^1, \xi^2) = (b^1, b^2)$, or equivalently $(\tilde{\xi}^1, \tilde{\xi}^2) = (b^1, b^2)$, so that both $(\partial_3)_\rho$ and $(\tilde{\partial}_3)_\rho$ belongs to $T_\rho (\sN_b)$ when $\rho\in \sN_b$. 
Since every point $\rho\in \sS$ belongs to $\sM_c \cap \sN_b$ for some $(c, b)$ for which 
$T_\rho (\sM_c) \perp T_\rho (\sN_b)$ holds, we obtain \eqref{global_orthogonal_qubit}. 
\end{remark}
} 

\medskip

\red{Finally, let us make several remarks on estimators for qubit manifolds.} 

\red{
\begin{remark}
\label{rem_estimator_qubit}
The following result is known for qubit models: 
for any model $(\sM, \xi)$, any state $\rho\in\sM$ and any positive-definite weight matrix $\wgt$, it holds that
\be
\min_{\Pi'\in\cU(\rho, \xi)}\tr (\wgt V_\rho(\Pi')) = \Bigl(\tr \Bigl(G_\rho^{-\frac{1}{2}} \wgt G_\rho^{-\frac{1}{2}}\Bigr)^{\frac{1}{2}}\Bigr)^2 =: \tau_\rho (\wgt ) 
\label{min_trWV_tau}
\ee
and that a locally $\wgt$-efficient estimator (see \eqref{eqn:G-efficient}) achieving the minimum in \eqref{min_trWV_tau} 
is given by a randomized procedure similar to the one introduced in Section~\!\ref{sec_efficient_estimator}. 
Historically, apart from the 1-parameter case which is trivial, the 2-parameter case  was established  in \cite{nagaoka_new_approach, nagaoka_generalization_simultaneous}, and the 
general case including $\sM = \sS (\cH)$  was established in \cite{hayashi_linear_programming} and \cite{Gill_Massar}. 
A  comprehensive and transparent exposition on this result can be found in \cite{yamagata_tomography}. 
While our estimator was chosen solely for the purpose of 
proving 
Prop.~\!\ref{prop_inf_uVu} and Theorem~\!\ref{thm:autoparallel_m-affine}, and 
was not expected to be optimal in any sense, 
it is meaningful to clarify its relation to \eqref{min_trWV_tau}. 
Recall that our estimator $\Pi = (\pi, f)$ is determined by 
$\{\bu^1, \ldots, \bu^n\}$, $(p_1, \ldots , p_n)$ and $\{\gamma_k^i\}$ with $n= \dim \sM$,  
and its variance matrix is represented as shown in item (2) of Lemma~\ref{lemma_Pi=(pi_f)}. 
The variance is minimized when we set $\gamma_k^i = \xi^i (\rho)$, 
and in this case we have
\be
\tr (\wgt V_\rho(\Pi)) = 
 \sum_{k} \frac{1}{p_k} \Bigl( \transpose{(\bu^k)} G_\rho^{-1} \bu^k \Bigr) \,  \Bigl(\transpose{(\bw_k)} \wgt \bw_k\Bigr). 
\ee
Since $\Pi$ belongs to $\cU(\rho, \xi)$, it should satisfy
\be
\tr (\wgt V_\rho (\Pi))  \geq \tau_\rho (\wgt). 
\label{trWV_tau}
\ee
A direct proof of this inequality is given in Appendix~\ref{sec_proof_equality_trWV_tau}. 
There,  it is shown that the equality holds iff there exists a positive-definite matrix $K$ such that
\be
 W = K G_\rho^{-1} K, 
\label{W_KGK} 
\ee
and
\be
K \propto \sum_k \frac{p_k}{\transpose{(\bu^k)} G_\rho^{-1} \bu^k}\,  \bu^k \transpose{(\bu^k)}.
\label{K_propto}
\ee
\\
\indent
When a weight $\wgt$ and a point $\rho$ of a model $(\sM, \xi)$ are given, 
a matrix $K >0$ satisfying \eqref{W_KGK} is determined from them and is represented as
\be
K = G_\rho^{\frac{1}{2}} \Bigl(G_\rho^{-\frac{1}{2}} W G_\rho^{-\frac{1}{2}}\Bigr)^{\frac{1}{2}} G_\rho^{\frac{1}{2}}. 
\ee
Then \eqref{K_propto} tells us that an estimator $\Pi$ achieving 
$\tr (\wgt V_\rho (\Pi)) = \tau_\rho (\wgt)$ is obtained by choosing  $(p_k)$ and $\{\bu^k\}$ so that  \eqref{K_propto} holds. There are infinitely many choices of such  $(p_k)$ and $\{\bu^k\}$. 
An example is given by using the spectral decomposition (diagonalization) of $K$, i.e., 
by letting $\{\bu^k\}$ be an orthonormal eigenvectors of $K$ and letting $(p_k)$ be a probability vector such that $p_k / \transpose{(\bu^k)} G_\rho^{-1} \bu^k$ is proportional to the corresponding eigenvalue.  Another example is given by using the spectral decomposition of 
$G_\rho^{- \frac{1}{2}} W G_\rho^{- \frac{1}{2}}$ instead of $K$ as follows. 
Let $\{\lambda_k\}$ and $\{\bv_k\}$ be the eigenvalues and 
the orthonormal eigenvectors of $G_\rho^{-\frac{1}{2}} \wgt G_\rho^{-\frac{1}{2}}$ so that
\be
G_\rho^{-\frac{1}{2}} \wgt G_\rho^{-\frac{1}{2}} = \sum_k \lambda_k \bv_k \transpose{(\bv_k)}
\;\; \text{with} \;\; 
\transpose{(\bv_k)}  \bv_l = \delta_{kl}. 
\label{GWG_lamda_vv}
\ee
Then we have
\be
K = G_\rho^{\frac{1}{2}}  \Bigl( \sum_k \sqrt{\lambda_k} \bv_k \transpose{(\bv_k)} \Bigr) G_\rho^{\frac{1}{2}} 
= \sum_k 
\frac{\sqrt{\lambda_k}}{\transpose{(\bv_k)} \bv_k} \bigl(G_\rho^{\frac{1}{2}} \bv_k\bigr) \transpose{\bigl(G_\rho^{\frac{1}{2}} \bv_k\bigr)}.
\ee
This shows that \eqref{K_propto} holds if we let
\be
\bu^k = G_\rho^{\frac{1}{2}}\, \bv_k\;\;\text{and}\;\; p_k \propto \sqrt{\lambda_k}.
\label{u_Gv_p_sqrt_lambda}
\ee
The estimator defined by \eqref{u_Gv_p_sqrt_lambda} 
corresponds to that of \cite{nagaoka_generalization_simultaneous, Gill_Massar, yamagata_tomography}; 
see item (1) of Remark~\ref{rem_supplement_estimator_qubit}. 
\flushright (End of Remark~\ref{rem_estimator_qubit})  
\end{remark}
} 

\red{
\begin{example}
\label{example_qubit_estimators}
Let us demonstrate the randomized estimators treated so far for a concrete example. 
Let $(\sM, \xi)$ be a qubit model with $n = \dim \sM (\leq 3)$ and  $\rho$ be a point of $\sM$. 
With appropriate choices of a triple of Pauli operators $\{\sigma_i\}$ satisfying \eqref{pauli_operators} and of a 
cordinate system up to affine transformation, we may assume without loss of generality that 
\be
\any i \in \{1, \ldots , n\}, \; \xi^i (\rho) =  \expect{\sigma_i}_\rho \;\;\text{and}\;\; 
L_\rho^i = \sigma_i -  \xi^i (\rho).
\label{xi_sigma_L}
\ee
In particular, when $\sM$ is e-{\ap} in $\sS$ and $\xi$ is m-affine, we may assume that 
\eqref{xi_sigma_L}  holds uniformly at all points $\rho$ of $\sM$ due to 
(i) $\Leftrightarrow$ (iii) of Prop.~\!\ref{prop_autopara_affine_coordinate_quantum}. 
When $n=1$ and when $n=2$, this is equivalent to redefining the operators $F=\vec{u}\cdot\vec{\sigma}$ 
and $F^i = \vec{u_i}\cdot\vec{\sigma}$ in Prop.~\!\ref{prop_qubit_e-geodesic} and Prop.~\!\ref{prop_qubit_e-autoparallel} as $\sigma_1$ and $\sigma_i$, respectively. 
Note that \eqref{xi_sigma_L} implies that
\be
g^{ij} (\rho) = \inprod{\sigma_i - \xi^i (\rho)}{\sigma_j - \xi^j (\rho)}_\rho = \delta_{ij} - \xi^i(\rho) \xi^j (\rho), 
\label{gij_qubit_example}
\ee
which means that 
\be
G^{-1}_\rho = I  - \xi(\rho) \transpose{\xi (\rho)}.
\label{G_qubit_example}
\ee
\\
\indent 
For simplicity, let us consider the special case when $\{\bu^k\}$ is the natural basis 
$\{{\mbox{\boldmath$e$}}_k\}$ 
of $\bR^n$ for our randomized estimators.  
Since $u^k_i = w_k^i = \delta_{ik}$ in this case, 
the estimation procedures \eqref{estimation_procedure_gamma_xi} and \eqref{estimation_procedure_filtration} are 
represented as
\be
\randest{(p_k)}{k}{ \fbox{$\sigma_k$}}{y}{\displaystyle \hat{\xi} = 
\xi (\rho) +\frac{y - \xi^k(\rho)}{p_k}\, {\mbox{\boldmath$e$}}_k}
\label{estimation_procedure_gamma_xi_qubit}
\ee
and
\be
\randest{(p_k)}{k}{ \fbox{$\sigma_k$}}{y}{\displaystyle \hat{\xi} = 
\frac{y}{p_k}\, {\mbox{\boldmath$e$}}_k,}
\label{estimation_procedure_filtration_qubit}
\ee
respectively. 
Letting $p_1 = 1 -\eps$, both of these estimation procedures give a \emph{locally} ${\mbox{\boldmath$e$}}_1 \transpose{({\mbox{\boldmath$e$}}_1)}$-efficient filtration 
at $\rho$, say $\filtPi = (\Pi_\eps)_{\eps>0}$, 
such that
\be
\lim_{\eps\downarrow 0} v^{11}_\rho (\Pi_\eps) = g^{11} (\rho) = 1 - \left(\xi^1(\rho)\right)^2, 
\ee
where $V_\rho (\Pi_\eps) = [v^{ij}_\rho (\Pi_\eps)]$.  
As noted in Remark~\ref{rem_variance_additional}, 
$v^{11}_\rho (\Pi_\eps)$ of \eqref{estimation_procedure_gamma_xi_qubit} is smaller than 
that of \eqref{estimation_procedure_filtration_qubit} by $\frac{\eps}{1-\eps} (\xi^1(\rho))^2$, while 
\eqref{estimation_procedure_filtration_qubit} has the advantage that the procedure does not depend on $\rho$ and gives an 
${\mbox{\boldmath$e$}}_1 \transpose{({\mbox{\boldmath$e$}}_1)}$-efficient filtration for an e-{\ap} model. 
\end{example}
} 

\red{
\begin{remark}
For a weight matrix $\wgt = [w_{ij}]$, the estimator $\Pi$ of the form \eqref{estimation_procedure_gamma_xi_qubit} achieves 
$\tr (\wgt V_\rho (\Pi)) = \tau_\rho (\wgt)$ iff the conditions \eqref{W_KGK} and \eqref{K_propto} hold, 
which now reduce to
\be
w_{ij} \propto p_i p_j \frac{\delta_{ij} - \xi^i(\rho) \xi^j(\rho)}{\bigl(1 - (\xi^i(\rho))^2\bigr)
\bigl(1 - (\xi^j(\rho))^2\bigr)}.
\label{wij_pipj}
\ee
When $(p_k)$ is the uniform distribution $p_k = 1/n$, in particular, this turns out to be
\be
w_{ij} \propto  \frac{\delta_{ij} - \xi^i(\rho) \xi^j(\rho)}{\bigl(1 - (\xi^i(\rho))^2\bigr)
\bigl(1 - (\xi^j(\rho))^2\bigr)}.
\label{wij_pipj_uniform}
\ee
This form of weight matrix appears in a main result of \cite{yamagata_tomography}, 
where a kind of efficiency of quantum state tomography is discussed. 
Futher details and the relation to our formulation are given in \rred{Appendix~\ref{sec_yamagata}}. 
\end{remark}
} 

\red{
\begin{remark}
In Section~\ref{sec_gaussian}, we showed that if a model $(\sM, \xi)$ has a $\wgt$-efficient estimator for every positive weight $\wgt$, then $\sM$ is e-autoparallel in $\sS (\cH)$ and $\xi$ is an m-affine coordinate system (Prop.~\!\ref{prop_cor_maintheorem}) and that a quantum Gaussian shift model gives an example of such a model. So it may be natural to ask whether  an e-autoparallel model of qubit states gives another example. The answer is negative except for the $1$-dimensional case. In fact, the following stronger proposition holds, whose proof is given in Appendix~\ref{sec_proof_prop_no_W-efficient_qubit} 
\end{remark}
} 

\red{
\begin{proposition}
\label{prop_no_W-efficient_qubit}
Let $\sM$ be an e-autoparallel submanifold of $\sS (\cH)$ with $\cH \simeq \bC^2$ and $n:= \dim \sM \in\{2, 3\}$, and 
$\xi$ be an m-affine coordinate system of $\sM$.  Then the model $(\sM, \xi)$ has no $\wgt$-efficient estimator for any positive-definite $\wgt\in \bR^{n\times n}$. 
\end{proposition}
} 

\red{
\begin{remark}
\label{rem_supplement_estimator_qubit}
We add some \rred{supplementary} remarks concerning Remark~\ref{rem_estimator_qubit}. 
\begin{itemize}
\item[(1)] 
The construction of an estimator achieving the minimum of \eqref{trWV_tau} 
given in \cite{nagaoka_generalization_simultaneous} for the two-parameter case may appear to differ from that in 
\cite{Gill_Massar, yamagata_tomography},  
but is in fact the same as shown below.
In the notation of the present paper, the construction in \cite{nagaoka_generalization_simultaneous}  can be described as follows: 
defining a linear operator $\phi$ on 
$T^{(\e)}_\rho (\sM) = {\rm span} \{L_{i, \rho}\}$  by
\be
\phi := \sum_{i, j} w_{ij}  \ket{L_\rho^i} \bra{L_\rho^j} \; :\; 
A \mapsto \sum_{i.j} w_{ij}  \inprod{L_\rho^j}{A}_\rho L_\rho^i, 
\label{phi_w_LL}
\ee
where $\wgt = [w_{ij}]$, and representing its spectral decomposition by
\be
\phi = \sum_{k} \lambda_k \ket{X^k} \bra{X^k} \;\;\text{with}\;\; 
\inprod{X^k}{X^l}_\rho = \delta^{kl}, 
\label{phi_lamda_XX}
\ee
an estimator achieving the minimum is realized by the random choice of 
the observable $X^k$ with probability $p_k =  \sqrt{\lambda_k} / \sum_l \sqrt{\lambda_l}$. 
Representing $X^k$ as $X^k = \sum_i u_i^k L^i_\rho$ by a vector $\bu^k = (u_i^k)$ 
and invoking  \eqref{phi_w_LL}, the spectral decomposition 
 \eqref{phi_lamda_XX} is rewritten as
\be
\wgt = \sum_k \lambda_k \bu^k \transpose{(\bu^k)} \;\;\text{with}\;\; 
\transpose{(\bu^k)} G_\rho^{-1} \bu^l = \delta^{kl}, 
\ee
which is equivalent to \eqref{GWG_lamda_vv} when  
$\bu^k = G_\rho^{\frac{1}{2}} \bv_k$. 
\item[(2)] 
The proof of \eqref{trWV_tau} with derivation of the equality condition 
given in Appendix~\ref{sec_proof_equality_trWV_tau} does not rely on the condition  $\dim \cH =2$. 
Therefore, the following identity holds for an arbitrary model regardless of $\dim \cH$: 
\be
\min_{\Pi'\in\hat{\cU}(\rho, \xi)}\tr (\wgt V_\rho(\Pi')) = 
\tau_\rho (\wgt ), 
\label{hatU_min_trWV_tau}
\ee
where $\hat{\cU}(\rho, \xi)$ denotes 
the totality of estimators in $\cU(\rho, \xi)$ that are realized by the randomized procedure 
introduced in Section~\!\ref{sec_efficient_estimator}. 
This identity appears to be closely related to the discussion in \cite{hayashi_linear_programming}, 
where it is claimed that the minimum in $\min_{\Pi'\in\cU(\rho, \xi)}\tr (\wgt V_\rho(\Pi'))$ is achieved by a certain class of randomized estimator iff the minimum is equal to $\tau_\rho (\wgt)$. 
\item[(3)] 
The identity \eqref{hatU_min_trWV_tau} proves  
Prop.~\ref{prop_inf_uVu} as follows. 
Given a vector  $\bu\in\bR^n$ and a positive number $\eps$, 
define  $\wgt_\eps := \bu \transpose{\bu} + \eps I$ 
and let $\Pi_\eps$ be an estimator in $\hat{\cU}(\rho, \xi)$ 
achieving the minimum for $\wgt_\eps$. Then we have
\begin{align}
\transpose{\bu} G_\rho^{-1} \bu 
\leq \transpose{\bu} V_\rho (\Pi_\eps) \bu & \leq 
\tr \wgt_\eps V_\rho (\Pi_\eps) 
\noret 
& = 
\Bigl(\tr \Bigl(G_\rho^{-\frac{1}{2}} \wgt_\eps\, G_\rho^{-\frac{1}{2}}\Bigr)^{\frac{1}{2}}\Bigr)^2
\; 
\rightarrow \transpose{\bu} G_\rho^{-1} \bu  
\;\; 
\text{as}\;\; \eps\downarrow 0.
\nonumber 
\end{align}
\end{itemize}
\end{remark}
} 


\section{Concluding remarks}
\label{sec_conclusion}

In this paper we study the {\apty} w.r.t.\ the e-connection for an information-geometric structure 
induced on $\sS (\cH)$. In particular, we focus  on the e-{\apty} for the SLD structure, 
for which  two different estimation-theoretical characterizations are given. We also investigate 
the existence conditions for e-{\ap} submanifolds by way of the involutivity of e-parallel distributions
and its relation to the torsion tensor. As a result, a specialty of the qubit case \rred{has been} revealed. 

Since the obtained estimation-theoretical characterizations of the e-{\apty} are complete in themselves, 
we do not see at this time what kind of development lies ahead. It is expected that 
the future development of quantum estimation theory and related fields may reveal new directions. 
The classical exponential family has a variety of important properties besides the existence of 
efficient estimator, some of which may present new materical to characterize certain 
geometric notions. 

For the {\apty} w.r.t.\ non-flat connections, our understanding is still very limited. 
For example, we do not yet have the whole picture about e-{\ap} submanifolds 
of $\sS (\cH)$ when $\dim \cH \geq 3$. 
We look forward to further research on this topic in information geometry and/or 
general differential geometry.  

It may also be a challenging problem to develop the infinite-dimensional quantum information geometry so that 
Theorem~\!\ref{thm:autoparallel_m-affine} is extended to 
the case when $\dim \cH =\infty$ and that the naive geometric consideration on the 
quantum Gaussian shift model presented in Section~\!\ref{sec_gaussian} 
is mathematically justified.

Geometry of quantum statistical manifolds in an asymptotic framework would also be an important subject to be addressed. 
For example, consider a sequence $\sM^{(n)}=\{\rho_\xi^{\otimes n}\}$, $n=1, 2, \dots$,  of  i.i.d.~\!extensions of a quantum statistical model $\sM=\{\rho_\xi\}$.
Recent progress in asymptotic quantum statistics has revealed that the sequence exhibits a desirable property called a quantum local asymptotic normality, which tells us that in a shrinking ($\sim 1/\sqrt{n}$) neighbourhood of a given point $\xi_0$, the sequence converges to a quantum Gaussian shift model \cite{{guta_qubit},{guta_qudit},{qlan_first},{qcontiguity},{qEfficiency}, {qLAN_IG}}. 
As pointed out in Section~\!\ref{sec_gaussian}, 
the limiting quantum Gaussian shift model has a characteristic feature in view of quantum information geometry. 
It would, therefore, be an interesting future project to extend the geometrical idea presented in this paper to an asymptotic framework so that the convergence of quantum statistical manifolds can be discussed under a suitably chosen topology.

\section*{Acknowledgments}
This work is partly supported by JSPS KAKENHI Grant Numbers 23H05492 (HN), 25K24611(HN), 17H02861 (HN, AF), 23H01090 (AF), 23K25787 (AF), and JST ERATO Grant Number JPMJER2402 (AF).

\begin{appendices}
\section{Proof of  \eqref{SLD_torsion}}
\label{sec_proof_SLD_torsion}

We first consider the general situation where the e-connection $\nabla^{(\e)}$ is determined by a family of inner products 
$\inprod{A}{B}_\rho = \inprod{A}{\Omega_\rho (B)}_{\rm HS}$, and show that  
 the torsion ${\torsion}^{(\e)}$ of  $\nabla^{(\e)}$ is represented as follows: 
for any $X, Y\in \vf (\sS)$,
\begin{align}
\iota_* ( {\torsion}^{(\e)}( X, Y)) &= (Y \Omega) (L_X) - (X \Omega) (L_Y), 
\label{characterize_e-torsion}
\end{align}
where $Y\Omega : \rho \mapsto Y_\rho \Omega$ denotes the derivative of 
the super-operator-valued map $\Omega : \rho\mapsto \Omega_\rho$ w.r.t.\ $Y$, and 
$(Y \Omega) (L_X)$ denotes \rred{the} map $\rho\mapsto (Y_\rho \Omega) (L_{X_\rho}) \in \cLh$.
In fact, 
invoking \eqref{def_LX_2} and \eqref{e-connection_operator_representation},
we have for any $X, Y, Z\in \vf (\sS)$, 
\begin{align}
Z = {\torsion}^{(\e)}( X, Y) \; & \Leftrightarrow \; Z = \nabla_X^{(\e)} Y - \nabla_Y^{(\e)} X - [X, Y] 
\noret 
& \Leftrightarrow \; 
L_Z = (X L_Y + g(X, Y)) - (Y L_X + g(Y, X)) - L_{[X, Y]} 
\noret 
& \Leftrightarrow \; 
L_Z =  X L_Y - Y L_X - L_{[X, Y]}
\noret 
& \Leftrightarrow \; 
\iota_* (Z) = \Omega (X L_Y - Y L_X - L_{[X, Y]}) 
\noret 
& \Leftrightarrow \; 
\iota_* (Z) = \Omega (X L_Y) -  \Omega (Y L_X) - \iota_* ([X, Y]). 
\nonumber
\end{align}
Noting that 
\begin{align}
\iota_* ([X, Y]) &= X \iota_* (Y) - Y \iota_* (X) 
\noret &= 
X (\Omega (L_Y))  - Y (\Omega (L_X)) 
\noret &= 
(X \Omega) (L_Y) + \Omega (X L_Y) 
- (Y \Omega) (L_X) - \Omega (Y L_X), 
\nonumber
\end{align}
we obtain \eqref{characterize_e-torsion}. 
For the SLD structure, we have 
$\Omega_\rho (A) = \frac{1}{2} (\iota (\rho)\, A + A\, \iota (\rho))$, 
which yields that 
for any point $\rho\in \sS$ and any tangent vectors 
$X_\rho, Y_\rho \in T_\rho (\sS)$, 
\begin{align*}
(Y_\rho \Omega) (L_{X_\rho}) &=
\frac{1}{2}\left\{
\iota_* (Y_\rho)  L_{X_\rho}+ L_{X_\rho}  \iota_* (Y_\rho)  \right\}
\\
&=
\frac{1}{4} 
(\rho L_{Y_\rho} L_{X_\rho}  + L_{Y_\rho} \rho L_{X_\rho} 
+
L_{X_\rho} \rho L_{Y_\rho}  +L_{X_\rho} L_{Y_\rho} \rho).
\end{align*} 
Similarly, we have
\begin{align*}
(X_\rho \Omega) (L_{Y_\rho}) &= 
\frac{1}{4} 
(\rho L_{X_\rho} L_{Y_\rho}  + L_{X_\rho} \rho L_{Y_\rho} 
+
L_{Y_\rho} \rho L_{X_\rho}  +L_{Y_\rho} L_{X_\rho} \rho).
\end{align*} 
Substituting these into \eqref{characterize_e-torsion}, we obtain \eqref{SLD_torsion}.

\red{
\section{On the attainability for the infimum in  \eqref{eq_inf_uVu}}
\label{sec_attainability_min_uVu}
} 

\red{
Let $(\sM, \xi)$ be an $n$-dimensional model in $\sS (\cH)$. Given a point $\rho\in \sM$ and a vector 
$\bu = (u_i)\in\bR^n \setminus \{0\}$, we present some necessary conditions for the infimum 
 in  \eqref{eq_inf_uVu} to be attained by a minimum. 
Define
\be
X_{\bu, \rho} :=  \sum_{i, j} u_i g^{ij}(\rho) (\partial_i)_\rho \,\in T_\rho (\sM),
\label{X_u_rho_def}
\ee
where $\partial_i = \frac{\partial}{\partial\xi^i}$ and $G^{-1} =[g^{ij}]$ is the inverse matrix of the SLD Fisher information matrix $G = [g_{ij}]$. 
Then 
the SLD of the tangent vector $X_{\bu, \rho}$ is given by
\be
L_{\bu, \rho} := \sum_i u_i L^i_\rho = \sum_{i, j} u_i g^{ij}(\rho) L_{i, \rho},
\label{L_u_rho_def}
\ee
where $\{L_i\}$ denotes the SLDs for $\partial_i = \frac{\partial}{\partial\xi^i}$.
Let the spectral decomposition of $L_{\bu, \rho}$ be denoted by
\be
L_{\bu, \rho} = \sum_k a_k \nu_k,
\label{L_u_rho_spectral}
\ee
where $\{a_k\}$ are the distinct eigenvalues and $\{ \nu_k\}$ are the projectors onto the eigenspaces.  
Then we have the following proposition. 
} 

\red{
\begin{proposition}
\label{prop_min_uVu}
If the infimum of \eqref{eq_inf_uVu} is attained by a minimum so that
\be
\min_{\Pi\in\cU(\rho, \xi)}\,\tbu V_\rho(\Pi)\bu
 =\tbu G_\rho^{-1}\bu, 
\label{eq_min_uVu}
\ee
\red{
then the operators $\{\sum_k \nu_k (\partial_i \rho) \nu_k\}_{i \in\{ 1, \ldots, n\}}$ are linearly independent, or equivalently, the linear map 
\be
T_\rho (\sM) \rightarrow \cLh (\cH), \;\; X_\rho \mapsto  \sum_k \nu_k \iota_*( X_\rho) \nu_k
\label{X_nu_iotaX_nu}
\ee
is injective, 
where  $\iota_*$ denotes the differential map of the natural embedding $\iota : \sS \rightarrow \cLhone$ and 
$\partial_i \rho := \iota_*( (\partial_i)_\rho )$ (cf.~\eqref{m-rep_partial_rho}). 
} 
\end{proposition}
} 

\red{
Before giving a proof, 
 let us observe some consequences of Prop.~\!\ref{prop_min_uVu}. 
We introduce the \emph{commutation operator}  (see \cite{Holevo}) $D_\rho : T_\rho (\sS) \rightarrow T_\rho (\sS)$ defined by 
\be
\any X_\rho \in  T_\rho (\sS), \; \iota_* (D_\rho (X_\rho)) = \sqrt{-1}\; [\rho, L_{X_\rho}], 
\ee
where $L_{X_\rho}$ is the SLD for $X_\rho$ and $[\cdot , \cdot]$ denotes the commutator of operators. 
Let $X'_{\bu, \rho}$ be a tangent vector 
of $\sS$ (not necessarily of $\sM$) whose SLD satisfies
\be 
L'_{\bu, \rho} \in {\rm span} \{\nu_k \}_{k=1}^n, 
\label{L_X_rho_spectral}
 \ee
including the case when $X'_{\bu, \rho} = X_{\bu, \rho}$. 
Then, as a corollary of Prop.~\!\ref{prop_min_uVu} we have:
} 

\red{
\begin{corollary}
\label{cor_prop_min_uVu}
If $D_\rho (X'_{\bu, \rho}) \in T_\rho (\sM)\setminus \{0\}$, then the infimum of \eqref{eq_inf_uVu} cannot be attained 
by a minimum. 
\end{corollary}
} 

\red{
\begin{proofs}
For any $k$, we have
\begin{align*}
\nu_k\, \iota_* (D_\rho (X'_{\bu, \rho} ))\, \nu_k & =  \nu_k\, \sqrt{-1}\,[\rho, 
L'_{\bu, \rho}  ]\, \nu_k \\
&= \sqrt{-1}\, \left(\nu_k \rho\, L'_{\bu, \rho}  \nu_k - \nu_k L'_{\bu, \rho}  \rho\, \nu_k\right) =0, 
\end{align*}
where the last equality follows since 
$L'_{\bu, \rho} \nu_k = \nu_k L'_{\bu, \rho} \propto \nu_k$ due to \eqref{L_X_rho_spectral}. 
Hence, if $D_\rho (X'_{\bu, \rho}) \in T_\rho (\sM)\setminus \{0\}$, then the linear map in \eqref{X_nu_iotaX_nu} has a nontrivial kernel, so that the corollary follows from Prop.~\!\ref{prop_min_uVu}. 
\end{proofs}
} 

\red{
\begin{corollary}
When $\sM = \sS (\cH)$, the infimum of \eqref{eq_inf_uVu} cannot be attained by a minimum 
if $[\rho , L_{\bu, \rho} ] \neq 0$. 
\end{corollary}
} 

\red{
\begin{proofs}
Obvious from Cor.~\!\ref{cor_prop_min_uVu} since $[\rho , L_{\bu, \rho} ] \neq 0 \;\Leftrightarrow\; 
D_\rho (X_{\bu, \rho} ) \in T_\rho (\sS)\setminus \{0\}$.
\end{proofs}
} 

\red{
Let us proceed to  proving Prop.~\!\ref{prop_min_uVu}. 
Assume \eqref{eq_min_uVu}, and let 
$\Pi = \Pi (d\xihat)$ 
be an estimator achieving the minimum; i.e., $\Pi$ is locally unbiased at $\rho$ for $\xi$ and satisfies 
\be
\tbu V_\rho(\Pi)\bu =  \tbu G_\rho^{-1}\bu.
\label{uVu=uG-1u}
\ee
We need the following lemma.
}

\red{
\begin{lemma} Letting $\{\nu_k\}$ be the projectors onto the eigenspaces of $L_{\bu, \rho}$ introduced in \eqref{L_u_rho_spectral}, 
we have
\be
\Pi (d\xihat) = \sum_k \nu_k\, \Pi(d\xihat)\, \nu_k.
\label{Pi_nu_Pi_nu}
\ee
\end{lemma}
}

\red{
\begin{proofs} 
Let 
 \[
H:= 
\sum_i u_i 
\int 
(\xihat^i - \xi^i (\rho))\, \Pi (d \xihat)  = \int  h(\xihat)\, \Pi (d \xihat)
\; 
\in \cLh (\cH), 
\]
where
\be
h(\xihat) := \sum_i  u_i (\xihat^i - \xi^i (\rho)). 
\nonumber
\ee
Then, just in parallel with the derivation of (ii) $\Ra$ \eqref{gij_Lj_2_var} in the proof of Theorem~\!\ref{thm:autoparallel_m-affine},  
we have 
\be
\tbu V_\rho(\Pi)\bu \geq \| H \|_\rho^2 \geq \| L_{\bu, \rho} \|_\rho^2 = \tbu G_\rho^{-1}\bu.
\label{uVu_H_uG-1u}
\ee
Here, the first inequality follows from the operator inequality
\be
\int h(\xihat)^2 \, \Pi (d\xihat ) - H^2 = 
\int  (h(\xihat) - H)\, \Pi (d\xihat ) \, (h(\xihat) - H) \geq 0,
\label{int_hPh_geq_0}
\ee
and the second inequality follows from 
\be
\any j, \; 
\inprod{H}{L_{j, \rho}}_\rho = u_j = \inprod{L_{\bu, \rho}}{L_{j, \rho}}_\rho 
\label{<H,L>=<L,L>}
\ee
implying
\be
 \| H \|_\rho^2 = \| L_{\bu, \rho} \|_\rho^2 +  \| H - L_{\bu, \rho} \|_\rho^2.
 \nonumber
\ee
Note that \eqref{<H,L>=<L,L>} is a consequence of the local unbiasedness condition
\be
\any i,  \any j, \;\; 
\int  (\xihat^i - \xi^i (\rho))\, \Tr\left( (\partial_j \rho)\, \Pi (d \xihat ) \right) 
= \delta^i_j, 
\label{locally_unbiased_f_pi}
\ee
where $\partial_j \rho$ denotes $\iota_* ((\partial_j)_\rho)$ (cf.~\eqref{m-rep_partial_rho}). 
(Actually, the above proof of \eqref{uVu_H_uG-1u} is a standard argument to prove the SLD Cram\'{e}r-Rao inequality \eqref{eqn:CR}.) 
We see that  \eqref{uVu=uG-1u} holds iff the operator inequality  
\eqref{int_hPh_geq_0} holds with equality  and $H = L_{\bu, \rho}$, which implies that
\be
\int  (h(\xihat) - L_{\bu, \rho})\, \Pi (d\xihat ) \, (h(\xihat) - L_{\bu, \rho} ) = 0.
\label{int_hPh_=_0}
\ee
From this it follows that the support of the POVM (Positive Operator-Valued \emph{Measure}) $\Pi$ satisfies 
\be
{\rm supp} (\Pi) \subset 
\Gamma := 
\bigcup_{k} h^{-1} (a_k) =  \{ \xihat\,|\, h(\xihat ) \in \{a_1, a_2, \ldots\}\}, 
\label{supp_Pi_Gamma}
\ee
where $\{a_k\} = \{a_1, a_2, \ldots\}$ denote the distinct eigenvalues of $L_{\bu, \rho}$ as in \eqref{L_u_rho_spectral}. 
To see this, let $U$ be an arbitrary open set of $\bR^n$ such that $U \cap \Gamma = \phi$.  From \eqref{int_hPh_geq_0} and \eqref{int_hPh_=_0}, we have
\be
\int_U  (h(\xihat) - L_{\bu, \rho})\, \Pi (d\xihat ) \, (h(\xihat) - L_{\bu, \rho} ) =0.
\ee
This means that the (unnormalized) POVM $\Lambda (d \xihat) =  (h(\xihat) - L_{\bu, \rho})\, \Pi (d\xihat ) \, (h(\xihat) - L_{\bu, \rho} )$ vanishes on $U$.  Since $h(\xihat) - L_{\bu, \rho}$ is invertible for all 
$\xihat \in U \subset \bR^n \setminus \Gamma$, 
the POVM $\Pi (d \xihat)$ also vanishes on $U$, i.e., $\Pi (U) =0$.  We thus have \eqref{supp_Pi_Gamma}.
\\
\indent 
Given an arbitrary Borel set $B \subset \bR^n$, it follows from \eqref{int_hPh_=_0} and \eqref{supp_Pi_Gamma} that
\begin{align}
0 & = \int_B (h(\xihat) - L_{\bu, \rho})\, \Pi (d\xihat ) \, (h(\xihat) - L_{\bu, \rho} ) 
\noret &= \sum_k \int_{B\cap h^{-1} (a_k)} (h(\xihat) - L_{\bu, \rho})\, \Pi (d\xihat ) \, (h(\xihat) - L_{\bu, \rho} ) 
\noret &= \sum_k  (a_k - L_{\bu, \rho})\, \Pi (B\cap h^{-1} (a_k)) \, (a_k - L_{\bu, \rho} ) ,
\nonumber
\end{align}
which leads to
\be
\any k, \;\;    (a_k - L_{\bu, \rho})\, \Pi (B\cap h^{-1} (a_k)) =0. 
\nonumber
\ee
This implies that, for any $k$, the image of $\Pi (B\cap h^{-1} (a_k))$ is contained in the eigenspace 
of $L_{\bu, \rho}$ corresponding to the eigenvalue $a_k$, so that 
\begin{align}
\any k,  \;\; 
\Pi (B\cap h^{-1} (a_k)) = \nu_k\,  \Pi (B\cap h^{-1} (a_k))  & = \nu_k \, \Pi (B\cap h^{-1} (a_k))\, \nu_k, 
\label{Pi_B_h_a_nu}
\end{align}
and hence 
\begin{align}
\any k,  \;\; 
\Pi (B\cap h^{-1} (a_k))
&= \sum_l  \nu_l \, \Pi (B\cap h^{-1} (a_k))\, \nu_l.
\nonumber
\end{align}
Then we have 
\begin{align}
\Pi (B) = \sum_k \Pi (B \cap h^{-1} (a_k)) &= \sum_{k, l}    \nu_l  \,\Pi (B \cap h^{-1} (a_k))\, \nu_l 
\noret &= \sum_l \nu_l \,\Pi (B) \, \nu_l.
\nonumber
\end{align}
Since $B$ is arbitrary, the lemma has been proved. 
\end{proofs}
}

\medskip

\red{
Now, assume that 
\be
\sum_{i=1}^n c_i \, 
\sum_k 
\, \nu_k (\partial_i \rho )\,  \nu_k =0
 \label{nu_c_partial_rho_nu}
\ee
holds for some $(c_i) \in \bR^n$. 
It follows from \eqref{locally_unbiased_f_pi}  that for any $i$, 
\begin{align}
c^i & = 
\sum_j c^j  \int  (\xihat^i - \xi^i (\rho))\, \Tr\left( (\partial_j \rho)\,  \Pi (d\xihat) \right) 
\noret 
&= \int (\xihat^i - \xi^i (\rho))\, \Tr\Bigl( \Bigl(\sum_j c^j \partial_j \rho \Bigr) \,  \Pi (d\xihat) \Bigr).
 \nonumber 
\end{align}
Since  \eqref{Pi_nu_Pi_nu} and \eqref{nu_c_partial_rho_nu} yield 
\begin{align}
\Tr\Bigl( \Bigl(\sum_j c^j \partial_j \rho \Bigr) \,  \Pi (d\xihat) \Bigr) 
& = \Tr\Bigl( \Bigl(\sum_j c^j \partial_j \rho \Bigr) \, \sum_k \nu_k\, \Pi (d\xihat)\,\nu_k \Bigr) 
\noret &= 
\Tr\Bigl(\sum_k \nu_k \Bigl( \sum_j c^j \partial_j \rho\Bigr) \nu_k\, \Pi (d\xihat) \Bigr) =0, 
\nonumber
\end{align}
we obtain $\any i, c_i =0$. This means that the operators $\{\sum_k \nu_k (\partial_i \rho) \nu_k\}_{i \in\{ 1, \ldots, n\}}$ are linearly independent, 
which completes the proof of Prop.~\!\ref{prop_min_uVu}. 
}

\medskip

\red{
\begin{remark}
Noting that  \eqref{Pi_B_h_a_nu} implies $\Pi (h^{-1} (a_k))= \nu_k \Pi (h^{-1} (a_k)) \nu_k \leq \nu_k$ 
and that $\sum_k \Pi (h^{-1} (a_k)) = I = \sum_k \nu_k$, we have
\be
\any k , \;\; \Pi (h^{-1} (a_k)) = \nu_k.
\label{Pi_h_a_nu}
\ee
This indicates that the measurement $\Pi$ is a refinement of the projective measurement $\nu = (\nu_k)$. 
In particular, when $\Pi$ is represented as 
$\Pi = (\pi, f)$ with a POVM 
 $\pi = (\pi_\omega)_{\omega\in \Omega}$ on a discrete set $\Omega$ and a function $f = (f^i) : \Omega \rightarrow \bR^n$, the LHS of \eqref{Pi_h_a_nu} is represented as
 \be
 \Pi  (h^{-1} (a_k)) = \sum_{\omega \in \hat{h}^{-1} (a_k)} \pi_\omega, 
 \nonumber
 \ee
 where $\hat{h} (\omega) := h (f(\omega)) = \sum_i u_i  (f^i (\omega) - \xi^i (\rho))$, 
 and \eqref{Pi_h_a_nu} implies that
 \be
 \any \omega, \; \some k, \;\; \pi_\omega \leq \nu_k.
 \nonumber
 \ee
According to  Theorem~5 of \cite{fujiwara_adaptive},  if the infimum of \eqref{eq_inf_uVu} is attained by a minimum, then an estimator attaining the minimum can be taken to have finite support with cardinality not exceeding $(\dim \cH)^2 + n (n+1)$.  
 Therefore, Prop.~\!\ref{prop_min_uVu} could have been proved under the assumption that $\Pi$ takes the form $\Pi = (\pi, f)$ with $\pi$ having finite support. 
\end{remark}
} 


\section{Proof of Lemma \ref{lemma_Pi=(pi_f)}}
 \label{sec_proof_lemma_Pi=(pi_f)}
 \begin{itemize}
\item[(1)] Let $A^i :=  \int \xihat^i  \Pi (d\xi) $. Then we have
\begin{align}
A^i &
= \sum_k p_k f^i (k, X^k) = \sum_k p_k (\gamma_i^k +\frac{w^i_k}{p_k} X^k)
\noret 
&= \xi^i(\rho) + \sum_k w_k^i X^k 
=  \xi^i(\rho) + \sum_k  \sum_j  w_k^i u^k_j\, L^j_\rho 
=  \xi^i(\rho) + L_\rho^i, 
\nonumber
\end{align}
where we have invoked \eqref{charactrize_Pi_fpi}, \eqref{constraint_gammaki}, \eqref{def_Xk} and \eqref{def_fi}. 
Now $\Pi \in \cU (\rho, \xi)$ follows from  \eqref{locally_unbiased_Ai_Lj}. 
\red{
\item[(2)] 
Invoking  \eqref{charactrize_Pi_fpi}, 
we have for each $i, j$ 
\begin{align}
B^{ij}  & := \int   (\xihat^i - \xi^i (\rho))  (\xihat^j - \xi^j (\rho))  \Pi (d\xihat ) \noret 
&= 
\sum_k p_k \Bigl(f^i (k, \rred{X^k}) - \xi^i (\rho)\Bigr) \Bigl(f^j (k, \rred{X^k}) - \xi^j (\rho)\Bigr) 
   \noret 
 &= \sum_k p_k \Bigl(\frac{w_k^i}{p_k} X^k + \gamma_k^i - \xi^i (\rho)\Bigr) \Bigl(\frac{w_k^j}{p_k} X^k + \gamma_k^j - \xi^j (\rho)\Bigr).
\nonumber
\end{align}
This leads to
\begin{align}
(V_\rho(\Pi))^{ij}  &= \Tr (\rho\, B^{ij}) \noret 
& = \sum_k p_k \Tr\, \Bigl[\rho\, \Bigl(\frac{w_k^i}{p_k} X^k + \gamma_k^i - \xi^i (\rho)\Bigr) \Bigl(\frac{w_k^j}{p_k} X^k + \gamma_k^j - \xi^j (\rho)\Bigr) \Bigr]
\noret 
&= \sum_k \frac{w_k^i w_k^j}{p_k}\, \Tr  (\rho (X^k)^2 ) + \sum_k p_k (\gamma_k^i - \xi^i (\rho))  (\gamma_k^j - \xi^j (\rho)) 
\noret 
&= \sum_k \frac{w_k^i w_k^j}{p_k}\,  \transpose{(\bu^k)} G_\rho^{-1} \bu^k  + \sum_k p_k (\gamma_k^i - \xi^i (\rho))  (\gamma_k^j - \xi^j (\rho)), 
\nonumber
\end{align}
where we invoked $\Tr ( \rho\, X^k) =0$ due to 
$X^k \in {\rm span} \{L_{i, \rho}\}_{i=1}^n$.  
\item[(3)] 
Obvious from (2) since  $[w_k^i] = [u_i^k]^{-1}$. 
} 
\end{itemize}

\medskip

\section{Proofs of Propositions \ref{prop_V(F)_|d<F>|^2}, \ref{prop_E(S)} and \ref{prop_E(M)}}
\label{sec_proofs_prop_E(S)_prop_E(M)}

\begin{proofsvar}{of Prop.~\!\ref{prop_V(F)_|d<F>|^2}} \ 
This proposition is essentially contained in Theorem~7.2 of \cite{amanag}.  Here we give a proof 
for the reader's convenience.  

Given $F\in \cLh$ and $\rho\in \sS$, there exists a tangent vector $X_\rho\in T_\rho (\sS)$ satisfying 
$L_{X_\rho} = F - \expect{F}_\rho$ by  \eqref{e-tangent_space}.  Applying \eqref{g(XY)=df(Y)} 
to the case $\sM = \sS$, we have $X_\rho = ({\rm grad}~\! \expect{F})_\rho$, and hence
\begin{align}
\|(d\expect{F})_\rho\|_\rho^2 = \| X_\rho\|_\rho^2 = \inprod{L_{X_\rho}}{L_{X_\rho}}_\rho 
= \bigl\langle(F- \expect{F}_\rho)^2\bigr\rangle_\rho = V_\rho (F).
\nonumber
\end{align}
\end{proofsvar}

\begin{proofsvar}{of Prop.~\!\ref{prop_E(S)}} \ 
Recalling \eqref{efficiency_F} and \eqref{def_E(M)}, we have
\be
\cE (\sS) = \{ f\in \func (\sS)\, |\, \some F\in \cLh, \; f = \expect{F} \;\;\text{and}\;\; 
\any \rho\in \sS, \; V_\rho (F) = \| (df)_\rho\|_\rho^2\, \}.
\ee
Since the condition $V_\rho (F) = \| (df)_\rho\|_\rho^2$ is always satisfied by Prop.~\!\ref{prop_V(F)_|d<F>|^2}, 
we have the first equality in \eqref{prop_E(S)}.  The second equality follows from Prop.~\!\ref{prop_efficient_f_1}, 
and the third follows since under the relation $X\stackrel{g}{\longleftrightarrow} \omega$ we have
\begin{align}
\text{$X$ is e-parallel} \; & \Leftrightarrow \; \any Y, Z\in \vf (\sS), \; 
g ( \nabla^{(\e)}_Y X, Z ) =0
\noret & \Leftrightarrow \; \any Y, Z\in \vf (\sS), \;  
Y g( X, Z) = g(X, \nabla^{(\m)}_Y Z)
\noret & \Leftrightarrow \; \any Y, Z\in \vf (\sS), \;  
Y \omega (Z) = \omega( \nabla^{(\m)}_Y Z)
\noret & \Leftrightarrow \; \text{$\omega$ is m-parallel}.
\label{X_e-parallel_omega_m-parallel}
\end{align}
\end{proofsvar}

\begin{proofsvar}{of Prop.~\!\ref{prop_E(M)}} \ 
By Propositions \ref{prop_V(F)_|d<F>|^2} and \ref{prop_E(S)}, 
the condition imposed on $f$ in \eqref{eq_prop_E(M)} is equivalent to the existence of 
$F\in \cLh$ satisfying \eqref{efficiency_F}. 
\end{proofsvar}

\section{\red{Proof of Prop.~\!\ref{prop_totally_geodesic_autoparallel}} 
}
\label{sec_proof_prop_totally_geodesic_autoparallel}


We present a proof of Prop.~\!\ref{prop_totally_geodesic_autoparallel} 
below, which is almost parallel to  the proof of Theorem~\!8.4 in Chap.~\!VII of \cite{KobayashiNomizu} 
cited as a result due to E. Cartan.

Let $\dim \mS = n+r$, and $\mM$ be a $\nabla$-totally geodesic submanifold 
with $\dim \mM =n$.  We take a coordinate system $\tilde{\xi} = (\tilde{\xi}^i)$ of $\mS$ such that 
$\mM$ is represented as 
\be \mM = \{ p\in \mS\,|\, \any i\in \{ n+1 , \ldots , n+r\}, \; \tilde{\xi}^i (p) =0\}
\nonumber
\ee
and that 
$(\xi^1, \ldots , \xi^n) := (\tilde{\xi}^1|_{\mM}, \ldots , \tilde{\xi}^n|_{\mM})$ forms a coordinate system of $\mM$. 
Let $\tilde{\partial}_i := \frac{\partial}{\partial\tilde{\xi}^i}$, $\partial_i := \frac{\partial}{\partial\xi^i}$, and 
denote the connection coefficients of $\nabla$ w.r.t.\ $\tilde{\xi}$ by $\{\Gamma_{ij}^k\}$: 
$\nabla_{\tilde{\partial}_i} \tilde{\partial}_j = \sum_k \Gamma_{ij}^k\, \tilde{\partial}_k$ for $i, j \in \{ 1, \ldots , n+r\}$. 
For arbitrary $i, j$, it follows from the assumption \eqref{torsionXY_spanXY} that $\torsion^{(\nabla)} (\tilde{\partial}_i, \tilde{\partial}_j) = \sum_k (\Gamma_{ij}^k - \Gamma_{ji}^k) \, \tilde{\partial}_k \in {\rm span}_{\func (\mS)} \{ \tilde{\partial}_i, \tilde{\partial}_j\}$, which implies that $ \Gamma_{ij}^k - \Gamma_{ji}^k =0$ for any $k\notin \{i, j\}$. 
Hence we have
\be
\any i, j \in \{1, \ldots , n\}, \; \any k\in \{n+1, \ldots , n+r\}, \;  \Gamma_{ij}^k = \Gamma_{ji}^k.
\label{Gammaijk=Gammajik}
\ee
Given a point $p\in \mM$ and a tangent vector $X_p = \sum_{i=1}^n x^i\, (\partial_i)_p \in T_p (\mM)$ 
arbitrarily, let $\gamma : t \mapsto \gamma (t) $ be a $\nabla$-geodesic with an affine parameter $t$ 
satisfying $\gamma (0) = p$ and $\dot{\gamma} (0) := \frac{d}{dt} \gamma (t) |_{t=0} = X_p$. 
The geodesic should satisfy the differential equation $\nabla_{\dot{\gamma} (t)} \dot{\gamma} (t) =0$, which   
is represented as
\begin{align}
 \any k\in \{1, &\ldots , n+r\}, \;  
 \frac{d^2}{d t^2}\, \tilde{\xi}^k (\gamma (t)) + \sum_{i, j=1}^{n+r} \frac{d}{d t}\, \tilde{\xi}^i (\gamma (t))\, 
 \frac{d}{d t}\, \tilde{\xi}^j (\gamma (t)) \, \bigl(\Gamma_{ij}^k\bigr)_ {\gamma (t)} = 0. 
 \nonumber
\end{align}
Since $\mM$ is assumed to be $\nabla$-totally geodesic, $\gamma (t)$ stays in $\mM$ and hence 
$\tilde{\xi}^k ( \gamma (t)) = 0$ for $ k\in \{n+1, \ldots , n+r\}$. 
Therefore, the above equation yields
\be
\any k\in \{n+1, \ldots , n+r\}, \;  \sum_{i, j=1}^{n} \frac{d}{d t}\, \tilde{\xi}^i (\gamma (t))\, 
 \frac{d}{d t}\, \tilde{\xi}^j (\gamma (t)) \, \bigl(\Gamma_{ij}^k\bigr)_ {\gamma (t)} = 0, 
 \nonumber
\ee
and letting $t=0$, we obtain
\be
\any k\in \{n+1, \ldots , n+r\}, \;  \sum_{i, j=1}^{n}x^i 
x^j  \bigl(\Gamma_{ij}^k\bigr)_ {p} = 0. 
\nonumber
\ee
Since $p\in \mM$ and $X_p = \sum_{i=1}^n x^i (\partial_i)_p$ are arbitrary 
and $\Gamma_{ij}^k$ is symmetric w.r.t.\ $i\leftrightarrow j$ by \eqref{Gammaijk=Gammajik}, 
it follows that
\be
\any i, j \in \{1, \ldots , n\}, \; \any k\in \{n+1, \ldots , n+r\}, \;  
\Gamma_{ij}^k |_{\mM} =0.
\ee
Now, for arbitrary vector fields $X = \sum_{i=1}^n X^i \partial_i =  \sum_{i=1}^n X^i \tilde{\partial_i}|_{\mM}$ and $Y = \sum_{j=1}^n Y^j \partial_j$ $= \sum_{j=1}^n Y^j \tilde{\partial_j}|_{\mM}$ on $\mM$, 
where $\{X^i\}, \{Y^j\} \subset \func (\mM)$, we have
\begin{align}
\nabla_X Y & =   \sum_{i, j=1}^{n} \sum_{k=1}^{n+r} X^i Y^j \Gamma_{ij}^k |_{\mM}\,  \tilde{\partial}_k|_{\mM}  + \sum_{j=1}^n X (Y^j) \partial_j 
\noret 
&=  \sum_{i, j =1}^{n} \sum_{k=1}^{n} X^i Y^j \Gamma_{ij}^k |_{\mM}\,  \partial_k + \sum_{j=1}^n X (Y^j) \partial_j  
\; \in \vf (\mM), 
\nonumber
\end{align}
which concludes that $\mM$ is $\nabla$-{\ap} in $\mS$ and completes the proof. 

Note that the only difference from the proof of \cite{KobayashiNomizu} is whether 
\eqref{Gammaijk=Gammajik} is derived from $\torsion^{(\nabla)} =0$ or from the weaker assumption 
\eqref{torsionXY_spanXY}. 

\red{
\section{Proof of Prop.~\!\ref{prop_foliation}}
} 
\label{sec_proof_prop_foliation}

\red{
\begin{itemize}
\item[(1)] (a) is straightforward from condition (i) as already mentioned.  To show (b) and (c), 
let us introduce the distribution $\cV^{\perp} : \mS \ni p \mapsto V_p^{\perp}$, where 
$V_p^\perp$  denotes the orthogonal complement of $V_p$ in $(T_p (\mS), g_p)$. Then $\cV^{\perp}$ is $\nabla^*$-parallel since 
\[
\any p, q \in \mS, \;\; 
 \para^{(\nabla^*)}_{p,q} (V_p^{\perp}) =  \bigl (\para^{(\nabla)}_{p,q} (V_p) \bigr)^{\perp} 
 = V_q^{\perp}, 
\]
where the first equality follows from \eqref{duality_parallel_transport}. 
Since $\nabla^*$ is torsion-free, the $\nabla^*$-parallel distribution $\cV^{\perp}$ is integrable 
(cf. (iv) in Prop.~\!\ref{prop_integrability}) and generates a foliation $\mS =  \bigsqcup_{\beta} \mN_\beta$  satisfying (b).  Property (c) is obvious from the construction. 
\item[(2)] 
Define the 1-forms $\{\omega^i\} = \{\omega^1, \ldots, \omega^n\}$ by $X^i \stackrel{g}{\longleftrightarrow} \omega^i$. 
Then $\{\omega^i\}$ are $\nabla^*$-parallel due to \eqref{X_nabla-parallel_omega_nabla*-parallel}, and 
$\some \{\zeta^i\}$, $\omega^i = d\zeta^i$ due to \eqref{omega_nabla*-parallel_df}. Thus  we have 
$X^i \stackrel{g}{\longleftrightarrow} d\zeta^i$. 
For each point $p\in \mS$ and each $i$, the relation $X^i_p \stackrel{g_p}{\longleftrightarrow} (d\zeta^i)_p$ implies that 
${\rm Ker}\, (d\zeta^i)_p = (X^i_p)^{\perp}$, so that ${\rm Ker}\, (d\zeta)_p = V_p^{\perp}$.  
To show (a), let $\mM = \mM_\alpha$ for an arbitrary $\alpha$, and define $\xi = (\xi^i)$ by $\xi^i := \zeta^i|_{\mM}$. Then we have
\be
\any p\in \mM, \;\; {\rm Ker}\, (d\xi)_p = {\rm Ker}\, (d\zeta)_p \cap T_p (\mM) = V_p^{\perp} \cap V_p = \{0\}, 
\nonumber 
\ee
which implies that $\xi$ forms a coordinate system of $\mM$ (at least locally). 
Noting that $\mM$ is a $\nabla$-{\ap} submanifold of $\mS$ generated by the distribution 
$\cV : p\mapsto V_p = {\rm span} \{X^i_p\}$ and that $X^i \stackrel{g}{\longleftrightarrow} d\zeta^i$, we have
\be
\text{$X^i|_{\mM}$ is a $\nabla|_{\mM}$-parallel vector field on $\mM$} \;\;\text{and}\;\; X^i|_{\mM} \stackrel{g|_{\mM}}{\longleftrightarrow} d\xi^i.  
\nonumber
\ee
It then follows that $d\xi^i$ is $\nabla^*_{\mM}$-parallel for every $i$ (cf.\ \eqref{X_nabla-parallel_omega_nabla*-parallel}), which is equivalent to saying that the coordinate system $\xi = (\xi^i)$ is $\nabla^*_{\mM}$-affine  as 
\begin{align}
& \any i,\; \any Y\in \vf (\mM), \; (\nabla^*_{\mM})_Y\, d\xi^i = 0 \; 
\noret 
 \Leftrightarrow \;\; & 
\any i, j, \; \any Y \in \vf (\mM), \; \bigl( (\nabla^*_{\mM})_Y\, d\xi^i \bigr) (\partial_j) = 0, 
\;\;\text{where}\;\; \partial_j := \frac{\partial}{\partial \xi^j}
\noret \Leftrightarrow \;\;  & 
\any i, j, \; \any Y\in\vf (\mM), \; 
Y (d\xi^i (\partial_j)) - d\xi^i ( (\nabla^*_{\mM})_Y \partial_j) = 0 
\noret \Leftrightarrow \;\; &
 \any j, \; \any Y\in \vf (\mM), \; (\nabla^*_{\mM})_Y \partial_j =0, 
 \end{align}
 where the last equivalence follows from $d\xi^i (\partial_j) = \delta^i_j = {\rm const}$.  
 This proves \rred{(a)}. 
 Property \rred{(b)} is derived from 
 \[
 \any \beta, \; \any p\in \mN_\beta, \;\; 
 T_p (\mN_\beta) = V_p^{\perp} = {\rm Ker}\, (d\zeta)_p, 
 \]
 which means that each leaf  $\mN_\beta$ is represented by the constraint $\zeta$ = const. 
\end{itemize}
} 

\red{
\section{Proof of Lemma \ref{lemma_QFQC_RFRC_R=0}}
 \label{sec_proof_lemma_QFQC_RFRC_R=0}
} 
\red{
We can assume $Q$ to be a  diagonal matrix with no loss of generality, and let
\[
Q = [q_i \delta_{ij}], \; q_i > 0, \; R = [r_{ij}], \; r_{ij} = - r_{ji}. 
\]
Suppose that real matrices $F = [f_{ij}]$, $f_{ij} = - f_{ji}$ and $C = [c_{ij}]$, $c_{ij} = c_{ji}$ satisfy 
\[
[Q, F] = Q\circ C 
\;\;\text{and}\;\; [R, F] = R\circ C, 
\]
which are represented as
\[
(q_i - q_j) f_{ij} = \frac{q_i + \rred{q_j}}{2} c_{ij}
\]
and 
\[
\sum_k (r_{ik} f_{kj} - f_{ik} r_{kj}) = \frac{1}{2} \sum_k (r_{ik} c_{kj} + c_{ik} r_{kj}).
\]
Eliminating $\{c_{ij}\}$ from these equations, we have 
\be
\sum_k \frac{q_j}{q_j + q_k} r_{ik} f_{kj} = \sum_k \frac{q_i}{q_i + q_k} r_{kj} f_{ik}. 
\label{component_QFQC_RFRC}
\ee
Condition \eqref{QFQC_RFRC} means that this equation holds identically for variables 
$\{f_{ij}\}$ satisfying $f_{ij} = - f_{ji}$. For each $(i, j)$ with $i\neq j$, the variables 
$\{f_{kj}\}_{k\neq j}$ and $\{f_{ik}\}_{k\neq i}$ appearing in \eqref{component_QFQC_RFRC} 
are all independent except for $f_{ij} $, so that the identity \eqref{component_QFQC_RFRC}  
implies that 
\[
\any k\in \{1 , \ldots , d\}\setminus\{ i, j\}, \;\; \frac{q_j}{q_j + q_k} r_{ik} = 
\frac{q_i}{q_i + q_k} r_{kj} =0, 
\]
which leads to
\be
\any k\in \{1 , \ldots , d\}\setminus\{ i, j\}, \;\; r_{ik} = r_{kj} =0. 
\label{rik=rkl=0}
\ee
When $d\geq 3$, the condition that \eqref{rik=rkl=0} holds for $\any i \neq \any j$ 
implies that $r_{ij} = 0$ for $\any i \neq \any j$, i.e., $R=0$. 
} 

\medskip

\section{Proof of Lemma \ref{lemma_qubit_torsion}}
 \label{sec_proof_lemma_qubit_torsion}

 When $\Tr A = \Tr B =0$, \eqref{eq_lemma_qubit_torsion} is reduced to
\begin{align}
\frac{1}{2} [[A, B], \rho] = & 
 - 2 \expect{A}_\rho  (\rho\circ B )
+ 2 \expect{B}_\rho (\rho \circ A), 
\label{eq_lemma_qubit_torsion_Tr=0}
\end{align}
which we prove first.  Letting $A = \avec\cdot \sigmavec$, $B = \bvec\cdot \sigmavec$ 
and $\rho = \frac{1}{2} (I + \rvec\cdot \red{\sigmavec} 
)$, 
it immediately follows from 
 \eqref{qubit_AcircB} and \eqref{qubit_[A,B]} that
\be
\frac{1}{2} [[A, B], \rho] = (\rvec \times (\avec\times \bvec) )\cdot \sigmavec
\nonumber
\ee
and
\be
 - 2 \expect{A}_\rho  (\rho\circ B )
+ 2 \expect{B}_\rho (\rho \circ A )
= \bigl\{
(\bvec\cdot \rvec) \avec - (\avec\cdot \rvec) \bvec
\bigr\} \cdot \sigmavec.
\nonumber
\ee
Hence, the well-known formula for the vector triple product proves 
\eqref{eq_lemma_qubit_torsion_Tr=0}. 

Remove the assumption $\Tr A = \Tr B =0$, and let
$A' := A - \frac{\Tr A}{2}  I$ and $B' := B - \frac{\Tr B}{2}  I$.  Then we have
\begin{align}
\frac{1}{2} [[A, B], \rho]  = & \frac{1}{2} [[A', B'], \rho]  
\noret 
= & - 2 \expect{A'}_\rho  (\rho\circ B' )
+ 2 \expect{B'}_\rho (\rho \circ A')
\noret 
= &
(\Tr A - 2 \expect{A}_\rho ) (\rho\circ B) - 
(\Tr B - 2 \expect{B}_\rho ) (\rho \circ A )
\noret 
& + \bigl\{
(\Tr B) \expect{A}_\rho - (\Tr A) \expect{B}_\rho\bigr\} \,\rho, 
\nonumber
\end{align}
where the second equality follows from \eqref{eq_lemma_qubit_torsion_Tr=0}. 
Thus we obtain \eqref{eq_lemma_qubit_torsion}.

\medskip

\red{
\section{Proof of \eqref{trWV_tau} and the condition for equality}
\label{sec_proof_equality_trWV_tau}
}

\red{
Let $A = [a_{kl}]$ and $B = [b_{kl}]$  be 
$n\times n$ positive-definite real matrices defined by
\[
a_{kl} := \transpose{(\bu^k)} G_\rho^{-1} \bu^l \;\;\text{and}\;\; 
b_{kl}:=   \transpose{(\bw_k)} \wgt \bw_l.
\]
Then we have
\begin{align}
\tr (\wgt V_\rho (\Pi)) & = \sum_k \frac{1}{p_k} a_{kk} b_{kk} 
= \sum_k \left(\frac{\sqrt{a_{kk} b_{kk}}}{\sqrt{p_k}}\right)^2 
\cdot
\sum_k (\sqrt{p_k})^2 
\noret 
&\geq \left(\sum_{k} \sqrt{a_{kk} b_{kk}}\right)^2 
\geq 
\Bigl(\tr \Bigl(A^{\frac{1}{2}} B A^{\frac{1}{2}}\Bigr)^{\frac{1}{2}}\Bigr)^2,
\label{aux1_trWV_tau}
\end{align}
where the last inequality can be regarded as a variant of 
the monotonicity of the fidelity or of the Bures distance \cite{nielsen_chuang}. 
Letting $U := ( \bu^1, \ldots, \bu^n) \in \bR^{n\times n}$, we have
$
A = \transpose{U} G_\rho^{-1} U$, so that 
there exists an orthogonal matrix $Q$ such that $A^{\frac{1}{2}} = Q G_\rho^{-\frac{1}{2}} U$. 
Noting that  $U^{-1} = \transpose{(\bw_1, \ldots, \bw_n)}$ and that 
$
B = U^{-1} \wgt \transpose{(U^{-1})}, 
$
we have
\[
A^{\frac{1}{2}} B A^{\frac{1}{2}} = Q G_\rho^{-\frac{1}{2}} \wgt G_\rho^{-\frac{1}{2}} \transpose{Q}, 
\]
which yields
\be
\Bigl(A^{\frac{1}{2}} B A^{\frac{1}{2}}\Bigr)^{\frac{1}{2}} = 
Q 
\Bigl(G_\rho^{-\frac{1}{2}} \wgt G_\rho^{-\frac{1}{2}}\Bigr)^{\frac{1}{2}} 
\transpose{Q} 
\nonumber 
\ee
and 
\be
\tr \Bigl(A^{\frac{1}{2}} B A^{\frac{1}{2}}\Bigr)^{\frac{1}{2}} = \tr \Bigl(G_\rho^{-\frac{1}{2}} \wgt G_\rho^{-\frac{1}{2}}\Bigr)^{\frac{1}{2}}.
 \label{aux3_trWV_tau}
\ee
Now \eqref{trWV_tau} follows from \eqref{aux1_trWV_tau} and \eqref{aux3_trWV_tau}. 
} 

\red{
Equality in  \eqref{trWV_tau} holds iff both of the two inequalities in \eqref{aux1_trWV_tau} hold with equality, 
which occurs iff
\be
p_k \propto \sqrt{a_{kk} b_{kk}} 
\label{1st_equality_trWV_tau} 
\ee
and
\be
\exists D = {\rm diag} \{d_1, \ldots , d_n\}, \; d_k > 0, \; B = D A D.
\label{2nd_equality_trWV_tau}
\ee
Since $U B \transpose{U} = W$ and $(\transpose{U})^{-1} A U^{-1} = G_\rho^{-1}$ we have
\begin{align}
B = DAD \; &  \Leftrightarrow \; W = K G_\rho^{-1} K, 
\label{B=DAD_W=KGK}
\end{align}
where 
\be
K := U D \transpose{U} = \sum_{k} d_k \bu^k \transpose{(\bu^k)}. 
\label{def_K_dk}
\ee
In addition, it follows from $B = DAD$ that 
$\sqrt{ a_{kk} b_{kk}} = d_k a_{kk}$. 
Hence,  the equality conditions \eqref{1st_equality_trWV_tau}  and \eqref{2nd_equality_trWV_tau} are equivalent to the existence of 
positive numbers $\{ d_k \}$ such that 
\[
d_k \propto \frac{p_k}{a_{kk}} \;\;\text{and}\;\; 
W = K G_\rho^{-1} K \;\;\text{with}\;\; \eqref{def_K_dk}, 
\]
which is also equivalent to the existence of a positive-definite matrix $K$ such that 
\[
W = K G_\rho^{-1} K 
\;\;\text{and}\;\;
K \propto \sum_k \frac{p_k}{a_{kk}} \bu^k \transpose{(\bu^k)}, 
\]
namely, \eqref{W_KGK} and \eqref{K_propto}. 
} 

\medskip

\red{
\section{On a result of Yamagata \cite{yamagata_tomography} and its relation to our results}
 \label{sec_yamagata}
 } 
 \red{
 In the situation of Example~\ref{example_qubit_estimators}, let $\pi$ be  the random measurement represened as
 \be
{\pi: \randmeas{(p_k)}{k}{\fbox{$\sigma_k$}}{y}}.
\label{random_measurement_gamma_xi_qubit}
\ee
Note that the estimators defined in \eqref{estimation_procedure_gamma_xi_qubit} and \eqref{estimation_procedure_filtration_qubit} are both represented in the form $\Pi = (\pi, f)$ by some functions $f$. 
Yamagata  \cite{yamagata_tomography}  treated the quantum tomography for qubit states as the 
MLE (maximum likelihood estimator) for the Stokes parameter $\xi = (\xi^1, \xi^2, \xi^3)$ of the full model $\sM = \sS (\cH)$ 
based on repetitive measurement of this $\pi$ with the uniform distribution 
$p_k = 1/3$.  According to the classical estimation theory, the asymptotic variance of the MLE 
for a state $\rho$ is equal to $(G_\rho^{(\pi)})^{-1}$, where $G_\rho^{(\pi)}$  denotes the Fisher information matrix of the classical model 
induced from $(\sS (\cH), \xi)$ and the measurement $\pi$. 
In addition, 
owing to the achievability of the classical Cram\'{e}r-Rao inequality, we can construct an 
estimator based on the measurement $\pi$ which is locally unbiased at $\rho$ and has the variance 
$(G_\rho^{(\pi)})^{-1}$. Hence, for an arbitrary weight matrix $\wgt = [w_{ij}] >0$,  it follows from \eqref{min_trWV_tau} that
\be
\tr (\wgt (G_\rho^{(\pi)})^{-1}) \geq \tau_\rho (\wgt).
\label{tr_WG_tau}
\ee
Yamagata showed that the necessary and sufficient condition for 
a weight field ${\mathcal \wgt} = \{\wgt_\rho = [w_{ij} (\rho)] \; |\; \rho \in \sS\}$ 
to satisfy 
\be
\any \rho\in \sS (\cH), \;\; \tr (\wgt_\rho (G_\rho^{(\pi)})^{-1}) = \tau_\rho (\wgt_\rho)
\ee
is that  \eqref{wij_pipj_uniform} identically holds; i.e., 
\be
\any \rho \in \sS (\cH), \;\; 
w_{ij}(\rho) \propto  \frac{\delta_{ij} - \xi^i(\rho) \xi^j(\rho)}{\bigl(1 - (\xi^i(\rho))^2\bigr)
\bigl(1 - (\xi^j(\rho))^2\bigr)},
\label{wij_rho_pipj_uniform}
\ee
where the proportionality constant may depend on $\rho$. 
\\
\indent
This result is readily extended to the situation of Example~\ref{example_qubit_estimators} 
where an e-{\ap} model $(\sM , \xi)$ such that \eqref{xi_sigma_L} holds for every $\rho\in\sM$ is given 
and $(p_k)$ in \eqref{random_measurement_gamma_xi_qubit} is an arbitrary positive probability vector,  implying that 
for a weight field ${\mathcal \wgt} = \{\wgt_\rho = [w_{ij} (\rho)] \; |\; \rho \in \sM\}$ on the model, we have
\begin{align}
& \any \rho\in \sM, \;\; \tr (\wgt_\rho (G_\rho^{(\pi)})^{-1}) = \tau_\rho (\wgt_\rho) 
\noret 
\Leftrightarrow \;\; &
\any \rho \in \sM, \;\; 
w_{ij}(\rho) \propto  p_i p_j \frac{\delta_{ij} - \xi^i(\rho) \xi^j(\rho)}{\bigl(1 - (\xi^i(\rho))^2\bigr)
\bigl(1 - (\xi^j(\rho))^2\bigr)}.
\label{wij_rho_pipj_G}
\end{align}
On the other hand, \eqref{wij_pipj} means that 
\begin{align}
& \any \rho\in \sM, \;\; \tr (\wgt_\rho V_\rho (\Pi_\rho)) = \tau_\rho (\wgt_\rho) 
\noret 
\Leftrightarrow \;\; &
\any \rho \in \sM, \;\; 
w_{ij}(\rho) \propto  p_i p_j \frac{\delta_{ij} - \xi^i(\rho) \xi^j(\rho)}{\bigl(1 - (\xi^i(\rho))^2\bigr)
\bigl(1 - (\xi^j(\rho))^2\bigr)},
\label{wij_rho_pipj_V}
\end{align}
where $\Pi_\rho$ denotes the estimation defined by \eqref{estimation_procedure_gamma_xi_qubit}. 
Here,  $\Pi_\rho$ depends on $\rho$ in the form $\Pi_\rho = (\pi, f_\rho)$, where 
\be
f_\rho (k, y) := 
\xi (\rho) +\frac{y - \xi^k(\rho)}{p_k}\, {\mbox{\boldmath$e$}}_k.
\ee
The equivalence of \eqref{wij_rho_pipj_G} and \eqref{wij_rho_pipj_V} may be understood as a consequence of the identity
\be
V_\rho (\Pi_\rho) = (G_\rho^{(\pi)})^{-1}, 
\ee
which follows from the classical Cram\'{e}r-Rao inequality $V_\rho (\Pi_\rho) \geq (G_\rho^{(\pi)})^{-1}$ and the existence of $\wgt >0$ such that 
$\tr (\wgt V_\rho (\Pi_\rho) ) = \tau_\rho (\wgt) \leq \tr (\wgt  (G_\rho^{(\pi)})^{-1})$ (see \eqref{tr_WG_tau}). 
\\
\indent
The conditions in \eqref{wij_rho_pipj_G} and \eqref{wij_rho_pipj_V} are expressed as
\be
\any \rho\in \rred{\sM}, \;\; \tr (\wgt_\rho (G_\rho^{(\pi)})^{-1}) 
= \min_{\pi': \text{POVM}} \tr (\wgt_\rho (G_\rho^{(\pi')})^{-1})
\label{tr_WG_pi_min}
\ee
and 
\be
\any \rho\in \sM, \;\; \tr (\wgt_\rho V_\rho (\Pi_\rho)) = 
\min_{\Pi'\in\cU(\rho, \xi)} \tr (\wgt_\rho V_\rho (\Pi')).
\label{tr_WV_Pi_rho_min}
\ee
It is worthwhile to compare these conditions with the definition of ${\mathcal \wgt}$-efficiency of an estimator given in Section~\ref{sec_efficient_estimator}; 
an estimator $\Pi$ for the model $(\sM, \xi)$ is said to be ${\mathcal \wgt}$-efficient when it satisfies $\any \rho\in \sM, \Pi \in \cU(\rho, \xi)$, which is equivalent to the unbiasedness 
$\any \rho\in \sM, E_\rho (\Pi) = \xi (\rho)$, and 
\be
\any \rho\in \sM, \;\; \tr (\wgt_\rho V_\rho (\Pi)) = 
\min_{\Pi'\in\cU(\rho, \xi)} \tr (\wgt_\rho V_\rho (\Pi')).
\label{tr_WV_Pi_min}
\ee
Since $f_\rho$ in $\Pi_\rho = (\pi, f_\rho)$ depends on $\rho$, \eqref{tr_WV_Pi_rho_min} does not mean the existence of a ${\mathcal \wgt}$-efficient estimator.  On the other hand, if we take $f$ to be $f_0(k, y) = \frac{y}{p_k} {\mbox{\boldmath$e$}}_k$ as in \eqref{estimation_procedure_filtration_qubit}, we obtain an unbiased estimator $\Pi_0 = (\pi, f_0)$, but its variance for a state $\rho\in\sM$ does not achieve the minimun in \eqref{tr_WV_Pi_min} except when $f_0 = f_\rho$, i.e. $\xi(\rho) =0$. 
} 

\medskip

\red{
\section{Proof of Prop.~\!\ref{prop_no_W-efficient_qubit}}
 \label{sec_proof_prop_no_W-efficient_qubit}
} 
\red{
Suppose that an $n$-dimensional  e-{\ap} qubit model $(\sM, \xi)$ and a positive weight $\wgt= [w_{ij}] \in \bR^{n\times n}$ are 
arbitrary given as in the statement of the proposition.  We assume the existence of a $\wgt$-efficient estimator for  $(\sM, \xi)$ and derive a contradiction. As noted in Example~\ref{example_qubit_estimators}, 
we may assume without loss of generality that 
 \eqref{xi_sigma_L} holds for any $\rho\in\sM$; i.e., 
\be
\any \rho\in \sM, \;\; 
\any i \in \{1, \ldots , n\}, \; \xi^i (\rho) =  \expect{\sigma_i}_\rho \;\;\text{and}\;\; 
L_\rho^i = \sigma_i -  \xi^i (\rho),
\nonumber
\ee
which implies \eqref{gij_qubit_example}; i.e., 
\be
g_\rho^{ij} = \delta_{ij}  - \xi^i(\rho) \xi^j(\rho).
\label{gij_qubit_example_appendix}
\ee
Now, assume that an estimator  $\Pi = \Pi(d\xihat)$ for the model $(\sM, \xi)$ is $\wgt$-efficient. 
Namely, we assume  that 
\be
\any \rho\in \sM, \;\; E_\rho (\Pi) = \xi (\rho) \;\;\text{and}\;\; \tr (\wgt V_\rho (\Pi)) = \tau_\rho (\wgt), 
\nonumber
\ee
which is  expressed as
\be
 \any i \in \{1, \ldots, n\}, \; \any \rho \in \sM, \;\; \expect{A^i}_\rho = \xi^i (\rho)
 \label{expect_A_xi}
\ee
and 
\be
\any \rho \in \sM, \;\; \sum_{i, j} w_{ij} \expect{B^{ij}}_\rho - \sum_{i, j} w_{ij} \xi^i (\rho) \xi^j (\rho) = \tau_\rho (\wgt), 
\label{tau_w_B}
\ee
where
\be
A^i := \int{\xihat}^i \Pi (d\xihat) \;\;\text{and}\;\; B^{ij} := \int{\xihat}^i {\xihat}^j \Pi (d\xihat), 
\nonumber
\ee 
Letting $\{\lambda_i(\rho) \}$ be the eigenvalues of $G_\rho^{-\frac{1}{2}} \wgt G_\rho^{-\frac{1}{2}}$, we have
\be
\tau_\rho (\wgt) = \Bigl( \sum_i \sqrt{\lambda_i (\rho)} \Bigl)^2 = \sum_i \lambda_i (\rho) + 2\sum_{i < j} \sqrt{\lambda_i (\rho) \lambda_j (\rho)}.
\ee
Since \eqref{gij_qubit_example_appendix} yields
\be
 \sum_i \lambda_i (\rho) = \tr (\wgt G_\rho^{-1}) = \sum_i w_{ii} -  \sum_{i, j} w_{ij} \xi^i (\rho) \xi^j (\rho), 
 \nonumber
\ee
\eqref{tau_w_B} is expressed as
\be
\any \rho\in\sM, \;\; \expect{B}_\rho = \tr \wgt + 2\sum_{i < j} \sqrt{\lambda_i (\rho) \lambda_j (\rho)},
\label{B_W_lambda}
\ee
where
\be
B:= \sum_{i, j} w_{ij} B^{ij}.
\nonumber
\ee
\\
\indent
Consider the case $n=2$ first. Then there exists a real constant $c$ for which the model $(\sM, \xi)$ is represented as \eqref{prop_qubit_e-autoparallel_Q} and \eqref{prop_qubit_e-autoparallel_r} in the Bloch ball.  
In the present situation, this means that
\be
\any \rho \in\sM, \;\;  \expect{\sigma_3}_\rho = c \sqrt{1 - \| \xi(\rho)\|^2}. 
\nonumber
\ee
Representing $B$ as
\be
B = b_0 + \sum_{k=1}^3 b_k\, \sigma_k 
\label{B_b_sigma}
\ee
and noting that
\be
\lambda_1 (\rho) \lambda_2 (\rho) 
= \det (\wgt G_\rho^{-1})  = (\det \wgt) ( 1 - \| \xi(\rho)\|^2), 
\nonumber
\ee
\eqref{B_W_lambda} is expressed as
\begin{align}
\any \rho\in\sM, \;\; & b_0 + \sum_{i=1}^2 b_i \xi^i (\rho) + b_3 c \sqrt{1 - \| \xi(\rho)\|^2} 
\noret 
& = \tr W + 2 \sqrt{\det \wgt}  \sqrt{1 - \| \xi(\rho)\|^2} ,
\nonumber
\end{align}
which implies that
\be
b_0 = \tr \wgt, \;\; b_1 = b_2 = 0, \;\; \text{and}\;\;b_3 c = 2 \sqrt{\det \wgt}.
\label{cond_b}
\ee
If $c=0$, the third condition of \eqref{cond_b} leads to a contradiction. So we assume $c\neq 0$, which yields
\be
B = \tr \wgt + \frac{2\sqrt{\det \wgt}}{c} \sigma_3. 
\label{B_trW_sigma} 
\ee
On the other hand, letting $A^i$ be represented as 
\be
A^i = a_0^i + \sum_{k=1}^3 a_k^i \sigma_k, 
\nonumber
\ee
the unbiasedness condition \eqref{expect_A_xi} is expressed as
\be
\any \rho\in \sM, \;\; a_0^i + \sum_{j=1}^2 a_j^i \xi^j (\rho) + a_3^i c \sqrt{1-\|\xi (\rho)\|^2} = \xi^i (\rho), 
\nonumber
\ee
which implies that
\be
a_0^i =0, \;\; a_j^i = \delta_j^i, \;\; \text{and}\;\; a_3^i = 0, 
\nonumber
\ee
or equivalently, 
\be
A^i = \sigma_i.
\ee
From the operator inequality 
\begin{align}
0 & \leq\,  \sum_{i, j} w_{ij} \int (\xihat^i - A^i) \, \Pi (d \xihat)\,  (\xihat^j - A^j) = B - A, 
\nonumber
\end{align}
where 
\be
A := \sum_{i, j} w_{ij} A^i A^j = \sum_{i, j} w_{ij} \sigma_i \sigma_j = \tr \wgt, 
\nonumber
\ee
we have
\be
B \geq  \tr \wgt. 
\ee
This contradicts with \eqref{B_trW_sigma}.
}

\red{
Consider the case $n=3$ next; i.e. $\sM = \sS$. 
Representing $B$ as \eqref{B_b_sigma}, condition \eqref{B_W_lambda} is expressed as
\be
\any \rho\in \sS, \;\; 
b_0 + \sum_{i=1}^3 b_i \xi^i (\rho) = \tr \wgt +  2 \sum_{i < j} \sqrt{\lambda_i (\rho) \lambda_j (\rho)}. 
\ee
This  implies that $\sum_{i < j} \sqrt{\lambda_i (\rho) \lambda_j (\rho)}$ is an affine function of the Stokes parameters $(\xi^i)$ of $\rho$, which is clearly impossible.  Hence, we arrive at a contradiction.
 } 
\end{appendices}

\end{document}